\documentclass[apj]{emulateapj}
\usepackage{apjfonts}

\begin{document}


\title{Bulges Of Nearby Galaxies With Spitzer: Scaling Relations in Pseudobulges and Classical Bulges}

\shorttitle{Structure of Bulges}
\shortauthors{Fisher \& Drory}


\author{David~B.~Fisher}
\affil{Department of Astronomy, The University of Texas at Austin,\\
  1 University Station C1400, Austin, Texas 78712}
\email{dbfisher@astro.as.utexas.edu}

\author{Niv~Drory}

\affil{Max-Planck-Institut f\"ur
  Extraterrestrische Physik, Giessenbachstra\ss e, 85748 Garching, Germany}

\slugcomment{Submitted to ApJ}


\begin{abstract}
  We investigate scaling relations of bulges using bulge-disk
  decompositions at 3.6~$\mu$m and present bulge classifications for
  173 E-Sd galaxies within 20~Mpc. Pseudobulges and classical bulges
  are identified using S\'ersic index, HST morphology, and star
  formation activity (traced by 8~$\mu$ emission). In the near-IR
  pseudobulges have $n_b<2$ and classical bulges have $n_b>2$, as
  found in the optical. S\'ersic index and morphology are essentially
  equivalent properties for bulge classification purposes.  We
  confirm, using a much more robust sample, that the S\'ersic index of
  pseudobulges is uncorrelated with other bulge structural properties,
  unlike for classical bulges and elliptical galaxies. Also, the
  half-light radius of pseudobulges is not correlated with any other
  bulge property.  We also find a new correlation between surface
  brightness and pseudobulge luminosity; pseudobulges become more
  luminous as they become more dense. Classical bulges follow the well
  known scaling relations between surface brightness, luminosity and
  half-light radius that are established by elliptical galaxies.  We
  show that those pseudobulges (as indicated by S\'ersic index and
  nuclear morphology) that have low specific star formation rates are
  very similar to models of galaxies in which both a pseudobulge and
  classical bulge exist. Therefore, pseudobulge identification that
  relies only on structural indicators is incomplete.  Our results,
  especially those on scaling relations, imply that pseudobulges are
  very different types of objects than elliptical galaxies.
\end{abstract}



\section{Introduction}\label{sec:intro}

A preponderance of observational evidence suggests that there are at
least two types of bulges. Reviews of evidence for a dichotomy of
bulges can be found in \cite{kk04}, \cite{athan05},
\cite{kormendyfisher2005}, and for a review of more recent literature
see \cite{combes2009}. The dichotomy in bulge properties can be
summarized as follows: many bulges have properties as described by
\cite{renzini99} of ``little elliptical galaxies surrounded by a
disk'', while other bulges are similar in many properties to disk
galaxies. We call those bulges that are similar to E-type galaxies
``classical bulges'', and those that are more like disk-galaxies are
referred to as ``pseudobulges.''

Many bulges in the nearby Universe are filled with young stars
\citep{peletier1996, gadotti2001,carollo2002,macarthur2003}, and many
bulges are gas rich \citep{regan2001bima, helfer2003,
  sheth2005,jogee2005}. Also, \cite{peletier1996} find that the ages
and stellar populations of bulges vary greatly from galaxy-to-galaxy;
however, the stellar population of the outer disk is an excellent
predictor of that of the bulge. \cite{fisher2006} shows that bulge
morphology is also a good predictor of star formation in the bulge. He
finds that pseudobulges are actively forming stars, and have ISM
properties that are like that of their outer disks. \cite{fdf2009}
study star formation in bulges and their outer disks using Spitzer and
GALEX data.  They show that pseudobulges are presently increasing the
bulge-to-total ratio of stellar light (here after $B/T$) via internal
star formation. If pseudobulges have similar star formation histories
as their outer disks, then the mean historic star formation rate (over
the past $\sim$10~Gyr) ought to be 1-3 times that of present day star
formation rate \citep{kennicutt94}.  The results of \cite{fdf2009}
thus indicate that present day star formation rates are high enough to
account for the entire stellar mass of most pseudobulges.  Also,
\cite{fdf2009} find positive correlations between bulge mass and star
formation rate density suggesting that long term {\em in situ} growth
may have formed pseudobulges. Furthermore, those pseudobulges with
highest star formation rate density are exclusively found in the most
massive disks, suggesting that pseudobulge growth is connected to
outer disk properties.  They also find that classical bulges, in contrast,
are not forming significant amounts of stars today. Furthermore, data
from \cite{regan2001bima,helfer2003} show that classical bulges are
gas poor when compared to their outer disk.

The shape of stellar density profiles is also thought to participate
in the bulge dichotomy.  \cite{andredak95} and, more recently,
\citealp{scarlata2004} show that the distribution of surface
brightness profile shapes of bulges is bimodal.  Also,
\cite{carollo1999} finds that many of the the 'amorphous nuclei' of
intermediate-type disk galaxies are better fit by exponential
profiles, rather than the traditional $r^{1/4}$ profile. This leads
\cite{kk04}, among others, to suggest that surface brightness is tied
to the pseudobulge -- classical bulge dichotomy.
\cite{fisherdrory2008} directly test this hypothesis. They identify
pseudobulges with bulge morphology.  Using $V$-band bulge-disk
decompositions they find that that 90\% of pseudobulges have $n<2$ and
all classical bulges have $n>2$.  \cite{fisherdrory2008} also find
that the S\'ersic index of pseudobulges does not correlate with bulge
luminosity, half-light radius or bulge-to-total ratio as it does for
classical bulges and elliptical galaxies.

It is well known that elliptical galaxies follow a ``fundamental
plane'' in parameter space that relates size, surface density, and
velocity dispersion
\citep[e.g.][]{djorgovski1987,dressler1987,faber1989}.  To lowest
order, these correlations are a consequence of the virial theorem;
small deviations from virial predictions in slopes of these
correlations represent variation in mass-to-light ratios and
non-homologous density profiles. Disks correlate differently than
ellipticals in fundamental plane parameter space
\citep{bbf92,kfcb}. Yet, the location of pseudobulges compared to that
of classical bulges, elliptical galaxies, and disks in structural
parameter correlations is poorly understood. Thus, knowing where
pseudobulges lie with respect to fundamental plane correlations would
help in interpreting their structural properties.

\cite{graham2001},\cite{macarthur2003}, and \cite{dejong2004} each
investigate the scaling relations of bulges. All find parameter
correlations with high scatter.  The \cite{sersic1968} function is
highly degenerate \citep{graham1997}, and the dynamic range available
to fit bulges is limited. Indeed, \cite{balcells2003} and
\cite{gadotti2008} show that high spatial resolution is necessary to
accurately determine fit parameters when using the S\'ersic
function. Accordingly, \cite{balcells2007} and \cite{fisherdrory2008}
use composite HST and ground-based data to calculate bulge-disk
decompositions. The central isophotes are calculated with HST data,
providing adequate spatial resolution, and outer isophotes are
calculated with ground based data. However, this method is more labor
intensive, and the the resulting data set is smaller.  They find a
high probability of correlation among many structural parameters of
bulges. Nonetheless, the scatter remains high.

\cite{carollo1999} find that exponential bulges are systematically
lower in effective surface brightness than those better fit by an
$r^{1/4}$-profile. \cite{falcon2002} find that bulges which deviate
from the edge-on projection of the fundamental plane are found in late
type (Sbc) galaxies, and it is generally assumed that pseudobulges are
more common in late-type galaxies \citep{kk04}.  \cite{gadotti2009}
finds a large population of pseudobulges that are much fainter in
surface brightness for a given half-light radius than predicted by a
correlation fit to elliptical galaxies.

There is a great variety of properties which participate in the
dichotomy of bulges and motivate the association of pseudobulges with
disk-like objects. \citet{kk04} suggest that kinematics dominated
by rotation; flattening similar to that of their outer disk; nuclear
bar, nuclear ring and/or nuclear spiral; near-exponential surface
brightness profiles are all features of pseudobulges and not classical
bulges. \cite{fisherdrory2008} use morphology to identify
bulge-type and find that morphology correlates with S\'ersic index.
\cite{gadotti2009} uses the position in $\mu_e-r_e$ parameter
space to study the distribution of bulge properties from SDSS data.

Yet, there is no reason to think that all pseudobulges must have a
small S\'ersic index, nor is it necessary that no pseudobulges overlap
in $\mu_e-r_e$ parameter space with classical bulges. In this paper,
we will identify pseudobulges using multiple methods including
morphology, star formation, and structural properties. We will present
a quantitative prescription for identifying pseudobulges using these
multiple methods, and apply that method to all non-edge on bulge-disk
galaxies within 20~Mpc that have been observed by the Spitzer Space
Telescope (SST). We will also use this more robust method for
identifying bulge types to study the behavior of pseudobulges in
photometric projections of the fundamental plane.

\section{Sample}

Our aim in this paper is to investigate differences in parameter
correlations between pseudobulges and classical bulges.  In this
section, we outline our sample selection process and investigate
properties of that sample.  We use data from SST
to perform bulge-disk decompositions at 3.6~$\mu$m, where the data are
less affected by gradients in dust and stellar populations than in the
optical bands. Also, we use 8~$\mu$m mid-IR data to estimate star
formation rates. Although star formation rates estimated this way are
less reliable than using, e.g., the 24~$\mu$m band, the increase in
spatial resolution allows us to isolate star formation in smaller
bulges and we feel that the benefits from increase in dynamic range
and sample size offsets the loss in star formation rates accuracy.  We
can thus revisit correlations in previous optical work
\citep[e.g.][]{fisherdrory2008}, and those with significantly smaller
samples \citep[e.g.][]{fdf2009}. To minimize effects due to selection
we analyze all bulge-disk galaxies in both the RC3 and Spitzer archive
sufficiently nearby to accurately decompose the bulge.

\subsection{Sample Selection}

Many studies have found that bulges are never smaller than a few
hundred parsecs \citep[e.g.][]{graham2001,balcells2007,fdf2009}. Thus,
we set a distance limit of 20~Mpc such that all bulges in our sample
will be resolved to $r\lesssim150$~pc with IRAC on SST.

We cross reference the RC3 \citep{rc3} with the Spitzer Archive. We
select all galaxies between Hubble types S0 and Sd (as given in RC3)
available in the SST archive within our distance
limit (20~Mpc). We do not analyze bulges later type than Sd because
bulges are extremely rare in Sdm - Irr galaxies

We restrict our sample to exclude significantly inclined galaxies.  We
only keep galaxies that satisfy $R25<0.5$. ($R25$ is the mean decimal
logarithm of the ratio of the major isophotal diameter at isophotal
diameter, $D25$, to that of the minor axis.) This corresponds to an
inclination of about 75$\degr$ for an Sc spiral. This cut yields 331
galaxies.  We visually identify 9 galaxies that are highly inclined,
but not excluded by the $R25$ criterion because of a very large
bulge-to-total ratio.  These galaxies are removed from the sample.

We only select galaxies that are free of tidal-tails, warps and
asymmetries in IRAC1 images. This is done to attempt to reduce the
number of galaxies significantly affected by interactions or mergers
with other galaxies. For each galaxy, we also consider comments in the
Carnegie Atlas of Galaxies \citep{carnegieatlas}; galaxies described
as 'amorphous' and/or 'peculiar' are scrutinized for evidence of
recent or ongoing interactions. We remove 34 galaxies due to peculiar
morphology or tidal tails.

The total list that meets all of the above selection criteria is 297
galaxies. Of these 297 galaxies, 146 have adequate data in the spitzer
archive. Note that a handful of galaxies have over-exposed centers,
or the galaxy is not well covered in the Spitzer image. Those galaxies
are not included in the 146 disk galaxy sample.

\begin{deluxetable}{lccccc}
  \tablewidth{0pt} \tablecaption{Sample Galaxy Properties}
  \tablehead{\colhead{Galaxy} & \colhead{Morphology\tablenotemark{a}} & \colhead{Distance\tablenotemark{b}} 
    & \colhead{M$_B$\tablenotemark{a}} & \colhead{$B-V$\tablenotemark{a}} & \colhead{log(D25)\tablenotemark{a}}\\
    \colhead{ } & \colhead { } & \colhead{[Mpc]} & \colhead{[mag]} &
    \colhead{[mag]} & \colhead{log([pc])}} 
\startdata
IC342 & .SXT6.. & 2.58 (1) & -17.96 & 1.10 & 3.90 \\
IC749 & .SXT6. & 13.24 (1) & -17.69 & 0.63 & 3.65 \\
NGC0300 & .SAS7.. & 2.69 (1) & -18.43 & 0.59 & 3.93 \\
NGC0404 & .LAS-*. & 3.27 (2) & -16.36 & 0.85 & 3.22 \\
NGC0628 & .SAS5.. & 11.14 (1) & -20.28 & 0.64 & 4.23 \\
NGC0672 & .SBS6.. & 7.52 (1) & -17.91 & 0.59 & 3.90 \\
NGC0925 & .SXS7.. & 10.30 (1) & -19.37 & 0.58 & 4.20 \\
NGC1023 & .LBT-.. & 11.53 (2) & -19.96 & 1.01 & 4.16 \\
NGC1058 & .SAT5.. & 10.30 (1) & -18.24 & 0.62 & 3.66 \\
NGC1097 & .SBS3.. & 17.82 (1) & -21.02 & 0.83 & 4.38 \\
NGC1313 & .SBS7.. & 3.99 (1) & -18.80 & 0.47 & 3.72 \\
NGC1317 & .SXR1..  & 18.41 (1) & -19.42 & 0.98 & 3.87 \\
NGC1433 & PSBR2.. & 11.41 (1) & -19.59 & 0.85 & 4.03 \\
NGC1512 & .SBR1.. & 11.41 (1) & -19.16 & 0.87 & 4.17 \\
NGC1543 & RLBS0.. & 13.78 (1) & -19.24 & 0.97 & 3.99 \\
NGC1559 & .SBS6.. & 13.91 (1) & -19.72 & 0.64 & 3.85 \\
NGC1566 & .SXS4.. & 13.78 (1) & -20.37 & 0.68 & 4.22 \\
NGC1617 & .SBS1.. & 13.78 (1) & -19.32 & 0.96 & 3.93 \\
NGC1637 & .SXT5.. & 8.40 (1) & -18.15 & 0.73 & 3.69 \\
NGC1672 & .SBS3.. & 17.82 (1) & -20.97 & 0.67 & 4.23 \\
NGC1744 & .SBS7.. & 14.14 (1) & -19.15 & 0.53 & 4.22 \\
NGC1808 & RSXS1.. & 11.39 (1) & -19.54 & 0.82 & 4.03 \\
NGC2403 & .SXS6.. & 3.31 (1) & -18.67 & 0.55 & 4.02 \\
NGC2500 & .SBT7.. & 7.49 (1) & -17.17 & 0.58 & 3.50 \\
NGC2655 & .SXS0..  & 20.54 (1) & -20.60 & 0.91 & 4.17 \\
NGC2685 & RLB.+P. & 13.45 (1) & -18.52 & 0.93 & 3.94 \\
NGC2775 & .SAR2.. & 14.42 (1) & -19.76 & 0.96 & 3.95 \\
NGC2841 & .SAR3*. & 8.96 (1) & -19.67 & 0.93 & 4.03 \\
NGC2903 & .SXT4.. & 8.16 (1) & -19.88 & 0.71 & 4.17 \\
NGC2950 & RLBR0.. & 20.62 (1) & -19.73 & 0.92 & 3.91 \\
NGC2964 & .SXR4*. & 19.90 (1) & -19.50 & 0.75 & 3.92 \\
NGC2976 & .SA.5P. & 3.31 (1) & -16.78 & 0.66 & 3.45 \\
NGC2997 & .SXT5.. & 10.98 (1) & -20.14 & 0.93 & 4.15 \\
NGC3031 & .SAS2.. & 3.91 (2) & -20.07 & 0.99 & 4.18 \\
NGC3032 & .LXR0.. & 19.66 (1) & -18.29 & 0.63 & 3.76 \\
NGC3156 & .L...*. & 14.82 (1) & -17.78 & 0.75 & 3.62 \\
NGC3184 & .SXT6.. & 9.11 (1) & -19.44 & 0.66 & 3.99 \\
NGC3185 & RSBR1.. & 16.53 (1) & -18.10 & 0.82 & 3.75 \\
NGC3190 & .SAS1P/ & 16.53 (1) & -18.97 & 0.98 & 4.02 \\
NGC3198 & .SBT5.. & 9.11 (1) & -18.93 & 0.62 & 4.05 \\
NGC3319 & .SBT6. & 9.11 (1) & -18.32 & 0.48 & 3.91 \\
NGC3344 & RSXR4. & 8.16 (1) & -19.11 & 0.67 & 3.92 \\
NGC3351 & .SBR3. & 8.57 (1) & -19.13 & 0.85 & 3.97 \\
NGC3368 & .SXT2. & 8.57 (1) & -19.55 & 0.94 & 3.98 \\
NGC3384 & .LBS-*. & 8.57 (1) & -18.81 & 0.95 & 3.84 \\
NGC3412 & .LBS0.. & 8.57 (1) & -18.21 & 0.92 & 3.66 \\
NGC3486 & .SXR5. & 8.62 (1) & -18.63 & 0.61 & 3.95 \\
NGC3489 & .LXT+.. & 8.57 (1) & -18.54 & 0.84 & 3.65 \\
NGC3511 & .SAS5. & 12.32 (1) & -18.92 & 0.59 & 4.01 \\
NGC3521 & .SXT4. & 8.15 (1) & -19.73 & 0.88 & 4.11 \\
NGC3593 & .SAS0*.  & 8.76 (1) & -17.85 & 0.98 & 3.83 \\
NGC3621 & .SAS7. & 7.88 (1) & -19.20 & 0.62 & 4.15 \\
NGC3675 & .SAS3. & 10.68 (1) & -19.14 & 0.94 & 3.96 \\
NGC3726 & .SXR5. & 13.24 (1) & -19.70 & 0.49 & 4.07 \\
NGC3906 & .SBS7. & 19.30 (3) & -17.94 & ... & 3.72 \\
NGC3938 & .SAS5. & 13.24 (1) & -19.71 & 0.62 & 4.01 \\
NGC3941 & .LBS0.. & 12.85 (1) & -19.29 & 0.93 & 3.81 \\
NGC3945 & RLBT+.. & 18.96 (1) & -19.59 & 0.97 & 4.16 \\
NGC3953 & .SBR4. & 13.24 (1) & -19.77 & 0.86 & 4.12 \\
NGC3982 & .SXR3* & 13.24 (1) & -18.84 & ... & 3.65 \\
NGC3990 & .L..-*/ & 13.24 (1) & -17.18 & 0.91 & 3.43 \\
NGC4020 & .SB.7? & 11.15 (1) & -16.85 & ... & 3.53 \\
NGC4117 & .L..0*. & 12.10 (3) & -16.31 & ... & 3.50 \\
NGC4136 & .SXR5. & 12.48 (1) & -18.43 & ... & 3.86 \\
NGC4138 & .LAR+.. & 13.24 (1) & -18.45 & 0.86 & 3.69 \\
NGC4150 & .LAR0\$. & 13.74 (1) & -18.25 & 0.82 & 3.67 \\
NGC4203 & .LX.-*. & 15.14 (2) & -19.10 & 0.90 & 3.87 \\
NGC4237 & .SXT4. & 14.28 (1) & -18.26 & 0.87 & 3.64 \\
NGC4254 & .SAS5..  & 14.28 (1) & -20.33 & 0.65 & 4.05 \\
NGC4258 & .SXS4. & 8.17 (1) & -20.46 & 0.77 & 4.34 \\
NGC4267 & .LBS-\$. & 14.28 (1) & -18.91 & 0.97 & 3.83 \\
NGC4274 & RSBR2. & 12.48 (1) & -19.14 & 0.97 & 4.09 \\
NGC4293 & RSBS0..  & 14.63 (1) & -19.57 & 0.94 & 4.08 \\
NGC4294 & .SBS6. & 14.28 (1) & -18.24 & 0.51 & 3.83 \\
\enddata

 \tablenotetext{a}{Taken from RC3}
 \tablenotetext{b}{Distance References: (1)- \cite{tully1998} (2)- \cite{tonry2001} (3)-\cite{rc3} }

 \end{deluxetable}

\begin{deluxetable}{lccccc}
\setcounter{table}{1}
  \tablewidth{0pt} \tablecaption{Table 1 Continued}
  \tablehead{\colhead{Galaxy} & \colhead{Morphology\tablenotemark{a}} & \colhead{Distance\tablenotemark{b}} 
    & \colhead{M$_B$\tablenotemark{a}} & \colhead{$B-V$\tablenotemark{a}} & \colhead{log(D25)\tablenotemark{a}}\\
    \colhead{ } & \colhead { } & \colhead{[Mpc]} & \colhead{[mag]} &
    \colhead{[mag]} & \colhead{log([pc])}} 
\startdata
NGC4303 & .SXT4..  & 19.77 (1) & -21.30 & 0.63 & 4.27 \\
NGC4314 & .SBT1. & 12.48 (1) & -19.05 & 0.87 & 3.88 \\
NGC4321 & .SXS4..  & 14.28 (1) & -20.72 & 0.71 & 4.19 \\
NGC4371 & .LBR+.. & 14.28 (1) & -18.98 & 0.99 & 3.92 \\
NGC4380 & .SAT3* & 14.28 (1) & -18.11 & 0.89 & 3.86 \\
NGC4394 & RSBR3. & 14.28 (1) & -19.04 & 0.90 & 3.88 \\
NGC4413 & PSBT2* & 14.28 (1) & -18.52 & 0.66 & 3.69 \\
NGC4414 & .SAT5\$ & 12.48 (1) & -19.52 & 0.87 & 3.82 \\
NGC4419 & .SBS1. & 14.28 (1) & -18.69 & 0.95 & 3.84 \\
NGC4421 & .SBS0.. & 14.28 (1) & -18.30 & 0.89 & 3.75 \\
NGC4424 & .SBS1* & 14.28 (1) & -18.43 & 0.69 & 3.88 \\
NGC4442 & .LBS0.. & 14.28 (1) & -19.39 & 0.97 & 3.98 \\
NGC4448 & .SBR2. & 12.48 (1) & -18.48 & 0.88 & 3.85 \\
NGC4450 & .SAS2..  & 14.28 (1) & -19.87 & 0.90 & 4.04 \\
NGC4457 & RSXS0.. & 10.22 (1) & -18.29 & 0.88 & 3.60 \\
NGC4491 & .SBS1* & 5.90 (3) & -15.35 & 0.79 & 3.16 \\
NGC4498 & .SXS7..  & 14.28 (1) & -17.98 & ... & 3.79 \\
NGC4501 & .SAT3..  & 14.28 (1) & -20.41 & 0.87 & 4.16 \\
NGC4519 & .SBT7. & 14.28 (1) & -18.43 & 0.52 & 3.82 \\
NGC4526 & .LXS0*. & 14.28 (1) & -20.11 & 0.97 & 4.18 \\
NGC4548 & .SBT3. & 19.23 (2) & -20.46 & 0.89 & 4.18 \\
NGC4559 & .SXT6. & 12.48 (1) & -20.02 & 0.48 & 4.29 \\
NGC4569 & .SXT2. & 14.28 (1) & -20.51 & 0.79 & 4.30 \\
NGC4571 & .SAR7. & 14.28 (1) & -18.95 & 0.51 & 3.88 \\
NGC4578 & .LAR0*. & 18.54 (2) & -18.96 & 0.92 & 3.95 \\
NGC4580 & .SXT1P & 14.63 (1) & -19.00 & 0.83 & 3.65 \\
NGC4605 & .SBS5P & 5.53 (1) & -17.82 & 0.52 & 3.67 \\
NGC4612 & RLX.0.. & 14.28 (1) & -18.87 & 0.89 & 3.71 \\
NGC4639 & .SXT4. & 16.63 (1) & -18.86 & 0.78 & 3.82 \\
NGC4651 & .SAT5. & 14.28 (1) & -19.38 & 0.75 & 3.92 \\
NGC4654 & .SXT6. & 14.28 (1) & -19.67 & 0.65 & 4.01 \\
NGC4688 & .SBS6. & 10.28 (1) & -16.55 & ... & 3.67 \\
NGC4689 & .SAT4.. & 14.28 (1) & -19.17 & 0.65 & 3.95 \\
NGC4698 & .SAS2. & 14.28 (1) & -19.31 & 0.94 & 3.92 \\
NGC4701 & .SAS1. & 10.22 (1) & -17.25 & 0.52 & 3.61 \\
NGC4713 & .SXT7. & 17.91 (1) & -19.08 & 0.45 & 3.85 \\
NGC4725 & .SXR2P & 13.24 (1) & -20.50 & 0.90 & 4.31 \\
NGC4736 & RSAR2. & 4.23 (1) & -19.14 & 0.83 & 3.84 \\
NGC4808 & .SAS6* & 10.22 (1) & -17.70 & 0.68 & 3.61 \\
NGC4941 & RSXR2* & 5.53 (1) & -16.81 & 0.92 & 3.47 \\
NGC4984 & RLXT+.. & 16.20 (1) & -18.80 & 0.91 & 3.81 \\
NGC5005 & .SXT4. & 13.29 (1) & -20.01 & 0.75 & 4.05 \\
NGC5033 & .SAS5. & 13.29 (1) & -19.87 & 0.75 & 4.32 \\
NGC5055 & .SAT4. & 7.83 (1) & -20.16 & 0.80 & 4.16 \\
NGC5068 & .SXT6. & 6.99 (1) & -18.52 & 0.64 & 3.87 \\
NGC5128 & .L...P. & 4.21 (2) & -20.28 & 1.00 & 4.20 \\
NGC5194 & .SAS4P & 7.83 (1) & -20.51 & 0.69 & 4.11 \\
NGC5236 & .SXS5. & 7.83 (1) & -21.27 & 0.72 & 4.17 \\
NGC5248 & .SXT4. & 14.79 (1) & -19.88 & 0.65 & 4.12 \\
NGC5273 & .LAS0.. & 16.52 (2) & -18.65 & 0.86 & 3.82 \\
NGC5338 & .LB..*. & 10.79 (3) & -16.11 & ... & 3.60 \\
NGC5457 & .SXT6. & 5.03 (1) & -20.20 & 0.56 & 4.32 \\
NGC5474 & .SAS6P & 5.03 (1) & -17.23 & 0.53 & 3.54 \\
NGC5585 & .SXS7. & 5.03 (1) & -17.31 & 0.46 & 3.62 \\
NGC5643 & .SXT5. & 15.13 (1) & -20.16 & 0.81 & 4.00 \\
NGC5832 & .SBT3\$ & 12.51 (1) & -16.88 & 0.58 & 3.83 \\
NGC5879 & .SAT4* & 12.63 (1) & -18.29 & 0.71 & 3.88 \\
NGC5949 & .SAR4\$ & 12.51 (1) & -17.58 & ... & 3.61 \\
NGC6207 & .SAS5. & 14.88 (1) & -18.70 & 0.54 & 3.81 \\
NGC6300 & .SBT3. & 12.27 (1) & -19.46 & 0.86 & 3.90 \\
NGC6503 & .SAS6. & 5.53 (1) & -17.80 & 0.75 & 3.76 \\
NGC6684 & PLBS0.. & 10.06 (1) & -18.70 & 0.95 & 3.77 \\
NGC6744 & .SXR4. & 10.06 (1) & -20.87 & 0.86 & 4.47 \\
NGC7177 & .SXR3. & 19.75 (1) & -19.47 & 0.90 & 3.95 \\
NGC7217 & RSAR2. & 16.63 (1) & -20.08 & 0.98 & 3.97 \\
NGC7331 & .SAS3. & 15.50 (1) & -20.60 & 0.97 & 4.37 \\
NGC7457 & .LAT-\$. & 13.24 (2) & -18.52 & 0.90 & 3.91 \\
NGC7713 & .SBR7* & 9.32 (1) & -18.34 & 0.32 & 3.78 \\
NGC7741 & .SBS6. & 13.81 (1) & -18.86 & 0.56 & 3.94 \\
NGC7793 & .SAS7. & 2.69 (1) & -17.52 & 0.51 & 3.56 \\
UGC10445 & .S..6? & 12.90 (1) & -16.52 & 0.63 & 3.71 \\
\enddata

 \tablenotetext{a}{Taken from RC3}
 \tablenotetext{b}{Distance References: (1)- \cite{tully1998} (2)- \cite{tonry2001} (3)-\cite{rc3} }

 \end{deluxetable}

\subsection{The Control Set: Virgo Cluster Ellipticals}

We also wish to compare the location of pseudobulges and classical
bulges in parameter space to that of elliptical galaxies. We therefore
include a set of elliptical galaxies as a control sample. A problem
for calculating S\'ersic fits to elliptical galaxies is that resulting
parameters are highly dependent on both spatial resolution (for low
luminosity elliptical galaxies) and field-of-view (for high luminosity
ellipticals). Yet, near-IR HST archival data is not necessarily
available, and typical data in the Spitzer archive do not always cover
the entire galaxy. Furthermore, to assure an unbiased comparison, we wish to
have a complete set of elliptical galaxies.  

We use the elliptical galaxy sample and data from
\cite{kfcb}. \cite{kfcb} determines surface photometry for all
elliptical galaxies in the Virgo cluster. They construct $V$ band
surface brightness profiles of composite data sets of HST and ground
based data.  To reduce the effects of sky subtraction errors and
smearing due to point spread functions, \cite{kfcb} capitalized on the
wealth of published data available on Virgo ellipticals. Therefore
galaxies in their sample tend to have many different data sources
which are averaged together. Also, they add extremely deep new
observations from the McDonald 0.8~m telescope, frequently reaching
30~$V$~mag~arcsec$^{-2}$.

We use the half-light radius, $r_e$, and S\'ersic index, $n$, directly
from the fits in \cite{kfcb}.  For luminosity and surface-brightness
we shift the measurement in \cite{kfcb} to account for the $V-L$
color, $L$ designating the magnitude in IRAC channel~1.  This carries
an implicit assumption that color gradients in elliptical galaxies
average to zero. Though not exactly true, this assumption is supported
by relatively flat $g-z$ profiles in \cite{kfcb}. If a galaxy has
3.6~$\mu$m data in the Spitzer archive, we calculate the luminosity at
3.6~$\mu$m, and we integrate the $V$-band profile to the maximum
isophotal radius observed in the 3.6~$\mu$m profile; this yields $V-L$
for that galaxy. For those elliptical galaxies that do not have data
in the Spitzer archive, we use the average $V-L$ of the rest of the
elliptical galaxies in the sample. The root-mean-squared deviation of
$V-L$ about the mean is 0.5~mags. Therefore, differences of less than
half a magnitude will be considered less significant in our analysis.
It is well known that fainter elliptical galaxies are bluer,
however \cite{bower92} find that the correction is roughly $\pm$0.2
mag, which is small enough to be negligible for our purposes.  The
elliptical galaxy sample consists of 27 galaxies, therefore the total
sample in this paper is increased to 173 galaxies.

\begin{figure}[t!]
\includegraphics[width=.49\textwidth]{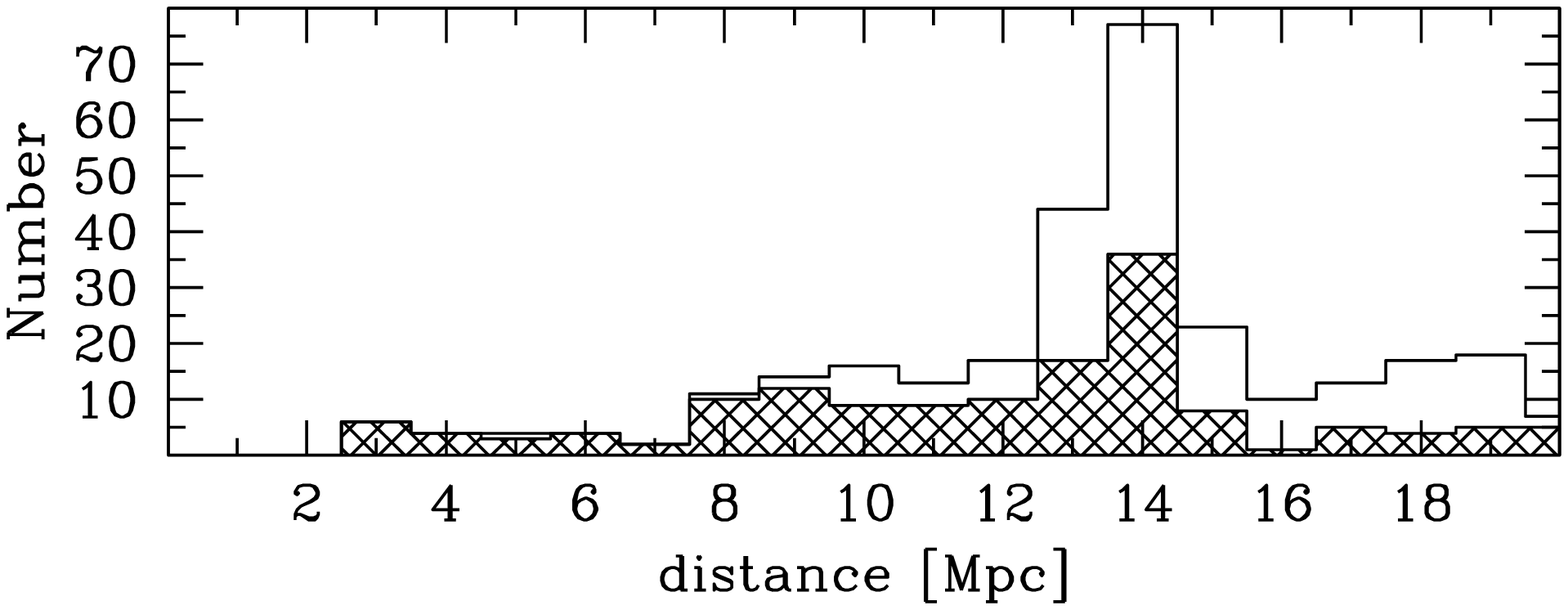}$\\$
\includegraphics[width=.49\textwidth]{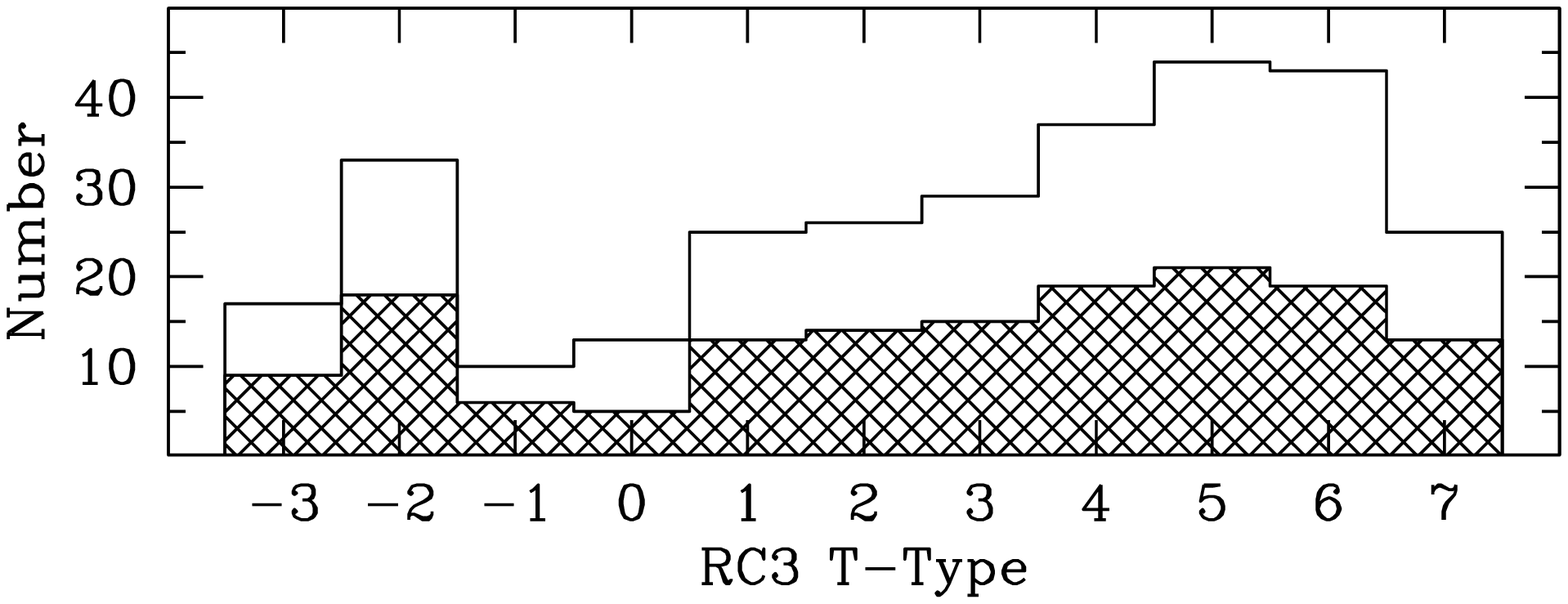}$\\$
\includegraphics[width=.49\textwidth]{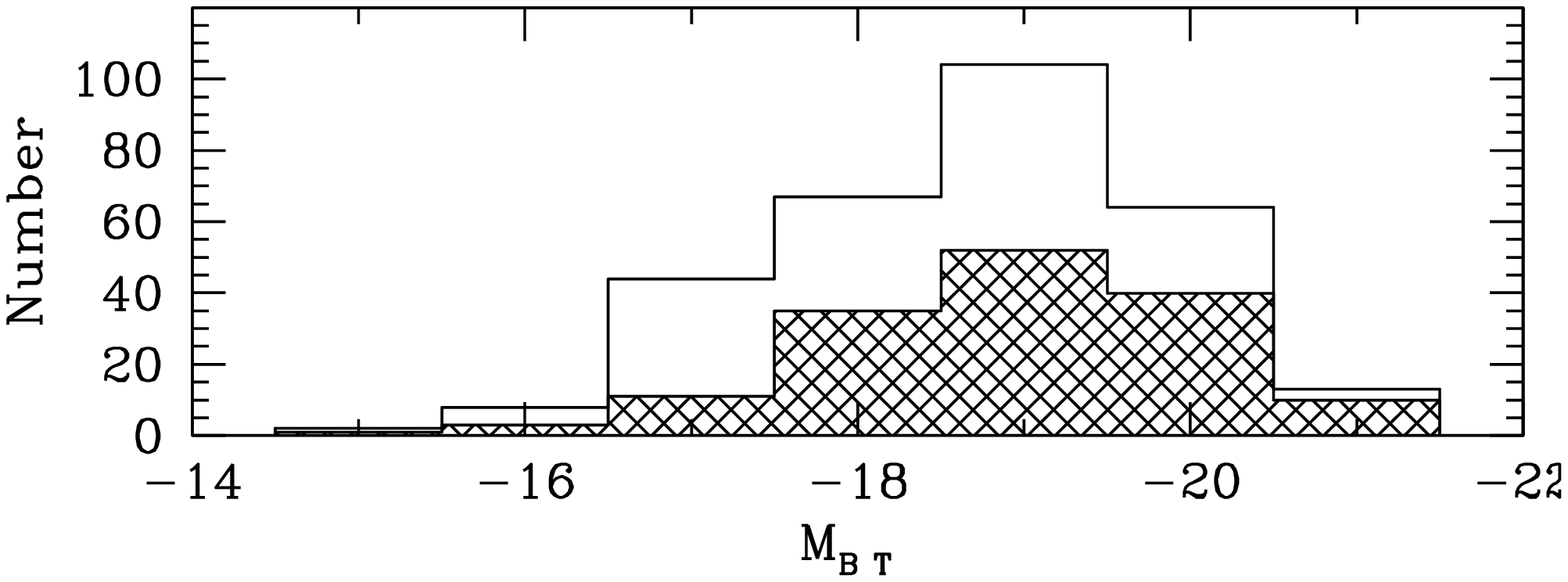}$\\$
\includegraphics[width=.49\textwidth]{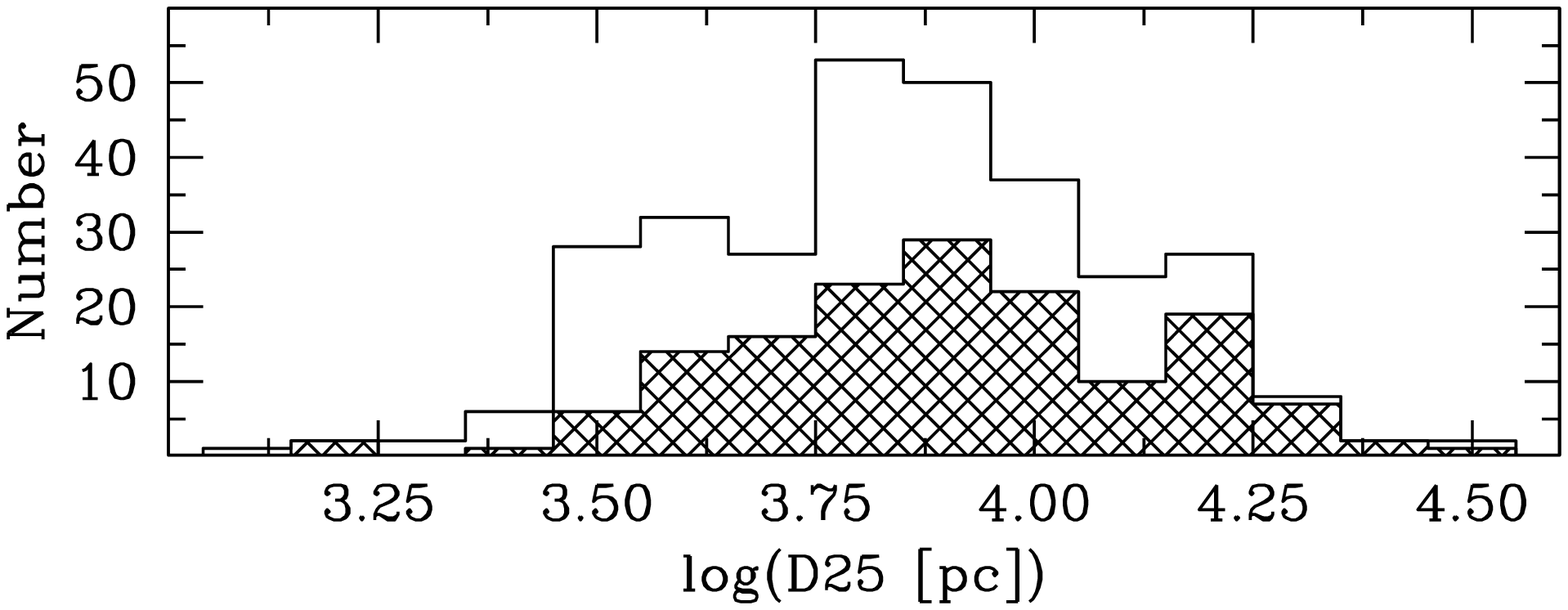}
\caption{Histograms (from top to bottom) of
  distance, T-type from RC3, absolute B$_T$ magnitude, and the
  radial size of the $\mu_B=25$~mag~arcsec$^{-2}$. The open histograms
  represent the complete RC3 data, the shaded histograms represent
  our sample having Spitzer data. \label{fig:sample_props}}
\end{figure}

\subsection{Sample Properties and Completeness}

Table~1 lists the properties of our sample galaxies. In
Fig.~\ref{fig:sample_props}, we compare our sample properties to those
of the complete RC3. (Note that at our distance limit the RC3 is
complete to M$_B\lesssim -16.5$ B-mag.)  From top to bottom, the
properties shown are distance, RC3 T-type, absolute $B_T$ magnitude,
and the physical radius in parsecs of the isophote at
25~B-mags~arcsec$^{-2}$.

Our sample is complete within 9~Mpc, and 75\% complete within
13~Mpc. Beyond 13~Mpc the Spitzer coverage becomes very
incomplete. Overall, the available data in the Spitzer archive yields
the inclusion of half the galaxies that meet our selection criteria in
the RC3.

Hubble type correlates with $B/T$ \citep{simien1986}, and thus
provides an important tool for identifying biases among bulge samples. The
second panel from the top in Fig.~\ref{fig:sample_props} shows the
distribution of T-types in our sample. Our sample distribution appears
very similar to that of the complete RC3.  Our  selection
is fairly independent of Hubble type, with a dip at S0/a
($T=0$). There is a slight bias in the sense that more galaxies are
missing at later Hubble types Sbc-Scd ($T=4-6$).

We are missing more galaxies that are smaller. Comparing our sample to
the RC3 we find that our sample includes 66\% of galaxies larger than
one standard deviation from the mean D25, 52\% of those within one
standard deviation, and only 29\% of galaxies below one standard
deviation. However, our sample includes galaxies in the extremes of
the distribution. For the purpose of this paper, our sample adequately
covers the properties of the complete RC3 distribution.

\subsection{Data Sources and the Higher Quality Data Set}

Including the entire Spitzer archive of bulge-disk galaxies out to
20~Mpc means that our data come from a variety of campaigns, each with
separate goals and often differing observational strategies. A few
notable surveys comprise of the vast majority of galaxies in our
sample: the SINGS survey \citep{sings}, the nearby bulge survey
\citep{fdf2009}, and the local volume survey \citep{localvolume}.

For the analysis in subsequent sections, we define a subset of our
sample of higher-quality data.  We do this so that we can make sure
that bulges with low signal-to-noise or bright Seyfert nuclei do not
cause us to miss correlations that would be seen with better data. The
criteria we use for creating the high-quality data sample are as
follows:

1. - Availability of optical images from HST for morphological
classification, as described below.  Of the complete sample, 24
galaxies do not have optical HST images.

2. - $B/T \geqq 1$\% and radius at which the S\'ersic bulge and the
exponential disk equal in surface brightness, $r_{b=d}$, be greater
than twice the PSF of the surface brightness profile. In this cut, 19
galaxies are removed.

3. - Finally we remove any galaxies with a prominent active galactic
nucleus (AGN); this is to ensure that the mid-IR color is not skewed
by non-thermal emission. Most of these galaxies do not make it into
the sample to begin with, because the AGN typically results in
overexposed IRAC images. To identify the remaining AGN, we use the
same method used by \cite{fdf2009}. We measure the cuspiness of the
8~$\mu$m profile by measuring the ratio of the flux within 2
arcseconds (200~pc at 20~Mpc) to that of the entire galaxy. We only
include those galaxies with flux ratio less than 15\%. Of the
remaining sample, 10 have $L_8(r<2")/L_8(total)<15$\%, leaving the
clean sample at 94 galaxies. We note that this criterion may bias our
sample against extremely cuspy and compact bulges. However, galactic
nuclei contain less than 1\% of galaxy stellar light
\citep{phillips1996}. An object containing more than 15\% of the
stellar light within 200~pc lies outside any known parameter
correlation of galactic components.

\section{Methods}

\subsection{Calculation Of Surface Brightness Profiles}\label{sec:sbprof}

We use post basic calibrated data images taken on the Spitzer Space
Telescope with IRAC channel~1, centered at 3.6~$\mu$m, to measure the
light profiles of galaxies in our sample. When available, we also use
HST NICMOS data, typically in the F160W filter.

We calculate sky values by fitting a 2$^{nd}$ order surface to the
areas of the image not covered by galaxy light. Then we subtract the
surface from the image. In some cases there is not enough area on the
image that is not affected by galaxy light to constrain the surface
fit. In these cases, we take a mean value in an unaffected region of
the image, and then we compare the resulting profile from the
3.6~$\mu$m with one calculated on 2MASS \citep{2mass} $H$-band images,
which allow for more accurate sky subtraction. If the 3.6~$\mu$m
profile is significantly different, we determine the sky again using a
different region of the image. At 3.6~$\mu$m the sky values are very
near zero, and typically of not much concern. For the 8.0~$\mu$m
channel, ``sky'' (which is most likely diffuse emission from the Milky
Way) is more important. Typical uncertainties from sky subtraction are
0.3~mag for the IRAC~CH1 and 0.1~mag for IRAC~CH4.

We use the isophote fitting routine of \cite{bender1987} to generate ellipse fits. First,
interfering foreground objects are identified in each image and masked
manually. Then, isophotes are sampled by 256 points equally spaced in
an angle $\theta$ relating to polar angle by $\tan \theta = a/b\,\tan
\phi$, where $\phi$ is the polar angle and $b / a$ is the axial ratio.
An ellipse is then fitted to each isophote by least squares. The
software determines six parameters for each ellipse: relative surface
brightness, center position, major and minor axis lengths, and
position angle along the major axis. We combine HST and SST data by
shifting the zero-point of the NICMOS data to match the IRAC
profile. The two profiles are then averaged, to construct a composite
HST+SST IR profile.

\subsection{S\'ersic Fitting} 

We determine bulge and disk parameters by fitting each
surface brightness profile with a one-dimensional S\'ersic function
plus an exponential outer disk,
\begin{equation}
I(r)=I_0\exp\left[-(r/r_0)^{1/n_b} \right ] + I_d\exp\left[-(r/h) \right ]\, ,
\end{equation}
where $I_0$ and $r_0$ represent the central surface brightness and
scale length of the bulge, $I_d$ and $h$ represent the central surface
brightness and scale length of the outer disk, and $n_b$ represents
the bulge S\'ersic index. The half-light radius, $r_e$, of the bulge is
obtained by converting $r_0$,
\begin{equation}
r_e=(b_n)^nr_0,
\end{equation}
where the value of $b_n$ is a proportionality constant defined such
that $\Gamma (2n)=2\gamma(2n,b_n)$. $\Gamma$ a\& $\gamma$ are the
complete and incomplete gamma functions, respectively. We use the
approximation $b_n\approx2.17n_b-0.355$. We restrict our range in possible S\'ersic indices to $n_b>0.33$ to ensure that the approximation is accurate. 
 A more precise expansion is
given in \citealp{macarthur2003}. The average surface-brightness within the
half-light radius, $<\mu_e>$, is given by
\begin{equation}
<\mu_e>=m_{3.6} + 2.5\log(\pi r_e^2)
\end{equation}
Where $m_{3.6}$ is the magnitude in 3.6~$\mu$m. We adjust all
luminosities by the average ellipticity in the region in which that
parameter dominates the light. (Also, see \citealp{graham2005} for a
review of the properties of the S\'ersic function.)

Despite its successes, the S\'ersic bulge plus outer exponential disk
model of bulge-disk galaxies does not account for many features of
galaxy surface brightness profiles. Disks of intermediate type
galaxies commonly have features such as bars, rings, and lenses (see
\citealp{k82} for a description of these). Further, \cite{carollo2002}
show that many bulges of early and intermediate-type galaxies contain
nuclei. Bars, rings, lenses, and similar features do not conform to
the smooth nature of Eq.~1, hence we carefully exclude regions of the
profile perturbed by such structures from the fit.  This is a
subjective procedure, as it requires selectively removing data from a
galaxy's profile, and undoubtedly has an effect on the resulting
parameters.  We are often helped to identify bars by the structure of
the ellipticity profile, as described in \cite{jogee2004} and also
\cite{marinova2007}. For a detailed description of this procedure, and
analysis of its effects see the appendix of
\cite{fisherdrory2008}. They find that not accounting for bars can
have the affect of increasing $n_b$ by at most 0.5. For those galaxies
in which a bar is present, it is our assumption that removing the bar
from the fit provides the best estimation of the properties of the
underlying bulge and disk.  All of our fits are given in tabular form
in Table A1 in the appendix A2, and also shown as figures in appendix
A2.

\subsection{Non-parametric Determination of Bulge Luminosity}

In addition to S\'ersic fitting we determine the bulge luminosity by a
non-parametric method to obtain a second estimate of bulge luminosity
and surface density.  To isolate the light from the bulge, we must
subtract underlying disk light. For this purpose, we fit an
exponential profile $I_{exp}(r)=I_0exp(-r/h)$ to the major-axis
surface brightness profile of the outer part of the galaxy. The range
included in the exponential fit is determined iteratively. Typically,
this exponential fit is determined on isophotes with radii larger than
the radius which includes 20\% of the light. Bars are masked as usual.
Note this fit is independent of the fit described in Eq.~1. To account
for the variable ellipticity profiles of these disk galaxies we take
the mean ellipticity for each component and adjust the luminosity
accordingly: $L_{disk}=(1-\bar{\epsilon})L_{\mathrm{exp}}$. We then
calculate the inward extrapolation of the fit to the central (or
bulge) region; thus $L_{bulge}=L(r<R_{XS}) - L_{disk}(r<R_{XS})$ where
$R_{XS}$ is the radius containing the light that is in excess of the
inward extrapolation of the outer exponential.

\subsection{The Effects of Including NICMOS F160W Data}

\begin{figure}[t]
\includegraphics[width=.4\textwidth]{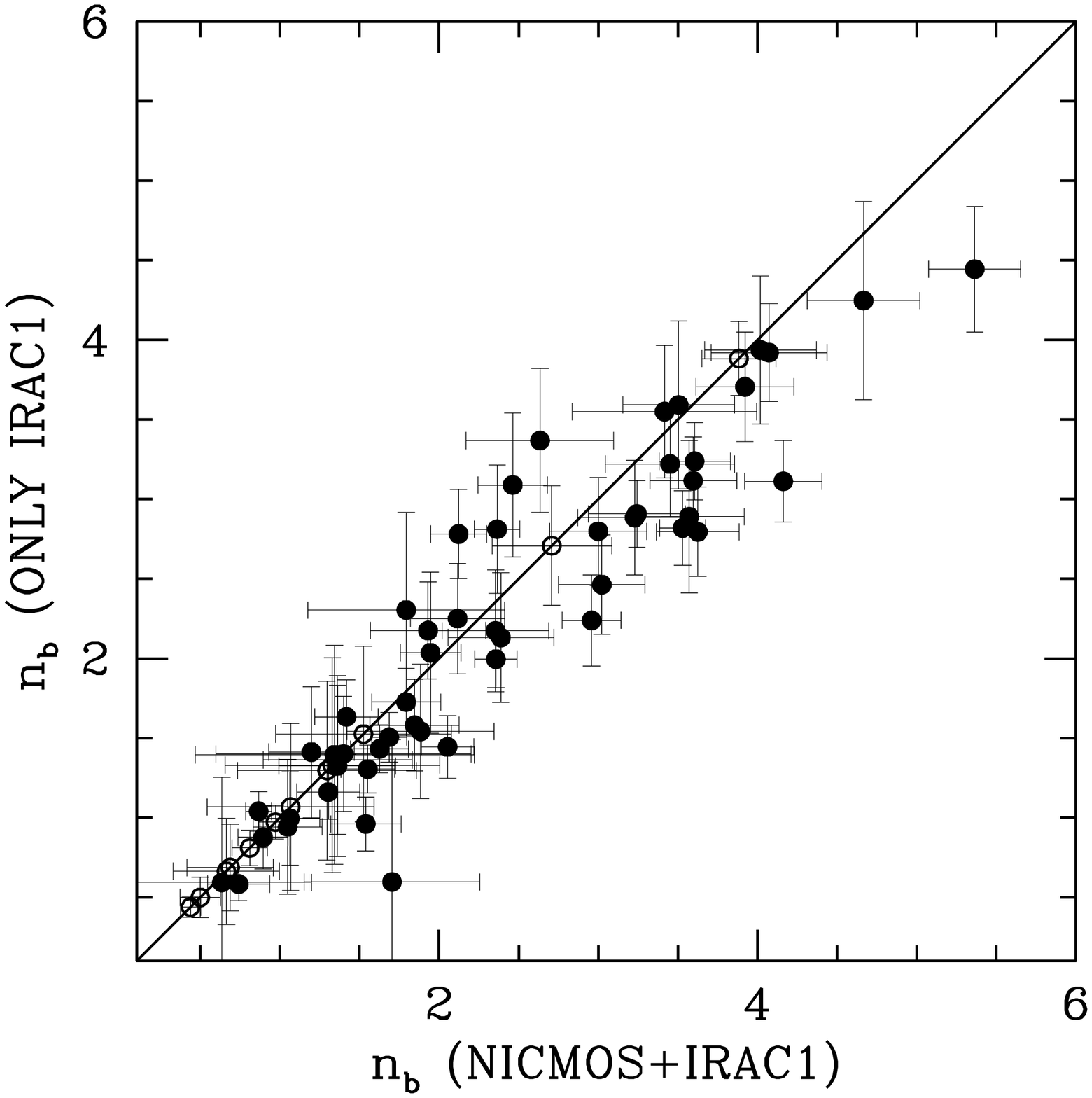}
\includegraphics[width=.4\textwidth]{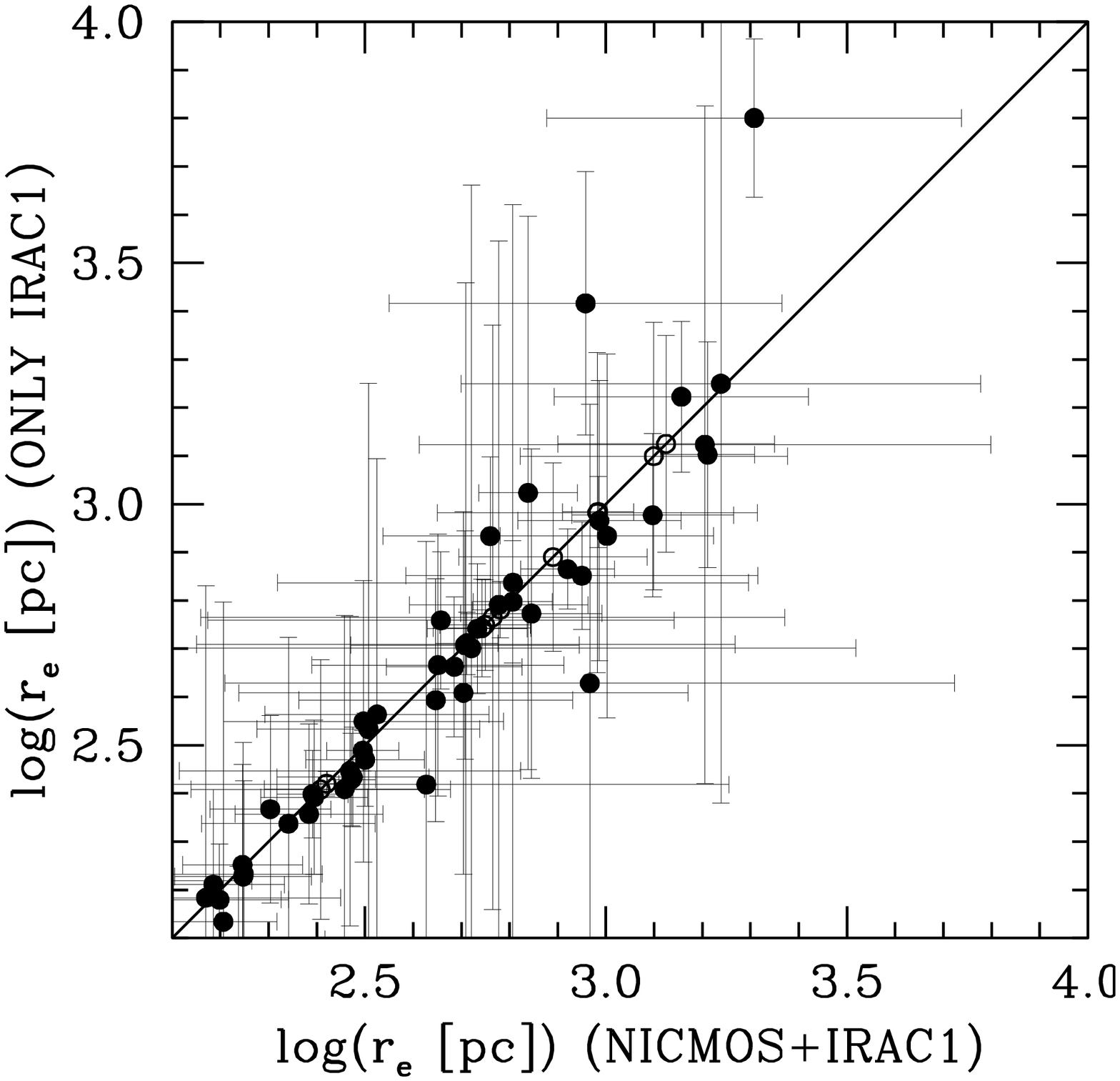}
\includegraphics[width=.4\textwidth]{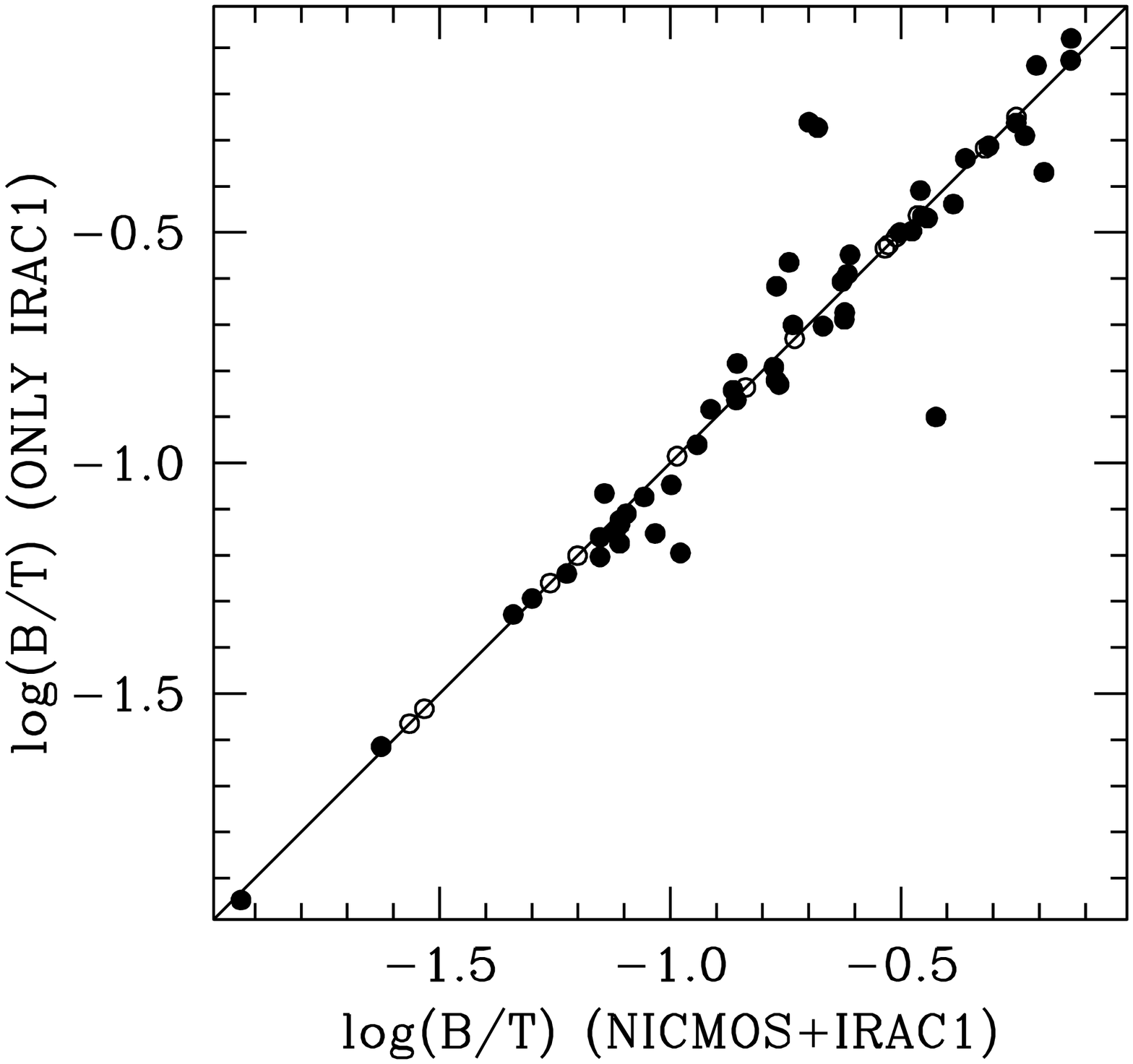}
\caption{Effect of including or excluding NICMOS data on S\'ersic fit
  parameters of light profiles of galaxies in our sample.  From top to
  bottom, the parameter comparisons shown are S\'ersic index,
  half-light radius, and bulge-to-total ratio. The open symbols
  indicate galaxies for which the inner radius at which fitting begins
  is larger than the resolution of IRAC at
  3.5~$\mu$m. \label{fig:nicmos}}
\end{figure}

We include HST/NICMOS F160W images in the analysis for all galaxies
with available archival data. This is done because it increases
dynamic fitting range on the bulge. The composite profile is NICMOS
data for r<1.22 arcsec, and IRAC 3.6~$\mu$m data at larger
radii. Inclusion of the NICMOS data may affect the results in two ways.

First, this procedure assumes a color gradient of zero from $L$-band
to $H$-band in our bulges. This assumption introduces a source of
uncertainty, yet allows for a more complete description of the stellar
light profile. To quantify this uncertainty we calculate the entire
radial surface brightness profile in $H$-band using NICMOS and 2MASS
data. We then shift that profile to have the same zero point as the
IRAC~3.6~$\mu$m profile, and then calculate the bulge luminosity,
which we call $L_{3.6(H)}$. The error in luminosity due to the
stitching is the difference $L_{3.6}-L_{3.6(H)}$ scaled by the the
fraction of light that comes from the shifted NICMOS F160W data:

\begin{equation}
\delta L_{\rm stitch} \equiv L_{3.6}-L_{3.6(H)} \times L_{3.6}(r<1.22")/L_{3.6}.
\end{equation}

This error is typically less than 5\% and rarely larger than errors
from other sources, such as fitting uncertainty. We use NICMOS data
because it is our experience that the high resolution data increases
accuracy, even if precision is compromised slightly.

Second, \cite{balcells2003} find that supplementing ground-based
profiles with HST data systematically affects S\'ersic index, making
it smaller. This effect is certainly distance dependent. Our distance
limit is 20~Mpc, closer than most galaxies in \cite{balcells2003}, and
likely is nearby enough to prevent this from affecting our
profiles. Nonetheless, we carry out a test to determine if any such
bias is present.
\begin{figure*}[t]
\begin{center}
\includegraphics[width=.99\textwidth]{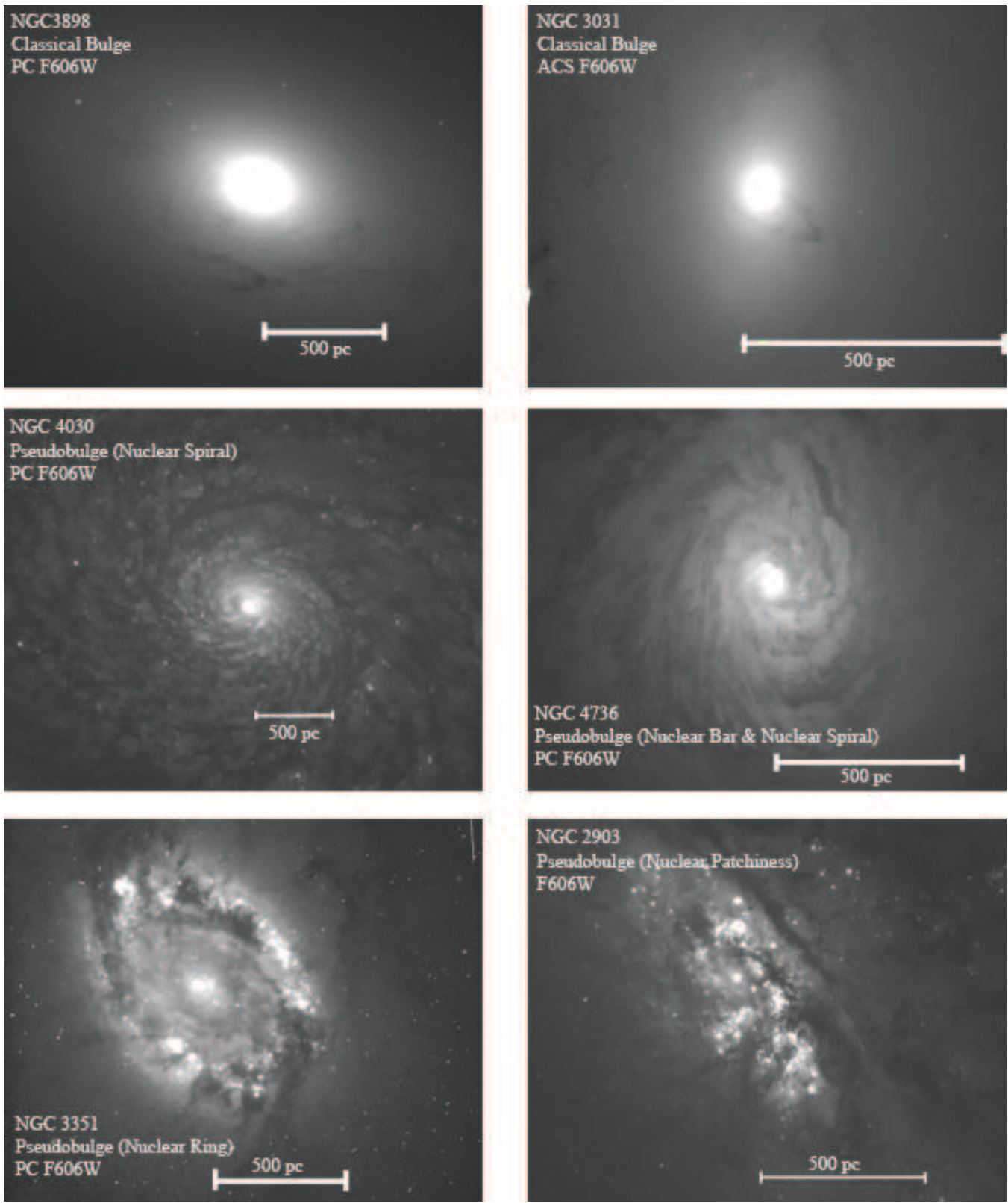} 
\end{center}
\caption{Examples of morphologies of classical
  bulges (top two panels) and pseudobulges (bottom four panels). All
  images are taken by HST in the F606W filter. On each panel we
  draw a line representing 500~pc.  \label{fig:eg_ims}}
\end{figure*}

\begin{figure*}[t]
\begin{center}
\includegraphics[width=.99\textwidth]{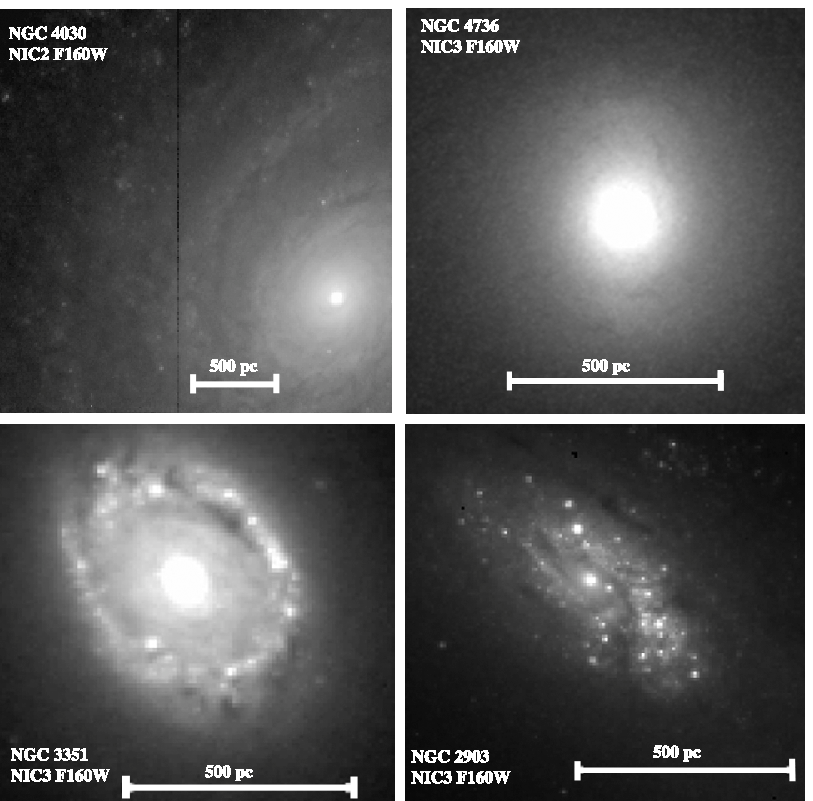} 
\end{center}
\caption{Examples of morphologies of pseudobulges shown in
  Fig~\ref{fig:eg_ims}. All images are taken from HST using the NICMOS
  F160W filter. On each panel we draw a line representing
  500~pc.  \label{fig:nic_ims}}
\end{figure*}

We re-fit all galaxies that have both HST and SST images. In this second
fit we do not include isophotes  that have smaller major axis than the
half width of the point spread function of  IRAC~3.6~$\mu$m, which is
1.5~arcsec. In  most galaxies, we keep all  other input parameters
exactly  the   same.  However,   the  $\chi^2$  space   of  bulge-disk
decomposition that  includes a S\'ersic bulge can  be very complicated
and  highly degenerate,  in  some  cases reduction  of  the fit  range
resulted in  independent minima in  parameter space. In those cases we
are forced to reduce  the allowable parameter range accordingly.

Fig.~\ref{fig:nicmos} shows the comparison of parameters from
decomposition of IRAC-only isophotes to those derived combining
NICMOS and IRAC data. The top panel shows S\'ersic index, the
middle panel half-light radius, and the bottom panel
bulge-to-total light ratio. In each panel a solid line representing the
line of equality is plotted. Bulges in which the inner cut in the
fitting range is larger than the resolution of IRAC are denoted as
open circles, all other bulges are denoted as closed circles. We do
not include galaxies with $B/T<0.01$, as there parameters are too
poorly constrained to be meaningful.
\begin{figure}[t]
\begin{center}
\includegraphics[width=.49\textwidth]{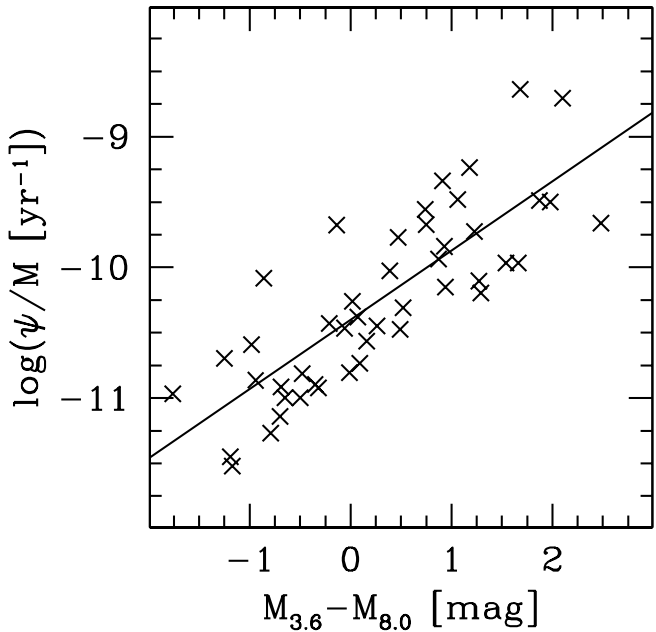} 
\end{center}
\caption{Specific star formation rates (star formation rate, $\psi$
  per unit stellar mass, $M$) from \cite{fdf2009} compared to mid-IR
  colors determined in this paper. \label{fig:ssfr}}
\end{figure}

In all properties there is good agreement between both methods of
decomposition. Bulge-to-total ratio appears to be the most robust
quantity. This is likely due to the degeneracy between surface
brightness, radius, and S\'ersic index with respect to bulge
magnitude. Also, we do not expect the inner truncation radius of the
fit to greatly alter fit parameters of outer disk properties in nearby
galaxies. Half-light radius also shows good agreement. The most
scatter appears to be in S\'ersic index, which also shows a slight
trend. In bulges with S\'ersic index higher than
about $n_b\sim3.5-4$, the S\'ersic index determined from IRAC-only
profiles is slightly smaller than that determined with composite
SST+HST data. The average of the absolute value of the difference
between the lower resolution and higher resolution data is in all
cases less than 10\%.

\section{Identifying Pseudobulges and Classical Bulges}

\cite{kk04} reviewed in great detail the observational evidence that
pseudobulges and classical bulges are two distinct classes of objects.
Furthermore, they go on to propose a preliminary list of general
observational properties of pseudobulges. These properties include:
dynamics that are dominated by rotation \citep{k93}, the bulge is
actively forming stars \citep{kennicutt98araa}, younger stellar
populations as indicated by optical colors \citep{peletier1996}, the
bulge has a nearly exponential surface brightness profile
\citep{andredak94}, flattening similar to that of their outer disk
\citep{fathi2003,k93}, nuclear bar \citep{erwin2002}, nuclear ring,
and/or nuclear spiral \citep{carollo97}.  On the contrary, classical
bulges are typically identified as having a relatively featureless
morphology, old stellar populations, and hot stellar dynamics.

For brevity we will refer to morphology that resembles disk phenomena
(spiral structure, rings and/or bars) as ``D-type'' morphology, and we
will refer to morphology that resembles that of dynamically hot
stellar systems (smooth isophotes with little-to-no substructure) as
``E-type'' morphology.

More recently, \cite{fisher2006} shows that in non-S0 disk galaxies
 bulge morphology correlates very well with bulge activity;
pseudobulges are actively growing. Also, \cite{fisherdrory2008} show
that in over 90\% of the galaxies in their sample bulges with
disk-like morphology have $n_b<2$, and all classical bulges have
$n_b>2$. 

\subsection{Identifying Pseudobulges with Morphology}

In Fig.~\ref{fig:eg_ims} we redraw a figure from \cite{fdf2009} which
shows examples of bulges with E-type morphology (top two panels) and
bulges with D-type morphology (bottom four panels). The striking
contrast between the exemplary D-type and E-type bulges is evident.
When the outer disk is not seen, the bulges bulges with E-type
morphology appear morphologically indistinguishable from elliptical
galaxies. On the other hand the bottom four bulges show pronounced
disk phenomena including: nuclear spirals that extend all the way to
the center of the galaxy (e.g.~NGC~4030 and NGC~4736), small bars that
are not much larger than a few hundred parsec across (e.g.~NGC~4736),
nuclear rings (e.g.~NGC3351) and somewhat chaotic nuclear patchiness
that is reminiscent of late-type disk galaxies (e.g.~NGC~2903).  We do
not show images of all of our bulges, rather we direct interested
readers to the Hubble Legacy Archive \footnote{Hubble Legacy Survey
  can be found at http://hla.stsci.edu/}

In Fig.~\ref{fig:nic_ims}, we show the images of the bulges with
D-type morphology from Fig.~\ref{fig:eg_ims} taken with NICMOS
F160W. In all of these near-IR images, the same features that
motivated D-type (hence pseudobulge) classification in the optical
bands are still present. Yet, the nuclear spiral in NGC~4736 is much
weaker, although the nuclear bar in this bulge is still visible. The
features found in the optical are not merely small perturbations
caused by dust inside a smooth light distribution, but likely features
of the stellar mass distribution.

\subsection{Using Mid-IR Color to Identify Star Forming Bulges}

Emission in the 8~$\mu$m channel on Spitzer Space Telescope is
dominated by dust, mostly re-radiated light from polycyclic aromatic
hydrocarbons \citep{leger1984}. Also, \cite{helou2004} shows that the
8~$\mu$m emission in the nearby disk galaxy NGC~300 traces the edges
of HII. Thus it is correlated with the number of ionizing photons from
young stars, and therefore with local star formation rate.

In Fig.~\ref{fig:ssfr} we show the correlation between mid-IR bulge
color (from this data set) with specific star formation rate (from
overlapping galaxies in \citealp{fdf2009}). A linear regression to the
data is plotted as a black line. There is an important distinction
between the two data sets, \cite{fdf2009} subtracts an inward
extrapolation of the exponential disk for the stellar mass, we do not
for this calculation. We find a correlation, with Pearson correlation
coefficient $r^2=0.65$, between mid-IR color, $M_{3.6}-M_{8.0}$, and
specific star formation rate.

\cite{calzetti2007} show that differences in inter-stellar medium
metallicity, and likely stellar populations of the galaxy, can
strongly affect the correlation between 8~$\mu$m emission and
luminosity of ionizing photons.  Given that we are specifically
looking at bulges that may have widely varying stellar populations
and metallicities, this is an important caveat. However, when studying
a large number of bulges, that may be difficult to resolve. The
increased resolution of IRAC data over MIPS data is very important.
Thus, we limit the interpretation of 8~$\mu$m emission to an on/off
metric of star formation. That is to say, we use the 8~$\mu$m emission
to determine if the bulge is active, and we refrain from using
8~$\mu$m emission as a means to quantify the star formation rate.  In
this paper, it is sufficient to merely know if the bulge is active or
inactive, and Fig.~\ref{fig:ssfr} indicates that mid-IR color is a
good estimate in 90\% of the bulges. For a detailed discussion of the
properties of star formation rates in pseudobulges the interested
reader should see \cite{fdf2009}.

\subsection{Using S\'ersic Index to Identify Bulge Types}

\cite{andredak95} showed that bulges in early-type galaxies have
larger S\'ersic index than those in late-type galaxies. This motivated
many authors \citep[e.g.][]{kk04} to assume that S\'ersic index of a
bulge is correlated with bulge-type.  \cite{fisherdrory2008} directly
tests this hypothesis and finds that in decomposition in the $V$-band
pseudobulges (bulges with D-type morphology) have $n_b\lesssim2$ and
classical bulges (with E-type morphology) have $n>2$. In this paper we
will also consider the possibility that S\'ersic index can be used to
identify bulge types. Also we will retest the connection between bulge
S\'ersic index and optical bulge morphology with near-IR
decompositions that more robustly reflect the stellar light density.

\subsection{Comparing Pseudobulge Diagnostics}

\begin{figure}[t]
\includegraphics[width=.48\textwidth]{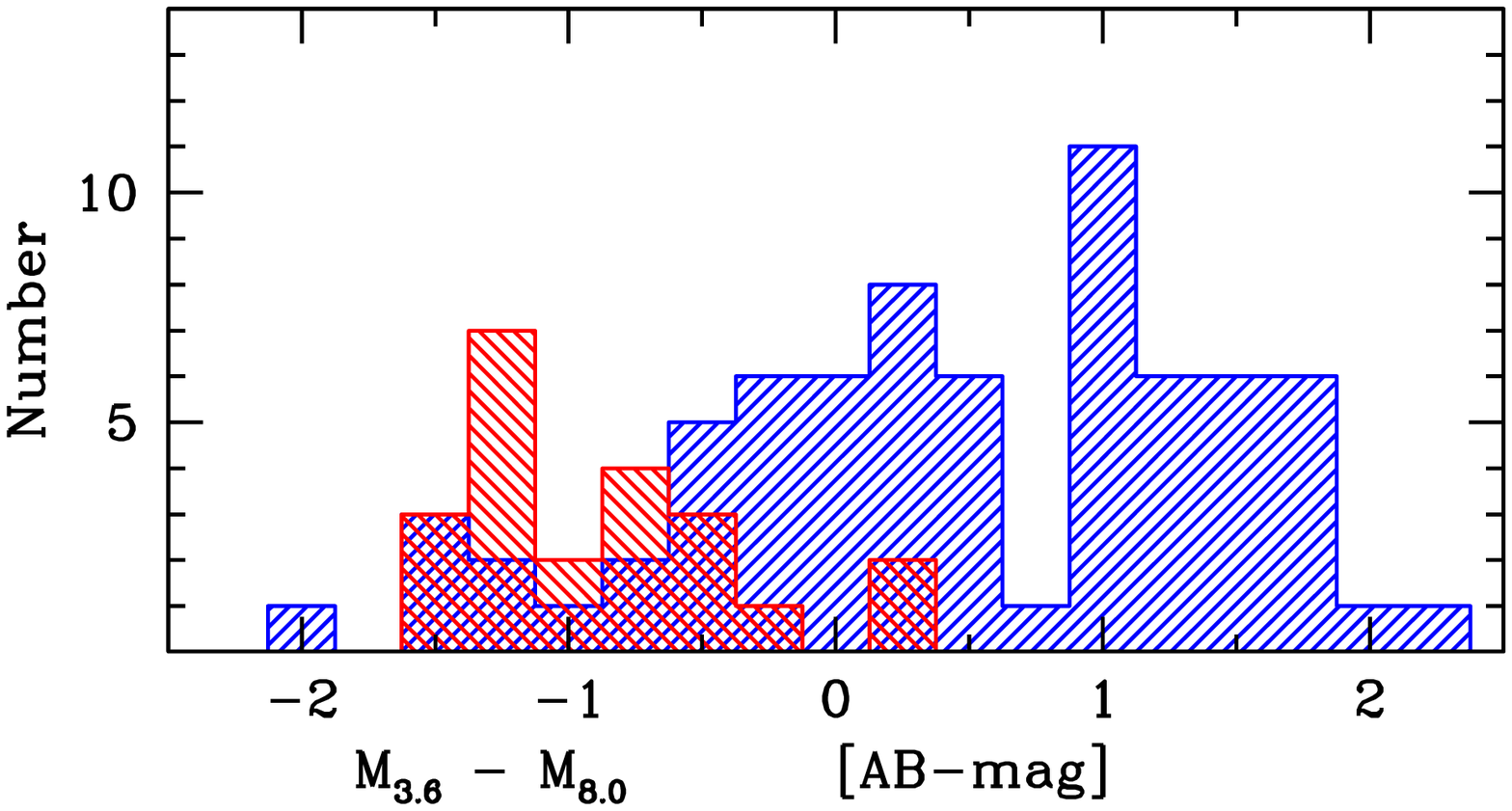} 

\includegraphics[width=.48\textwidth]{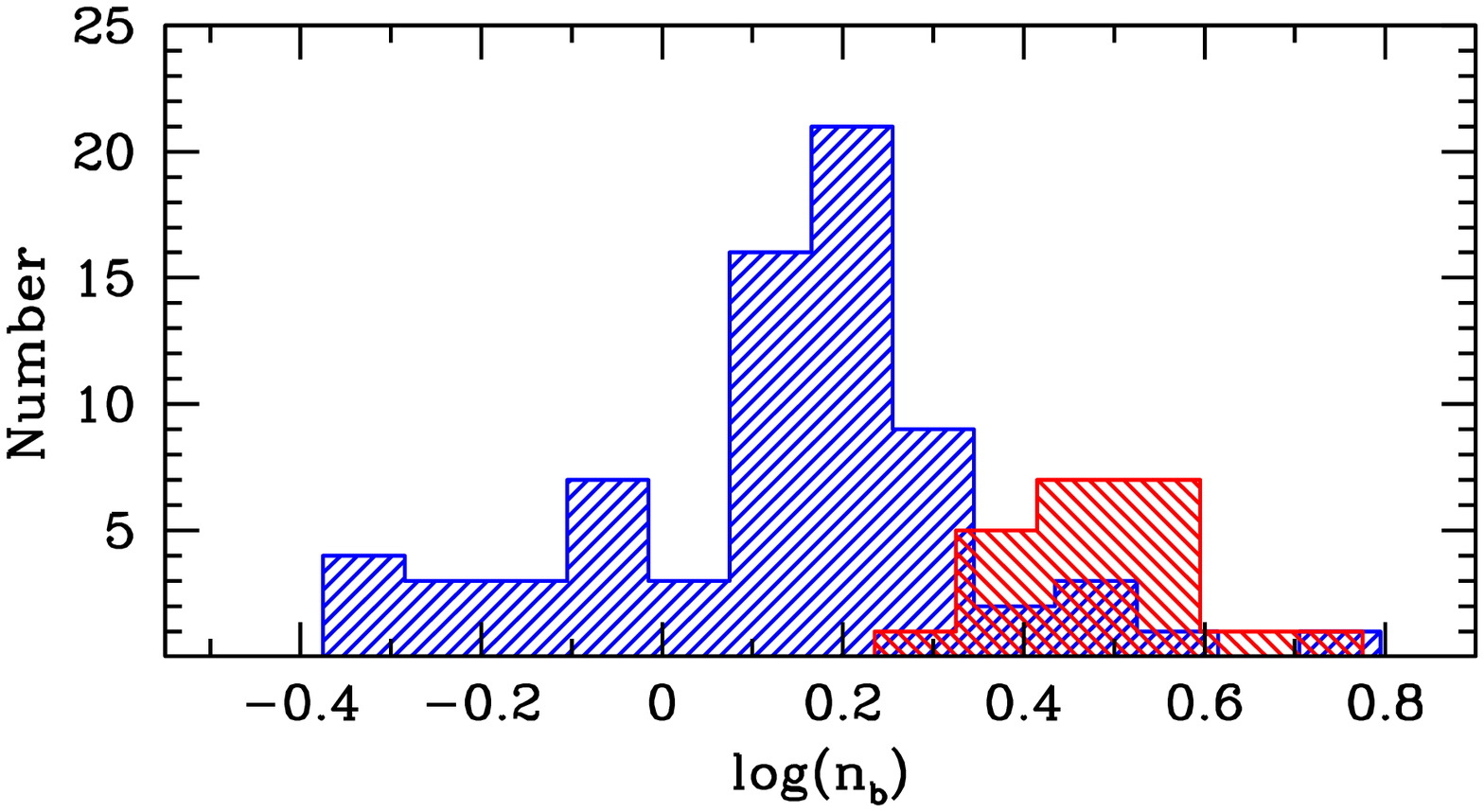}
\caption{The distribution of mid-IR color (top) and S\'ersic index
  (bottom) in our sample. In both panels galaxies with morphologically
  identified pseudobulges are represented by blue lines and those with
  classical bulges are represented by red
  lines.  \label{fig:morphcomp}}
\end{figure}



For the comparison of HST morphology to other bulge diagnostics we use
the higher quality data set. This is done to ensure that poorly
constrained S\'ersic fits or phenomena such as AGNs do not
artificially generate ambiguity in bulge diagnostics.

In Fig.~\ref{fig:morphcomp}, we show the distribution of mid-IR color
(top) and S\'ersic index (bottom) of bulges. In both panels those
bulges with D-type morphology (that is pseudobulges) are represented
by blue histograms filled with forward slanting lines, and bulges
with E-type morphology are represented by red histograms filled with
backward slanting lines.

In the top panel of Fig.~\ref{fig:morphcomp}, we show that no bulge
with E-type morphology has mid-IR color redder than
$M_{3.6}-M_{8.0}>0.3$, and very few (2 of 22) have
$M_{3.6}-M_{8.0}>0.0$. However, as found in \cite{fdf2009}, there is a
a broad distribution of star formation activity in
pseudobulges. Pseudobulges can be both active and inactive. When a
bulge is active, in almost every case (96\%) that bulge has D-type
morphology. Also the majority of bulges with D-type morphology (64\%)
have mid-IR colors that are redder than any classical bulge.
Therefore, use of mid-IR color alone to identify pseudobulges merely
finds one type of pseudobulge: active pseudobulges.
 
In the bottom panel of Fig.~\ref{fig:morphcomp}, we find the same
result as found in the V-band by \cite{fisherdrory2008}: the vast
majority (93\%) of bulges with D-type morphology have S\'ersic index
less than $n_b\leq 2$. Conversely, all bulges with E-type morphology
have S\'ersic index greater than two. Only one bulge with E-type
morphology has S\'ersic index $n_b\sim2$ (NGC~1617 has $n_b=2.1$).
This confirms that the correlation between S\'ersic index and bulges
morphology is independent of wavelength.

In the high quality sample, there are 53 galaxies with positive mid-IR
color (based on visual inspection of Fig.~\ref{fig:ssfr} this
translates very roughly to $SFR/M \gtrsim 40$~Gyr$^{-1}$) and only 6
of those galaxies have a bulge with S\'ersic index greater than two;
two of those six active high-$n_b$ bulges have E-type morphology. In the
high quality sample, we find 18 bulges with low S\'ersic index and low
specific SFR. The 8 pseudobulges with the lowest specific SFR are all
S0 galaxies.

If a bulge has a positive value of $M_{3.6}-M_{8.0}$, then that bulge
is 96\% likely to have low-S\'ersic index and D-type nuclear
morphology (hence be a pseudobulge). However, if the bulge is inactive
then there is little we can say about the type of bulge from mid-IR
color alone. As found by both \cite{droryfisher2007} and
\cite{fdf2009}, some pseudobulges are inactive and also reside in
inactive galaxies. The take away is that bulges with E-type morphology
are always inactive, and always have $n_b>2$. Bulges with D-type
morphology almost always have $n_b<2$, and usually are active.  We will
investigate the nature of those bulges that have disk-like
nuclear morphology, low S\'ersic index but are non-star forming in
subsequent sections.

\begin{figure*}[t]
\includegraphics[width=.99\textwidth]{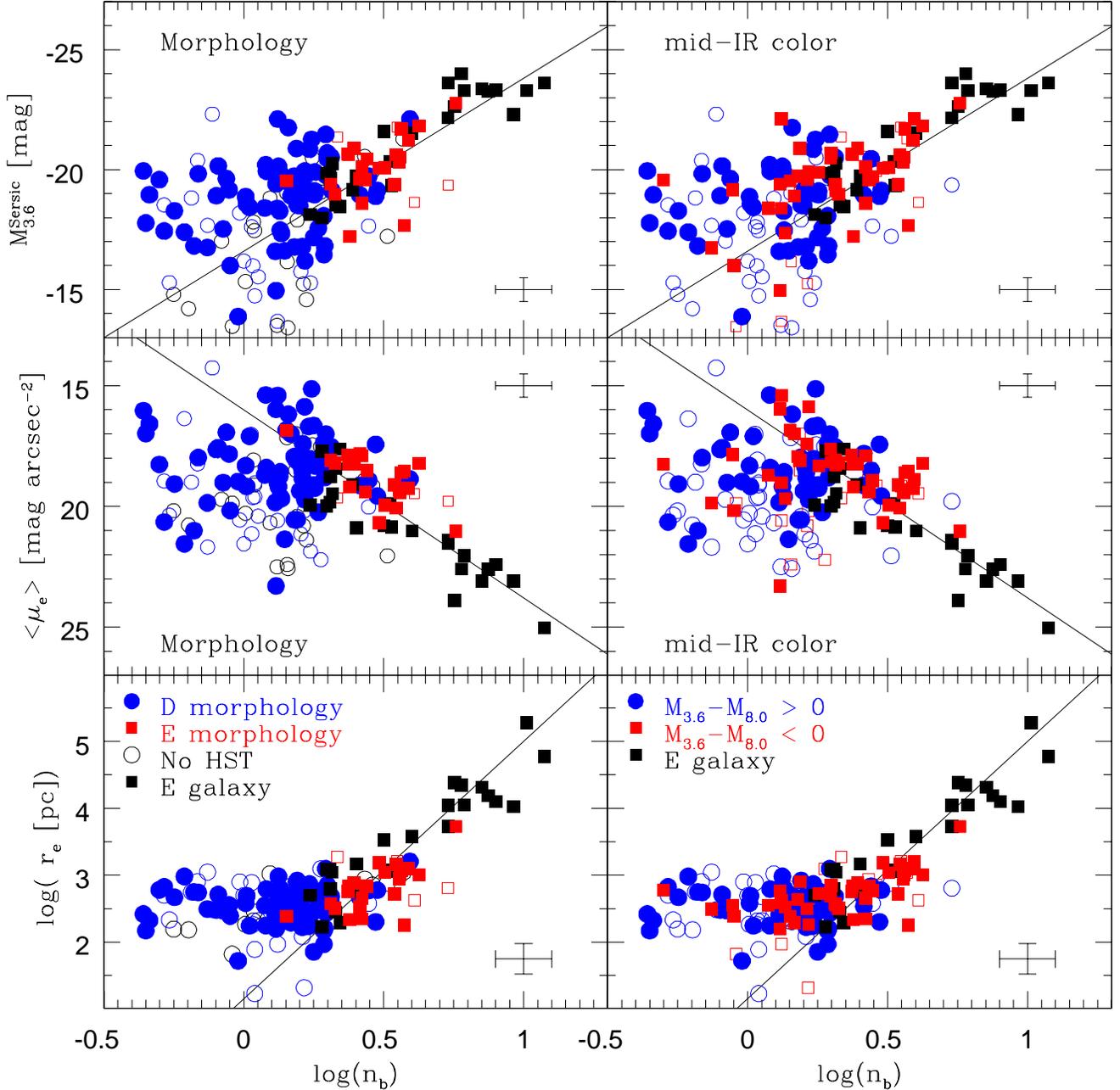} 
\caption{Correlations of the bulge S\'ersic index with (from bottom to
  top) half-light radius of the bulge (a), average surface brightness
  within the half-light radius (b), and magnitude of the bulge at
  $3.6$~$\mu$m (c). In all panels, pseudobulges are represented by
  blue circles, classical bulges by red squares, and elliptical
  galaxies by black squares. Galaxies in the high quality sub-sample
  are shown as filled symbols whereas galaxies that do not meet
  criteria for the high-quality data set (see text) are shown as open
  symbols. \label{fig:sersic_cors}}
\end{figure*}

\section{Scaling Relations Of Bulge Parameters}

\subsection{S\'ersic Index}

Here we turn our attention to the correlations of S\'ersic index with
other photometric parameters. \cite{fisherdrory2008} find with
$V$-band decompositions that it is not only the value of the S\'ersic
index that is different between pseudobulges and classical bulges, but
the way that S\'ersic index scales with other parameters. However
their sample did not contain a significant number of galaxies with low
$B/T$ or small $n_b$. Using our more representative sample, we observe
the same result at 3.6~$\mu$m where light better traces the stellar
mass and is less affected by dust.

In Fig.~\ref{fig:sersic_cors}, we show the correlations of the bulge
S\'ersic index with half-light radius of the bulge, surface brightness
at the half-light radius, and magnitude of the bulge at $3.6$~$\mu$m
(from bottom to top). In all panels we represent elliptical galaxies
as black squares. In each panel we also show a black line representing
a fit to the elliptical galaxies only.  Filled symbols represent
high-quality data and open symbols represent data that does not
fulfill the requirements of the high-quality dataset.  In the left panels
we identify bulges based on nuclear morphology; bulges with D-type
morphology are represented by blue circles, and those bulges with
E-type morphology are represented by red squares.  If no optical HST
images are available, we plot the bulge as an open black circle. Note
that all galaxies in the high quality sample have optical HST data. In
the right panels we identify bulges based on activity.
 Those bulges with mid-IR color $M_{3.6}-M_{8.0}>0$, indicating
higher specific star formation rate, are represented by blue
circles. Those bulges with mid-IR color $M_{3.6}-M_{8.0}<0$ indicating
low specific star formation rates are represented by red squares.

First, we notice the similarity in these scaling relations between
bulges with E-type morphology and elliptical galaxies. In each panel,
bulges with E-type morphology follow the correlations established by
elliptical galaxies. There is slightly more scatter in the bulges with
E-type morphology than in the elliptical galaxies alone; however, this
is not at all surprising since the presence of an outer disk means
that S\'ersic function fits to bulges are less well constrained than
those in elliptical galaxies.

\begin{figure*}[t]
\includegraphics[width=.99\textwidth]{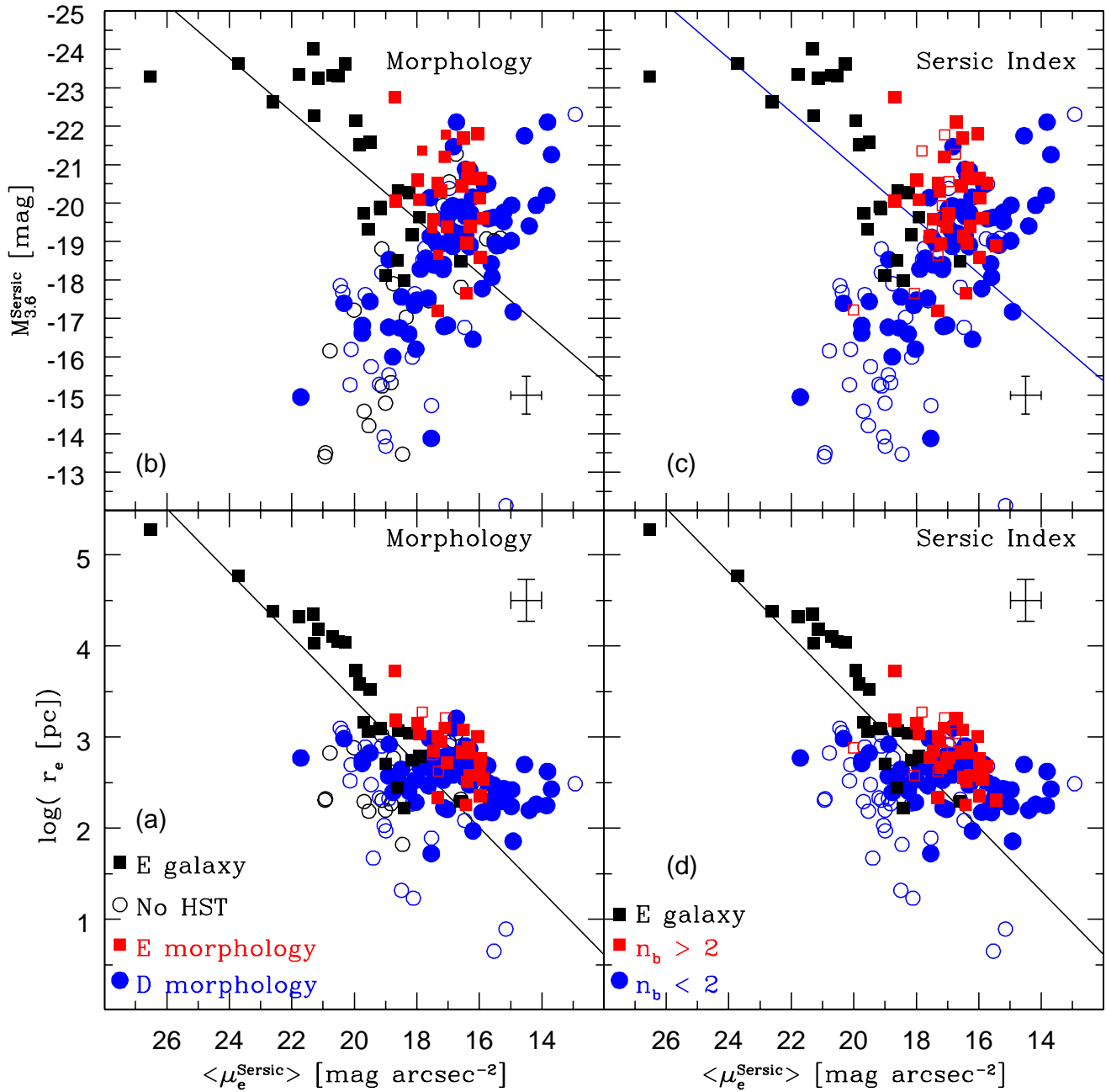} 
\caption{Correlations of the the surface brightness of the bulge with
  half-light radius of the bulge (bottom two panels, a \& d), and
  3.6~$\mu$m magnitude of the bulge (top two panels, b \&
  d). Elliptical galaxies are represented by black squares, classical
  bulges are represented by red squares, and pseudobulges are
  represented by blue circles. Galaxies in the high quality sub-sample
  are shown as filled symbols whereas galaxies that do not meet
  criteria for the high-quality data set (see text) are shown as open
  symbols. In the left two panels black open symbols are for those
  galaxies without available morphology on which to diagnose the bulge
  type.  \label{fig:fund}}
\end{figure*}

Also, \cite{kfcb} find that the distribution of S\'ersic indices in
elliptical galaxies is bimodal and that elliptical galaxies with
S\'ersic index higher than $n=4$ are almost exclusively giant
ellipticals with core profiles in their central region and that show
no signs of rotation. Conversely, they find that elliptical galaxies
with lower S\'ersic index have cuspy centers, are lower in luminosity,
and have dynamics indicative of an oblate spheroid that is flattened
by rotation. In the bottom panel of Fig.~\ref{fig:sersic_cors}, it is
clear that bulges with E-type morphology are like low-luminosity
ellipticals with S\'ersic index less than $n_b\sim4$. This restriction
is important for interpreting parameter correlations; the set of
classical bulges alone may not always reflect the distribution of all
elliptical galaxies.
\begin{figure}
\includegraphics[width=0.48\textwidth]{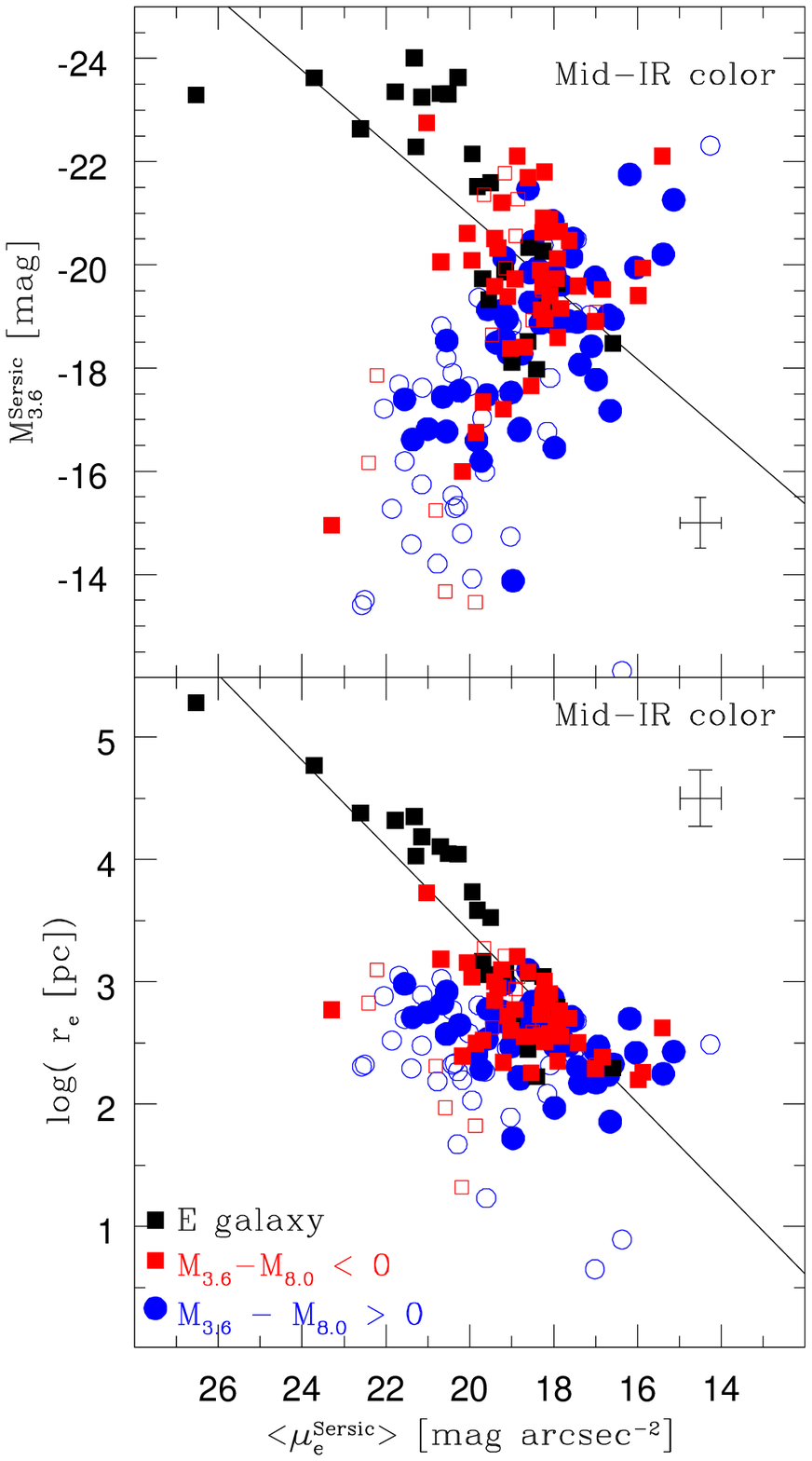}
\caption{Correlations of the the surface brightness of the bulge at
  the half-light radius with half-light radius (bottom panel) of the
  bulge, and 3.6~$\mu$m magnitude of the bulge (top panel). In both
  panels elliptical galaxies are represented by black squares, those
  bulges with $M_{3.6}-M_{8.0}<0$ are represented by red squares, and
  those bulges with $M_{3.6}-M_{8.0}>0$ are represented by blue
  circles. Galaxies in the high quality sub-sample are shown as filled
  symbols whereas galaxies that do not meet criteria for the
  high-quality data set (see text) are shown as open
  symbols.  \label{fig:fund_color}}
\end{figure}

\begin{figure*}[t]
\includegraphics[width=.99\textwidth]{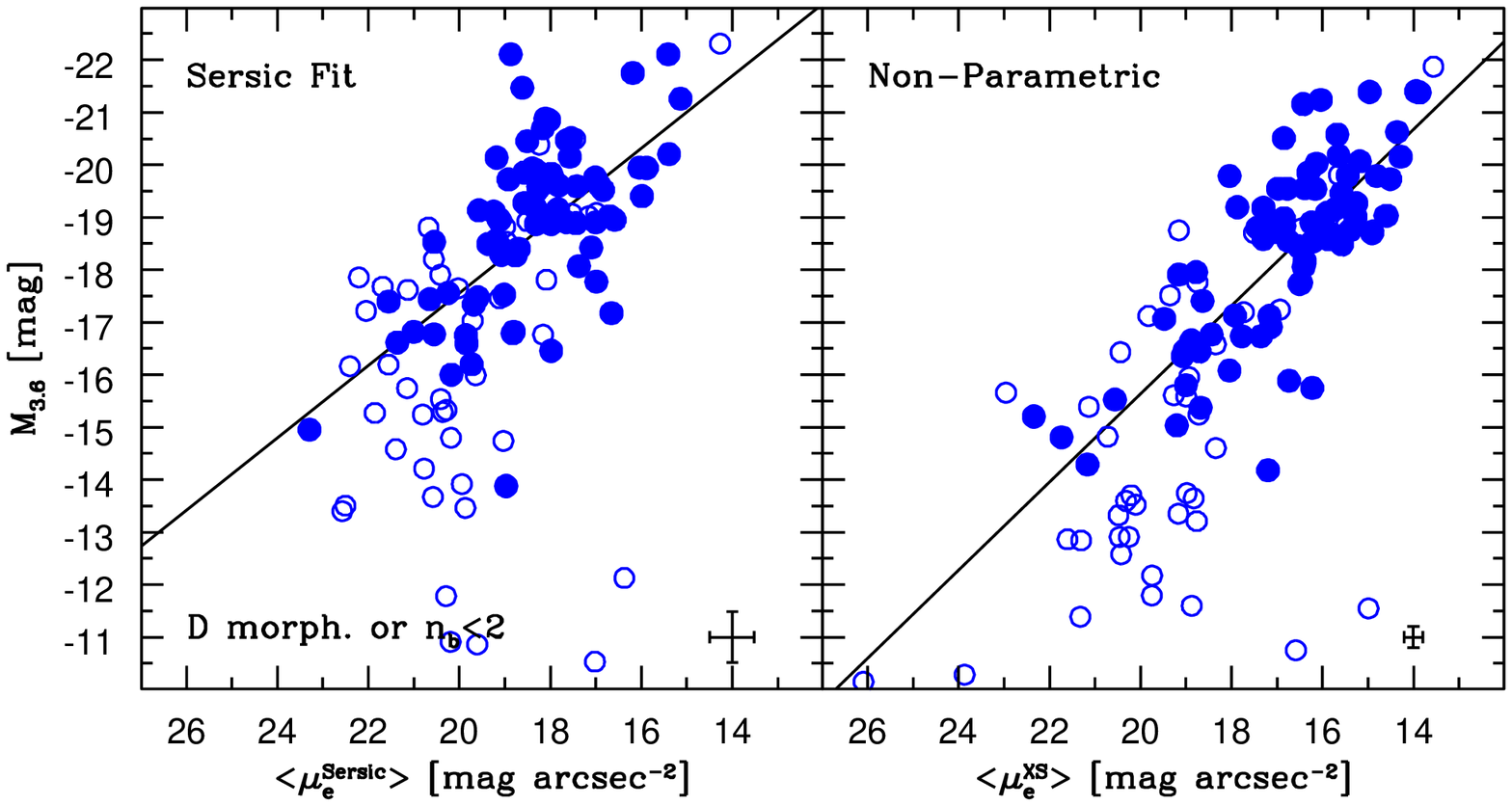} 
\caption{Correlations of the mean surface brightness of the bulge with
  3.6~$\mu$m magnitude of the bulge. The left panel shows $<\mu_e>$
  and $M_{3.6}$ derived from S\'ersic decomposition, and the right
  panel shows $<\mu_e>$ and $M_{3.6}$ derived non-parametrically.
  Only pseudobulges are shown in this figure. Here, pseudobulges are
  selected as the union of all three diagnostic methods (nuclear
  morphology reminiscent of disks, $n_b<2$, and
  $M_{3.6}-M_{8.0}>0$. We over-plot correlations fit to all bulges with
  $B/T>0.01$. Galaxies in the high quality sub-sample
  are shown as filled symbols whereas galaxies that do not meet
  criteria for the high-quality data set (see text) are shown as open
  symbols. \label{fig:mumag}}
\end{figure*}

In Fig.~\ref{fig:sersic_cors}, for those bulges with D-type morphology
it is clear that no correlation exists between the S\'ersic index and
half-light radius, surface-brightness, or bulge magnitude. In
magnitude and surface brightness, the bulges with D-type morphology
appear to be scattered with no regularity.  However, the half-light
radius of bulges with D-type morphology is limited to a very narrow
range in parameter space. We will discuss the half-light radii of
bulge types in more detail \S7.2.

In the right panels of Fig.~\ref{fig:sersic_cors} we select bulges
based on mid-IR color. Because the correlation between S\'ersic index
and morphology (described in the last section) is so strong, it is not
surprising that we see similar behavior here. Bulges with active star
formation ($M_{3.6}-M_{8.0}>0$) show no correlation between S\'ersic
index and any other bulge structural property shown here. Also,
selecting for bulges that are inactive ($M_{3.6}-M_{8.0}<0$) does not
uniquely select for bulges that correlate like elliptical galaxies in
parameter space shown in Fig.~\ref{fig:sersic_cors}.

These results provide confirmation that the differences in S\'ersic
index correlations of pseudobulges found in \cite{fisherdrory2008} are
due to real differences in the distribution of stellar mass. They are
not simply due to differences in internal extinction or mass-to-light
ratios.

\subsection{Photometric Projections of the Fundamental Plane}

How different should we expect pseudobulges and classical bulges do be
in photometric parameter space? First, pseudobulges and classical
bulges do not differ dramatically in bulge-to-total ratio, many of
which are in the range $B/T\sim0.1-0.4$ -- and in Sa to Sc galaxies,
disk size and mass does not follow a strong trend
\citep{robertshaynes1994} with Hubble Type.  Therefore, it follows
that bulges will be similar in size and luminosity, and thus surface
brightness as well.  At least some overlap ought to exist between
pseudobulges and classical bulges in luminosity, size and surface
brightness.

\cite{carollo1999} shows that pseudobulges deviate toward lower
density in the photometric projections of the fundamental
plane. \cite{gadotti2009} uses the \cite{k77} relation between
$<\mu_e>-r_e$ to define pseudobulges. They select pseudobulges as
those bulges that are more than 1$\sigma$ lower in surface density than
the correlation for elliptical galaxies.  However, they also find that
many of the bulges on the \cite{k77} relation have low S\'ersic index
and also have young stellar populations. It is not clear that using
only projections of the fundamental plane can uniquely identify a class
of bulges.

In Fig.~\ref{fig:fund}, we show the \cite{k77} relation between
$<\mu_e>-r_e$ (bottom panels a \& d) and the correlation between surface
brightness and bulge magnitude (top panels b \& c). The left panels
show the correlations using morphology as the method of identifying
pseudobulges and the right panels show these correlations using
S\'ersic index as a method of diagnosing bulge types. In each panel we
also plot a line that represents the corresponding scaling
relation fit to the elliptical galaxies only.

Those bulges with E-type morphology and those with $n_b>2$ are similar
to low-luminosity elliptical galaxies in both $<\mu_e>-r_e$ and the
$<\mu_e>-M_{3.6}^{Sersic}$ parameter correlation.  \cite{k77} shows
that elliptical galaxies follow a tight positive correlation between
size and surface density. Also, in elliptical galaxies more luminous
systems are less dense \citep{kormendy1985,lauer_cores1985}. We
observe similar behavior for classical bulges (identified both through
morphology and S\'ersic index) in our sample.

A few of the bulges with D-type morphology that also have high
S\'ersic index appear as red squares in the right two panels (c \& d)
in Fig.~\ref{fig:fund}, and blue circles in the left two panels (a \&
b). Two are clearly evident in the $<mu_e>-M_{3.6}$ parameter space as
low-luminosity outliers with respect to the correlation established by
elliptical galaxies.  For at least these cases the difference in
morphology appears to indicate that a physical difference in structure
exists, despite the high S\'ersic index.

Bulges with D-type morphology and/or those bulges with $n_b<2$ show
little-to-no correlation in the $<\mu_e>-r_e$ parameter space. Though
this is evident by simple inspection, we calculate Pearson's
correlation coefficient for the pseudobulges and find $r^2=0.08$, this
is in comparison to $r^2=0.87$ for the correlation of ellipticals and
classical bulges in the same parameters. (For this calculation, we
identify any bulge with disk-like nuclear morphology and/or $n_b<2$ as
a pseudobulge, and all other bulges are classical bulges.)

In Fig~\ref{fig:fund} (b) \& (c), it is clear that bulges with low
S\'ersic index and those with D-type nuclear morphology establish a
correlation that is opposite to the traditional correlation for
elliptical galaxies in the $<\mu_e>-M_{3.6}$ plane. As pseudobulges
(identified by morphology or S\'ersic index) become more luminous they
become more dense.

In Fig.~\ref{fig:fund_color} we show the same two photometric
projections of the fundamental plane, however in this figure bulges
are distinguished by the mid-IR color. Those bulges with
$M_{3.6}-M_{8.0}>0$ indicating active star formation are represented
by blue circles, and those bulges with $M_{3.6}-M_{8.0}<0$ indicating
low specific star formation rate are represented by red squares. As
with S\'ersic index and morphology the absence of star formation in a
bulge does not select for a unique type of bulge. Therefore, little
can be said of a bulge's position with respect to the fundamental
plane based solely on specific star formation rate.

In Fig.~\ref{fig:mumag}, we re-plot the $<\mu_e>-M_{3.6}$ parameter
space of pseudobulges only. Here pseudobulges are selected as bulges
with either D-type morphology or $n_b<2$. The black line in each panel
represents a fit to all bulges with $B/T>0.01$.  Bulges with D-type
morphology often have non-S\'ersic components (e.g.~nuclear rings and
nuclear star clusters), and thus S\'ersic fits to pseudobulges may be
poorly constrained. We therefore measure bulge magnitude and mean
surface brightness within the half-light radius non-parametrically, as
described in \S3.3. The left panel shows parameters derived from the
S\'ersic decompositions and the right panel shows parameters derived
non-parametrically.
\begin{figure}
\includegraphics[width=0.49\textwidth]{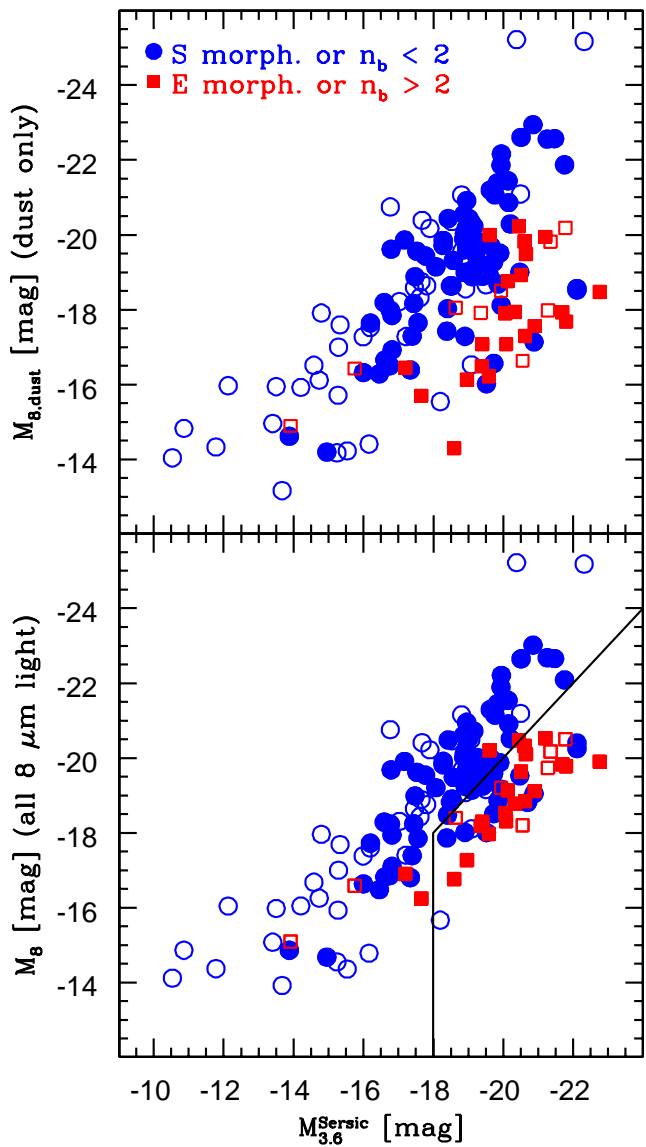}
\caption{Comparison of the 8.0~$\mu$m magnitude of the
  dust emission within bulges to the 3.6~$\mu$m mag of the bulges. The magniutes are obtained from surface brightness profile decompositions. Symbols are as follows: blue circles represent those bulges with either D-type nuclear morphology, or $n_b<2$; red squares represent those bulges with both E-type nuclear morphology and $n_b>2$. As before closed symbols represent the higher quality data set and open symbols represent data that does not meet these requirements (see text).
  \label{fig:pah}}
\end{figure}

Interestingly, the non-parametric $<\mu_e>-M_{3.6}$ correlation shows
lower scatter than that of the parametric fits; the Pearson
coefficients are $r^2=0.69$ for non-parametric and $r^2=0.44$ for the
S\'ersic derived quantities. The scaling relations for non-parametric
quantities is
\begin{equation}
M_{3.6}=0.84\pm0.01\times <\mu_e> - 32.43\pm0.08,
\end{equation}
and for the S\'ersic derived parameters
\begin{equation}
M_{3.6}=0.69\pm0.01\times <\mu_e> - 31.36\pm0.08.
\end{equation}
Both relations have a positive slope that is slightly more shallow
than a linear relationship. The difference between these is quite
possibly a result of the non-S\'ersic components (e.g. nuclear rings
and nuclear star clusters) which may contain higher fractions of
the light in lower-luminosity bulges.  We note that the lowest
magnitude bulges are not on the correlation at all. This may be
because they are so low in S/N that the parameters are somewhat
meaningless, or perhaps it is not accurate to treat these galaxies as
having a bulge at all. We include them on the plot for completeness.

\begin{figure}
\includegraphics[width=0.49\textwidth]{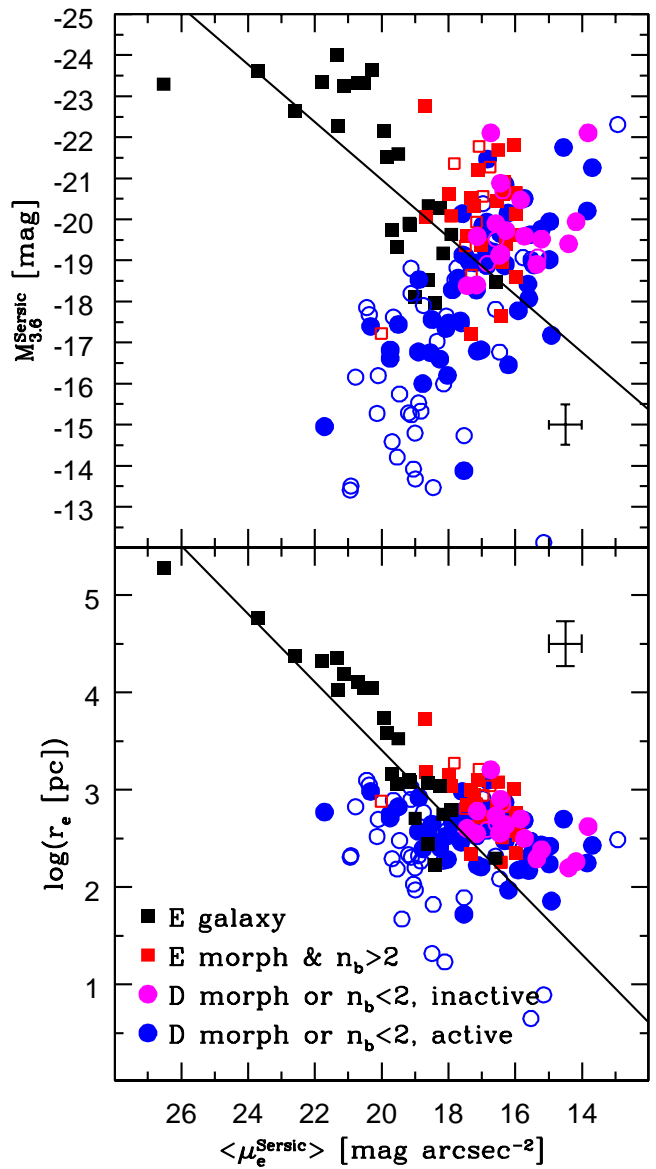}
\caption{The two photometric projections of the fundamental plane from
  Fig.~\ref{fig:fund}; however, in this figure we highlight inactive
  pseudobulges (magenta circles). All other symbols are the same as
  before: active pseudobulges are represented by blue circles,
  classical bulges by red squares, elliptical galaxies by red
  circles. Closed symbols represent the high quality sample, open
  symbols represent all other galaxies. \label{fig:fundip}}
\end{figure}

\section{Structural Properties of Active and Inactive Bulges}

\subsection{Inactive Pseudobulges}

In the top panel of Fig.~\ref{fig:pah}, we compare the magnitude of
bulges at 8~$\mu$m, dominated by dust emission, to the magnitude at
3.6~$\mu$m, which is almost completely stellar light.  We calculate
$M_{8,dust}$ using the relation from \cite{helou2004}, where
$M_{8,dust}=-2.5\times \log_{10}(L_8-0.232\times L_{3.6})$. For both
pseudobulges and classical bulges (diagnosed in this section by the
union of nuclear morphology and S\'ersic index) there is a positive
correlation between bulge the luminosity of the stars at 3.6~$\mu$m
and that of the dust at 8~$\mu$m. All bulges with 8~$\mu$m dust
emission brighter than $M_{8,dust}=-20.3$ are indeed pseudobulges.

It is clear from Figs.~\ref{fig:pah}~\&~\ref{fig:morphcomp} that in
those bulges with $n_b>2$ and E-type nuclear morphology (classical
bulges), the 8~$\mu$m dust emission is lower than the emission at
3.6~$\mu$m. The average classical bulge is 1.7$\pm$1~mags fainter in
8~$\mu$m dust flux than in the 3.6~$\mu$m band.  In most bulges with
low S\'ersic index or D-type nuclear morphology (pseudobulges), the
8~$\mu$m dust luminosity is typically as bright and often brighter than
the stellar luminosity at 3.6~$\mu$m.

However, a significant minority (16\%) of bulges with D-type nuclear
morphology or $n_b<2$ in our sample are bright at 3.6~$\mu$m but have
low dust luminosity.  \cite{fdf2009} notes the existence of bulges
with disk-like morphology and low S\'ersic index that have low
specific star formation rates, using the more precise UV+24~$\mu$m
star formation rate indicator; however, their sample was much smaller,
and they did not have as accurate bulge structural parameters. They
call these objects ``inactive pseudobulges.''  Fig.~\ref{fig:nic_ims}
shows NGC~4736 which turns out to be such a case. It is worth noting
that, unlike in the 3 other examples which are active pseudobulges,
the nuclear spiral that prompted classification of this galaxy as a
pseudobulge is much weaker in the near-IR image. However, a nuclear
bar (which is a disk feature) is visible, and also the S\'ersic index
of this bulge is below two.

In the bottom panel of Fig.~\ref{fig:pah}, we show the comparison of
bulge magnitude at 3.6~$\mu$m to bulge light at 8~$\mu$m without
subtracting the stellar contribution. We over-plot a line showing the
region in the $M_{3.6}$ versus $M_{8}$ plane which contains 90\% of
the classical bulges (identified by having both high S\'ersic index
and E-type nuclear morphology). This yields the boundaries
$M_{3.6}-M_{8.0} < 0$ and $M_{3.6}^{sersic} < -18$. We will refer to
those bulges with low S\'ersic index or disk-like optical morphology,
in this region of Fig.~\ref{fig:pah} as inactive pseudobulges.





\begin{figure}
\includegraphics[width=0.49\textwidth]{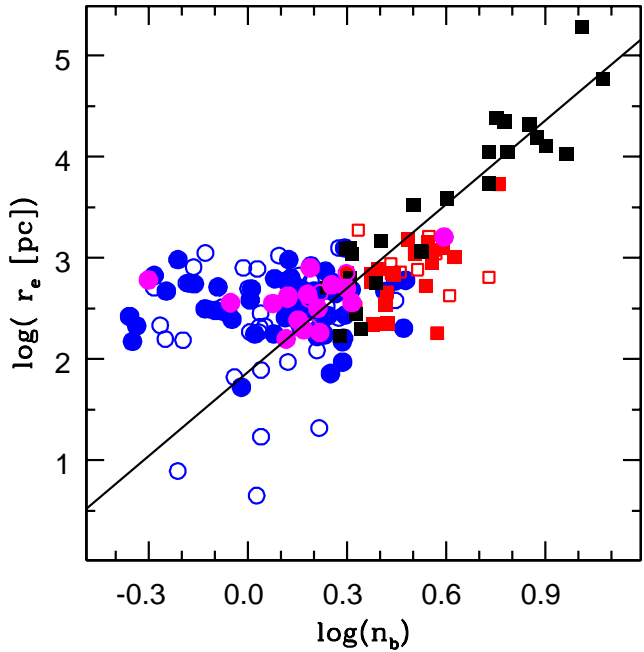}
\caption{The correlation of S\'ersic index with
  half-light radius. Symbols are same as in
  Fig.~\ref{fig:fundip}.  \label{fig:sersip}}
\end{figure}

In Fig.~\ref{fig:fundip} we show the location of inactive pseudobulges
on the $<\mu_e>-M_{3.6}$ and $<\mu_e> - r_e$ planes. Inactive
pseudobulges are represented by magenta circles, all other symbols are
the same as Fig.~\ref{fig:sersic_cors}.  Inactive pseudobulges occupy a region
of parameter space where both classical bulges and bright active
pseudobulges are found.

Classical bulges and inactive pseudobulges are slightly higher in
surface density at a given half-light radius than elliptical galaxies.
In fact, we are unable to fit a correlation that simultaneously
includes giant ellipticals and inactive pseudobulges. This may be a
real feature, however it could also be an effect of how we calculate
the 3.6~$\mu$m zero points for elliptical galaxies. As discussed
earlier, we use the S\'ersic fits from \cite{kfcb} for elliptical
galaxy parameters because this provides a volume limited sample of
very well determined S\'ersic fits. When the galaxy has Spitzer data
we shift the magnitude by the observed $V-L$ when the galaxy is
covered with Spitzer, when the galaxy does not we use the mean $V-L$
correction. The elliptical galaxies observed with Spitzer are
preferentially brighter, and the deviation we see is in low-luminosity
ellipticals. Given that the difference between classical bulges and
elliptical galaxies is so small, we cannot know with our data alone if
the difference in 3.6~$\mu$m luminosity between classical bulges and
elliptical galaxies is a real effect.

In Fig.~\ref{fig:sersip}, we show that inactive pseudobulges almost
always have $n_b<2$ (aside from NGC~4371 which has
$n_b=3.9\pm0.6$). Therefore, in S\'ersic index inactive pseudobulges
are similar to active pseudobulges and unlike classical bulges.  On
average, inactive pseudobulges have slightly higher S\'ersic index than
active pseudobulges. The mean S\'ersic index is $<n_b>=1.6\pm0.3$ for
inactive pseudobulges and $<n_b>=1.3\pm0.6$ for active
pseudobulges. Note that, to calculate the average of the inactive
pseudobulges we remove the value for NGC~4371, including the value
changes the average to $<n_b>=1.7\pm0.6$.  Also inactive pseudobulges
have similar half-light radii as active pseudobulges ($<r_e>=461$~pc
and $<r_e>=380$~pc respectively).

\begin{figure}
\includegraphics[width=0.49\textwidth]{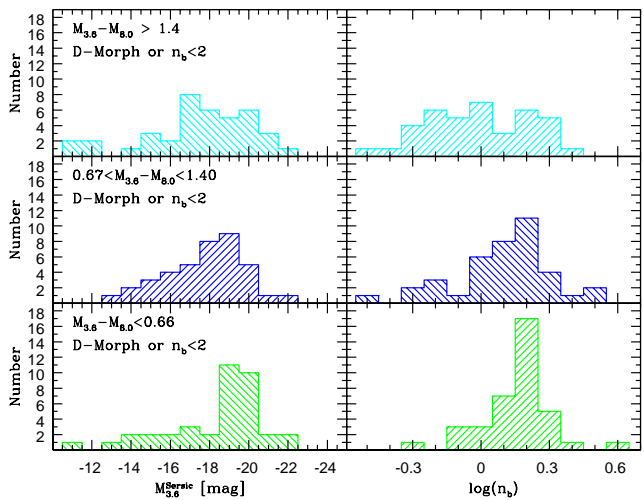}
\caption{Histograms of $M_{3.6}$ (left) and S\'ersic index (right)
  from bulge-disk decompositions. We divide the pseudobulges into
  three groups of equal numbers based on
  $M_{3.6}-M_{8.0}$.  \label{fig:acthists1}}
\end{figure}

\begin{figure}
\includegraphics[width=0.49\textwidth]{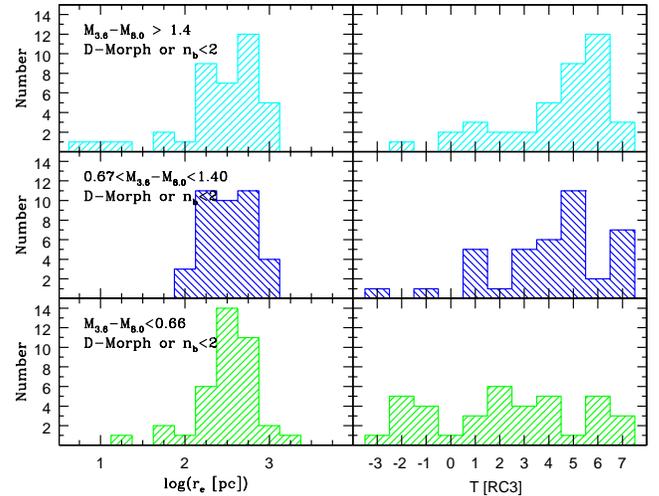}
\caption{Histograms of $r_e$ (left) from S\'ersic bulge-disk
  decompositions and and Hubble type (right) from \cite{rc3}.  We
  divide the pseudobulges into three groups of equal numbers based on
  $M_{3.6}-M_{8.0}$. \label{fig:acthists2}}
\end{figure}

\subsection{Active Pseudobulges}

In Figs.~\ref{fig:acthists1}~\&~\ref{fig:acthists2} we show the
distribution of absolute magnitude of the bulge in 3.6~$\mu$m
(Fig.~\ref{fig:acthists1} left), bulge S\'ersic index
(Fig.~\ref{fig:acthists1} right), half-light radius of the bulge
(Fig.~\ref{fig:acthists2} left) and Hubble type
(Fig.~\ref{fig:acthists2} right). We have divided the sample of
pseudobulges (both high-quality and non high-quality galaxies) into
three groups, of equal numbers, based on mid-IR color of the bulge.
The ranges of the groups from top to bottom are $M_{3.6}-M_{8.0}$ is
less than 0.66~mags, from 0.67~mags to 1.40~mags, and greater than
1.40~mags.

The least active group preferentially contains more luminous
pseudobulges. However, this is somewhat by design because at the
same 8~$\mu$m luminosity a bulge that is brighter in 3.6~$\mu$m will
have lower mid-IR color. However, from the mid to high range of mid-IR
color there appears to be little-to-no trend with bulge luminosity.
There appears to be some change in the distribution of S\'ersic
indices for active and inactive bulges. In the least active group
there is a strong pile up of pseudobulges with $n_b\sim1.6$ and very
few pseudobulges with low values of $n_b$. The high and intermediate
activity groups have a large fraction with very low values of
S\'ersic index.  There is only one bulge with $n_b \lesssim0.8$ in the
lowest specific star formation bin.  It appears that S\'ersic index of
the bulge is linked somehow to specific star formation rate, however,
that link is not so simple as a one-to-one correlation.

We find little-to-no correlation between mid-IR color and bulge
half-light radius. The median $r_e$ of inactive pseudobulges is not
very different than the median $r_e$ of the most active group. In
Hubble type there is minimal evolution that is as one expects: most of
the earlier type galaxies have low specific star formation rate.

\section{Discussion}

\subsection{Identifying Pseudobulges}
 
Morphological identification of pseudobulges is straight-forward and
simple. However it is also subjective, and is limited to those bulges
with HST images. Therefore, we need to move to more objective methods
of pseudobulge identification that are based on quantifiable
properties. The S\'ersic index has been shown here, and in
\cite{fisherdrory2008}, to be a very good tool for identifying
bulges. However, because inactive pseudobulges are so similar to
classical bulges in both specific star formation rate and location in
photometric projections of the fundamental plane, using S\'ersic index
alone with no information about stellar populations or star formation
appears too simplistic.

We find three important groups of bulges in our sample. Thus we
propose the following criteria for diagnosing bulges: Pseudobulges
have S\'ersic index $n_b<2$ or nuclear morphology that is similar to a
disk galaxy, as described by \cite{kk04}. Furthermore, S\'ersic index
and nuclear morphology are tightly correlated. Also, these bulges
follow very different structural parameter correlations than
elliptical galaxies do.  There is not such a clear distinction in
mid-IR color (and hence in specific star formation rate). If the bulge
is active ($M_{3.6}-M_{8.0}>0$ which very roughly corresponds to SFR/M
$\sim$ 40~Gyr$^{-1}$) it almost always has low S\'ersic index or
disk-like optical morphology. If the bulge is not active then it does
not necessarily have a large S\'ersic index nor E-type nuclear
morphology. We refer to the bulges with low S\'ersic index or
disk-like nuclear morphology, but are not actively forming stars
($M_{3.6}-M_{8.0}<0$) as inactive pseudobulges.  We suggest that
classical bulges are those bulges with {\em both} E-type nuclear
morphology and $n_b>2$. These bulges coincide in parameter space with
photometric projections of the fundamental plane, and are inactive.

\begin{figure*}[t]
\includegraphics[width=0.99\textwidth]{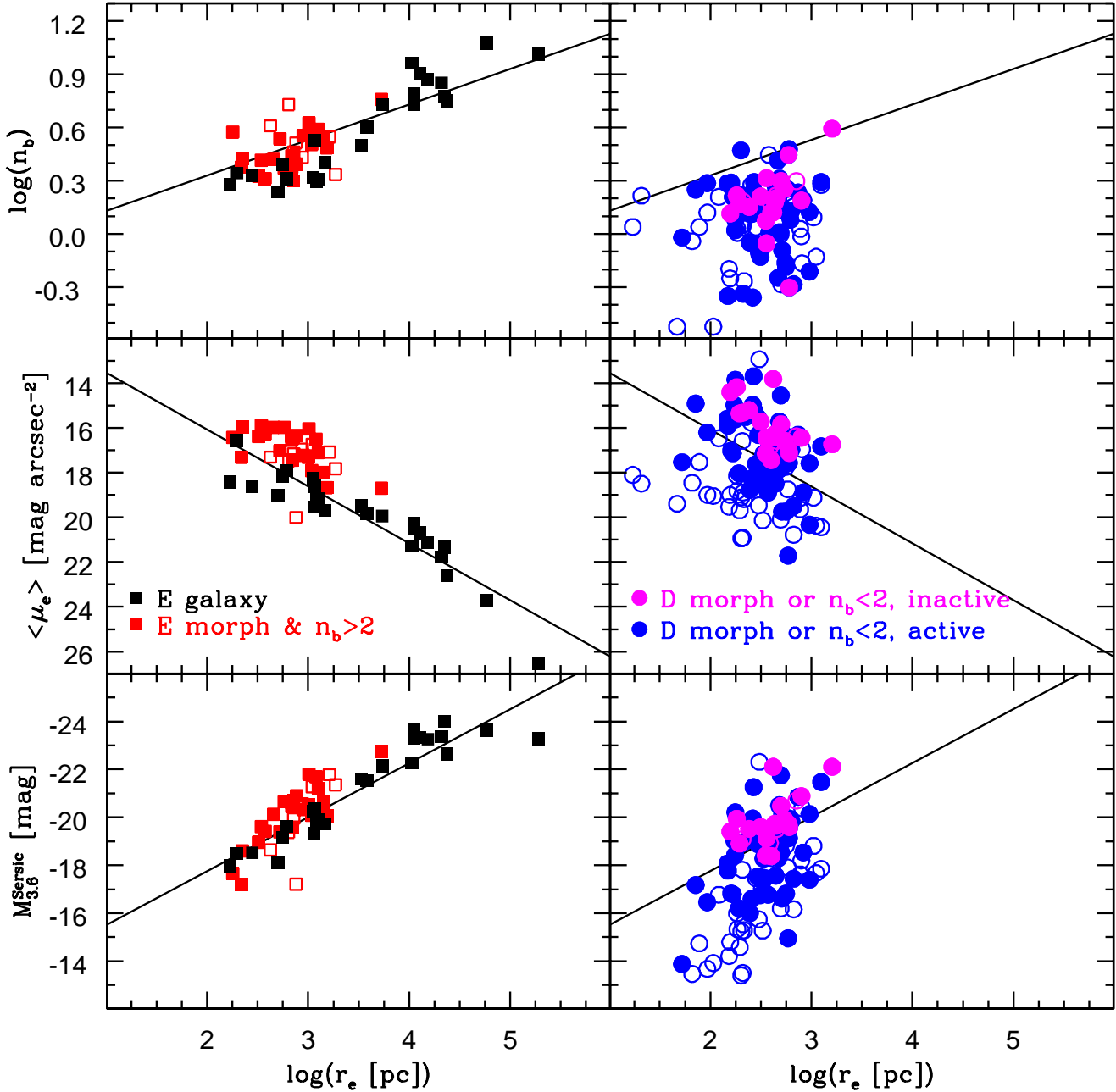}
\caption{Correlations of bulge magnitude
  (bottom), mean surface density (middle), and S\'ersic index (top)
  for classical bulges and elliptical galaxies (left) and pseudobulges
  (right). In both panels the black line represents a regression fit
  to the set of elliptical galaxies and classical bulges. Symbols are
  the same as in Fig.~\ref{fig:fundip}.  \label{fig:size}}
\end{figure*}


\subsection{What Do Structural Parameters Tell Us About Bulge Formation? }

Recently, progress has been made simulating the formation of
bulge-disk galaxies in hierarchical clustering scenarios through
additional channels such as ``clump'' instabilities that form in high
redshift disks and also secular evolution. The question remains what
type of bulges result in these simulations.  \cite{robertson2006} show
that rotationally supported disks can be formed in very gas rich
mergers which are more likely at high redshift. However, in their
suite of simulations the most common result was the formation of
bulge-dominated galaxies.  The formation of a disk galaxy in a late
merger by \cite{governato2009} is indeed remarkable, however the
resulting bulge-to-total ratio of the simulated galaxy of 0.65 is far
greater than any pseudobulge in our sample. Likewise, \cite{brook2007}
simulate a major merger which produces a kinematic profile indicative
of two nested disks, however the smaller disk is still larger than the
typical pseudobulge in our sample.  Indeed, the spatial resolution
limit in \cite{brook2007} is barely small enough to resolve the
typical pseudobulge in our sample.  \cite{elmegreen2008} finds that
clump instabilities in high redshift disk galaxies, as simulated by
\cite{noguchi1999} and more recently \cite{ceverino2009up}, are most
likely to form bulges that resemble classical
bulges. \cite{debattista2004} simulates bulge formation via internal
disk evolution with collisionless $N$-body simulations. They find that
simulated bulges built via secular evolution are able to maintain a
low S\'ersic index. However, \cite{fdf2009} show that present day star
formation leads to a significant addition of stellar mass to
pseudobulges, and therefore star formation and gas physics cannot be
neglected. \cite{eliche2006} simulate the growth of bulges in minor
mergers using collisionless $N$-body simulations, and finds that
merging increases the S\'ersic index of bulges from $n\sim1.5$ to
roughly $n\sim2$. Though, by neglecting star formation and dissipative
processes it becomes difficult to interpret their
results. Unfortunately, simulating galaxies with stars, gas, and star
formation is a difficult and computing-time-intensive process. It is
therefore rare that a simulation has a fine enough spatial resolution
to resolve the bulge for accurate S\'ersic decomposition. However, it
is fairly well accepted that elliptical galaxies are the result of
mergers \citep[recently][]{kfcb}. Thus, if bulges have different
scaling relations than elliptical galaxies, we can thus stipulate that
either they did not form in mergers, or something was significantly
different about the mergers (e.g.\ gas fraction or mass ratio) to
generate the different scaling relations.

Pseudobulges in our sample show little-to-no correlation between
half-light radius and any other bulge parameter.  In
Fig.~\ref{fig:size}, we re-illustrate the correlations of half-light
radius with bulge magnitude (bottom), mean surface brightness
(middle), and S\'ersic index (top). For clarity, in this figure we
plot classical bulges and elliptical galaxies (left) separate from
pseudobulges and inactive pseudobulges (right).  In each panel we also
show a black line representing the a regression fit to the elliptical
galaxies and classical bulges for the associated parameters. Over a 9
mag difference in bulge magnitude, pseudobulges remain roughly
constant in size. There is no change with surface brightness nor with
S\'ersic index.  Due to inherent parameter correlations in hot,
virialized systems (like the fundamental plane) it would be
essentially impossible for two pseudobulges to have roughly the same
half-light radius, and be 9~magnitudes different in luminosity.

In Figs.~\ref{fig:fund},~\ref{fig:mumag},~\&~\ref{fig:fundip} we show
that pseudobulges have a positive correlation in $<\mu>-M_{3.6}$. We
show this with luminosity and surface brightness derived from
S\'ersic fits and from non-parametric measurements of bulge luminosity
and density, thus it is not likely a relic of correlated errors in
S\'ersic fits.

How might a pseudobulge grow its mass without increasing size?
\cite{fdf2009} find that the star formation rate surface density of
pseudobulges correlates with stellar bulge mass and surface density of
stellar mass in such a way that more massive bulges have higher star
formation rate densities. If pseudobulges begin with very low mass,
but a variety of star formation rate densities, then after a period of
time those bulges with higher star formation rate density will have
higher stellar mass density.  In this scenario the total mass of the
bulge has not changed, but rather a mass element has changed state
from cold gas to stars. Therefore a star in orbit within the bulge
remains unaffected by the increase in stellar mass. Thus as {\em
  stellar} bulge mass increases, but the size of the bulge does not.
Furthermore, this means that the surface brightness and bulge
luminosity of the star light should be positively correlated, which
indeed is what we observe.

\begin{figure}[t]
\includegraphics[width=0.49\textwidth]{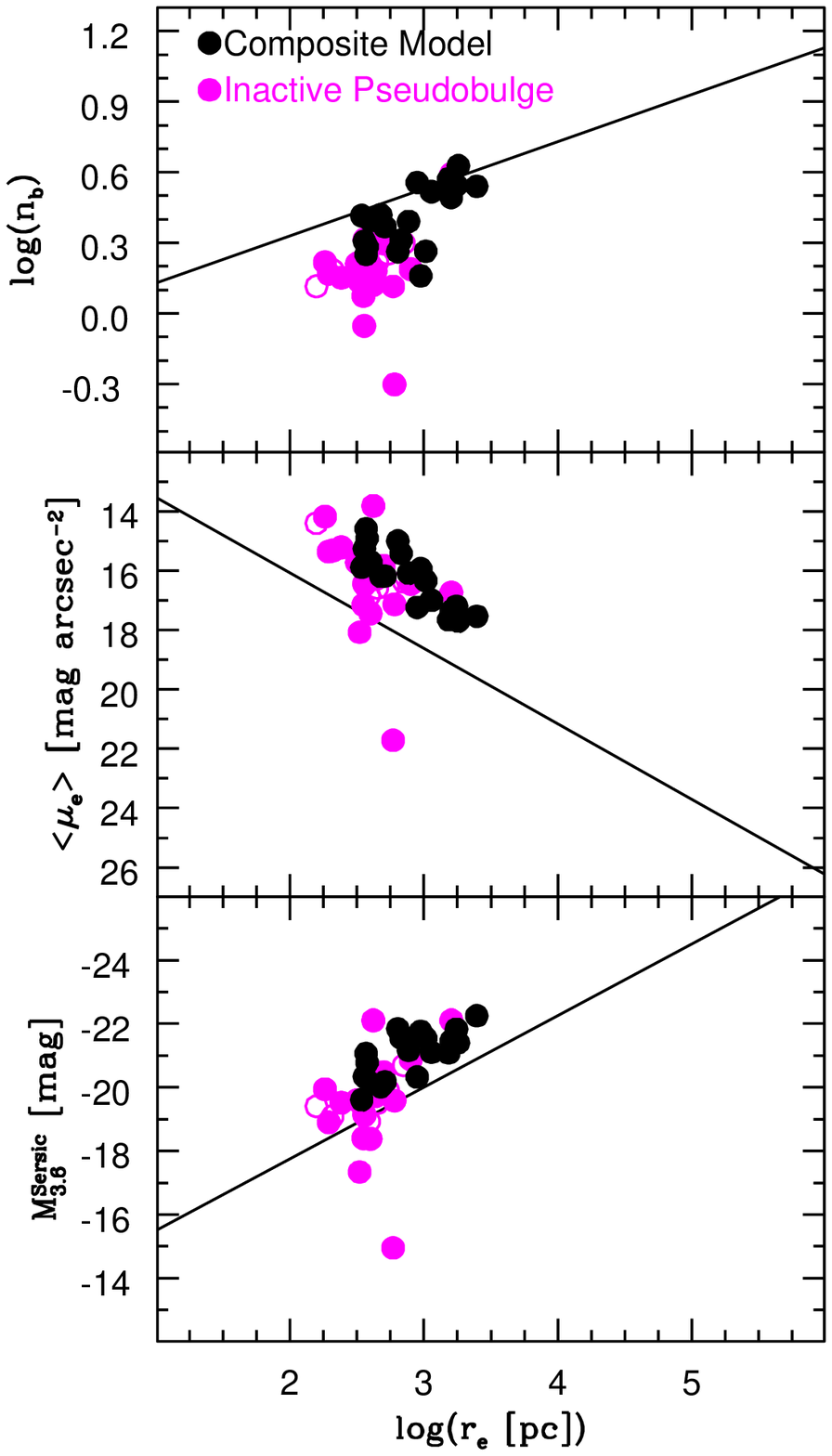}
\caption{Correlations of bulge magnitude
  (bottom), mean surface density (middle), and S\'ersic index (top)
  for inactive pseudobulges (magenta) and modeled composite
  profiles. In both panels the black line represents a regression fit
  to the set of elliptical galaxies and classical
  bulges.  \label{fig:composite}}
\end{figure}


\subsection{Are Inactive Pseudobulges Transition  Objects?}

Both \cite{droryfisher2007} and \cite{fdf2009} find pseudobulges that
are either residing in non-star forming galaxies or non-star forming
themselves. They are identified as pseudobulges by nuclear morphology
and S\'ersic index. It is worth noting that this finding illustrates
the necessity of combining information about the ISM or
stellar populations with other pseudobulge indicators, such as
S\'ersic index and morphology. In \S~6.1 we show that non-star forming
pseudobulges indeed exists in our sample as well. 

In Fig.~\ref{fig:fundip}, we show that inactive pseudobulges are not
inconsistent with the parameter correlations set by classical bulges
and elliptical galaxies. However, it is not obvious that they would
establish those correlations themselves. We have shown that inactive
pseudobulges are like active pseudobulges in two ways (morphology and
S\'ersic index) and like classical bulges in two ways (parameter
correlations and specific star formation rate), though there
similarity to classical bulges in activity is by definition. Therefore,
observationally speaking they are ambiguous objects.

\cite{fdf2009} find that in all bulges there is measurable star
formation in progress. However, in classical bulges that star
formation is unimportant when compared to the stellar mass of the
bulge. Furthermore, correlations between star formation rate density
and the stellar mass of the bulge and disk only exist for
pseudobulges, and the star formation is capable of doubling the
pseudobulge stellar mass in relatively short times. They conclude that
internal bulge growth is a nearly ubiquitous phenomenon.  It is simply
not important in classical-bulge galaxies. Also, in later-type
galaxies many studies \citep[e.g.][]{carollo2007,macarthur2009} show
that pseudobulges can have large populations of old stars within a
young bulge. Therefore there is room in the stellar population budget
for a low-luminosity classical bulge. If these are both true then it
would make sense that there be some cases where the pseudobulge and
classical bulge are similar mass.

What we wish to estimate is what would we observe if a low S\'ersic
pseudobulge were to exist in the same location as a classical
bulge. To study this we select two classical bulge galaxies, NGC~3521
and NGC~2841, and add a model pseudobulge to the profile. We then
decompose the resulting using Eq.~1. Both galaxies are chosen because
the original decomposition unaffected by non-S\'ersic components
(e.g.\ bars and/or rings), and the S\'ersic index is well
determined. The original S\'ersic indices of the real bulges are
$n_b=2.6$ and $3.6$, respectively.  We choose the model pseudobulge in
all cases to have $n_b=1.5$ and we fix the size according to the
scaling relation between pseudobulge size and disk scale length
\citep{courteau1996,macarthur2003,fisherdrory2008}, we choose
$r_e=0.2*h$. We then scale the pseudobulge luminosity to $0.5\times$,
$1\times$ , $2\times$, and $3\times$ that of the classical bulge. Note
that these relative weights are for the pseudobulges (and classical
bulge) 3.6~$\mu$m luminosity only. If these two different bulge
components have different mass-to-light ratios stemming from having
very different star formation histories, then a more sophisticated
method would need to be applied. Yet for our purposes, a
description of the luminosity is most important. We decompose the
resulting profile as usual, using Eq.~1, with a single S\'ersic
function for the bulge and an exponential outer disk. The resulting
profile of the composite bulge galaxy is well fit by a S\'ersic
function bulge and exponential outer disk.  In the two cases with a
high classical bulge S\'ersic index, and the
$L_{pseudo}/L_{classical}\geqq2$ we observe a small cusp in the
profile that looks similar to a nucleus. No other peculiarities are
visible in the constructed surface brightness profiles.

In Fig.~\ref{fig:composite} we investigate the location of the
composite pseudo+classical bulge models we describe above. The black
lines are from fits to classical bulges and elliptical galaxies. The
magenta circles represent the inactive pseudobulges from our survey,
and the black circles represent the composite model systems. The
composites are in exactly the same location as the inactive
pseudobulges. In every case, addition of a pseudobulge component to
the classical bulge galaxy results in lowering the S\'ersic
index. In the case in which the classical bulge has $n_b=2.6$ when
$L_{pseudo}\geqq L_{classical}$ the resulting S\'ersic index of the
composite system is less than two. For the case in which the classical
bulge has a larger S\'ersic index ($n_b=3.6$) the S\'ersic index of
the composite bulge only falls below $n_b=2$ for the most extreme case
in which $L_{pseudo}= 3\times L_{classical}$. Given that the mean
S\'ersic index of inactive pseudobulges is $<n_b>\sim1.6$, if this is
indeed what is occurring in inactive pseudobulges then
the underlying classical bulges in inactive pseudobulges would need to
have smaller S\'ersic index. 

\cite{fdf2009} find that in all bulge types there is bulge growth via
internal star formation. However, in classical bulges the specific
star formation is very low, unlike in pseudobulges where specific star
formation rates are higher. The results of \cite{fdf2009} and the
results of this experiment suggest that almost every bulge has a
component of new stars. In classical bulges the light of the new stars
is unimportant. In pseudobulges the new stars dominant the light of
the bulge component. The inactive pseudobulges may be systems in which
the two components (old and young) have similar luminosity and the
classical component has a low S\'ersic index. The questions remain: Is
there a low-luminosity cut off below which old components no longer
exists?  Also, is the old component of stars in pseudobulges (both
active and inactive) compatible with the properties of a typical
classical bulge?

\section{Summary \& Conclusions}

In this paper we develop a set of pseudobulge diagnostics that
incorporate morphology, S\'ersic index, and specific star formation
rate. To accomplish this we carry out bulge-disk decompositions on 173
E-Sd galaxies in the Spitzer~IRAC 3.6~$\mu$m band. We also measure the
3.6-8.0~$\mu$m color as a rough estimate of specific star formation
for all S0-Sd galaxies in the sample. We include in the appendix a
list of notes on bulge diagnosis for all S0-Sd galaxies in our sample.

We find that S\'ersic index and morphology are essentially
interchangeable at 3.6~$\mu$m as a means of identifying pseudobulges
and classical bulges, confirming the results of \cite{fisherdrory2008}
with near-IR data. Pseudobulges have $n_b<2$ and classical bulges have
$n_b>2$. Furthermore, with a much more robust sample we confirm that
the S\'ersic index of pseudobulges is uncorrelated with other bulge
structural properties.

A careful investigation of van~Albada's (\citeyear{vanalbada82})
result shows that simulated elliptical galaxies formed via more
violent collapses and lumpier initial conditions create resulting
surface brightness profiles with larger S\'ersic index. Also,
\cite{eliche2006} studies minor mergers onto disks with collisionless
$n$-body simulations and finds that mergers in general drive the
S\'ersic index of the bulge component up. Also,
\cite{hopkins2009_sersic_merging} finds that more merging generates
larger S\'ersic index. All three of these results seem fairly
straight-forward and quite intuitive. Surface brightness profiles with
larger S\'ersic index are characterized by having more light in the
tail of the distribution. Also, violent relaxation is a mechanism that
can easily convert ordered-rotational orbits to radial orbits, which
in turn puts more stellar light at larger radius. Thus more merging,
and hence more violent relaxation, puts more stellar mass at large
radius which in turn drives up S\'ersic index. Thus at the minimum the
now quite robust result that pseudobulges are marked by having very
low S\'ersic index seems to indicate that they have a more passive
history than classical bulges and elliptical galaxies.

We also study a group of pseudobulges (as indicated by S\'ersic index
and nuclear morphology) who are located in the same location in
$M_{3.6}-M_{8.0}$~vs.~$M_{3.6}$ parameter space. We call these bulges
``inactive pseudobulges''. We make a selection of $M_{3.6}<-18 AB~mag$
and $M_{3.6}-M_{8.0}<0$; note a similar (but much smaller) group of
bulges was found by \cite{fdf2009}).  We show that those inactive
pseudobulges are very similar to models of galaxies in which both a
pseudobulge and classical bulge exist in roughly equal parts and the
classical bulges has a low S\'ersic index. Additionally, star
formation in inactive pseudobulges must be suppressed.  Therefore,
pseudobulge identification that relies only on structural indicators
is unable to detect composite systems. Also pseudobulge identification
that relies only on stellar populations or star formation rates will
underestimate the number of pseudobulges significantly.

We also show that pseudobulges and classical bulges differ in
fundamental plane scaling relations. We find that the half-light
radius of pseudobulges does not correlate with any other bulge
parameter. Across 9~magnitudes in luminosity we find that the radial
size of pseudobulges changes only slightly. Thus it appears that the
only parameter in bulge-disk decompositions which is linked to
pseudobulge half-light radius is the scale length of the outer disk.
We find a positive correlation among pseudobulges between mean surface
brightness and bulge luminosity. Pseudobulges that are more luminous
are more dense. This relation is opposite that of classical bulges and
elliptical galaxies, and implies that pseudobulges are very different
types of objects than elliptical galaxies.

\acknowledgments

DBF wishes to thank J~Kormendy and the University of Texas at Austin
for support. ND and DBF thank the Max-Planck Society for support
during this project. We also thank J.~Kormendy and M.H.~Fabricius for
stimulating and valuable discussions. We also thank the anonymous
referee for helpful comments which contributed significantly to this
paper.

This work is based on observations made with the Spitzer Space
Telescope, which is operated by the Jet Propulsion Laboratory,
California Institute of Technology under a contract with NASA. Support
for this work was provided by NASA through an award issued by
JPL/Caltech. DBF acknowledges support by the National Science
Foundation under grant AST 06-07490.

Some of the data presented in this paper were obtained from the
Multi-mission Archive at the Space Telescope Science Institute
(MAST). STScI is operated by the Association of Universities for
Research in Astronomy, Inc., under NASA contract NAS5-26555. Support
for MAST for non-HST data is provided by the NASA Office of Space
Science via grant NAG5-7584 and by other grants and contracts.

This research has made use of the NASA/IPAC Extragalactic Database
(NED) which is operated by the Jet Propulsion Laboratory, California
Institute of Technology, under contract with the National Aeronautics
and Space Administration.


 \appendix
 \section{Diagnosing Bulge Types: Notes on Individual Galaxies}
 Here, we briefly describe the bulge diagnosis for all
 bulge-disk galaxies in our sample. For each galaxy we give the Hubble
 type from the RC3 \citep{rc3} and the Carnegie Atlas
 \citep{carnegieatlas} (RC3 first, then Carnegie Atlas separated by a
 slash). Our diagnosis of the bulge is then stated, followed by a brief
 motivation for that diagnosis.
  \\
 {\bf IC~342} .SXT6../... Pseudobulge. The bulge shows a strong nuclear spiral and near the very center breaks into large clumps, that are almost a ring shape. The decomposition yields $n_b=1.8$, and the bulge is actively forming stars with $M_{3.6}-M_{8.0}=1.9$~mag. \\
 {\bf IC~749} .SXT6../SBc(rs) Pseudobulge. The decomposition yields $n_b=1.4$, and the bulges is actively forming stars with $M_{3.6}-M_{8.0}=1.0$~mag.  There is no optical image in the HST archive.  \\
 {\bf NGC~300} .SAS7../Sc(s) No Bulge/Pseudobulge. The ``bulge'' is probably better described as a nuclear star cluster. Nonetheless, a decomposition yields $n_b=1.6$, and the bulges is inactive with $M_{3.6}-M_{8.0}=-1.4$~mag. \\
 {\bf NGC~404} .LAS-*./S0 Classical bulge. There is a very small
 nuclear spiral, however that is embedded within a much larger bulge
 that shows little-to-no substructure. The luminosity of the galaxy
 contained within the small nuclear spiral only accounts for 0.8\% of
 the total light, as compared to the larger bulge which is roughly 1/4
 of the total light. The decomposition yields $n_b=3.7$, and the bulges is inactive with $M_{3.6}-M_{8.0}=-0.5$~mag.\\
 {\bf NGC~628} .SAS5../Sc(s) Pseudobulge. The spiral arms of the outer disk extend all the way to the central $r\sim1$\", whereas the bulge begins to dominate the surface brightness profile at $r\sim14$\". The decomposition yields $n_b=1.6$, and the bulges is actively forming stars with $M_{3.6}-M_{8.0}=0.2$~mag.\\
 {\bf NGC~672} .SBS6../SBc(s) Pseudobulge. The bulge is extremely elongated, and shows patchiness that is similar to what is seen in late-type disk galaxies. The decomposition yields $n_b=1.1$, and the bulges is actively forming stars with $M_{3.6}-M_{8.0}=2.2$~mag.\\
 {\bf NGC~925} .SXS7../SBc(s) Pseudobulge. The decomposition yields
 $n_b=0.7$, and the bulges is actively forming stars with
 $M_{3.6}-M_{8.0}=0.3$~mag. The optical HST images are not well
 centered, making the morphology uncertain. Nonetheless, there appears
 to have patchiness that is similar to what is seen in late-type disk
 galaxies.\\
 {\bf NGC~1023} .LBT-../SB0 Classical Bulge. The bulge looks very much like and E type, there is no sign of spiral structure of any disk-like morphology. The decomposition yields $n_b=2.5$, and the bulges is inactive with $M_{3.6}-M_{8.0}=-1.1$~mag.\\
 {\bf NGC~1058} .SAT5../Sc(s) Pseudobulge. There is a prominent, nearly
 face-on spiral, that extends to the central $lesssim1$\" .  The
 decomposition yields $n_b=1.1$, and the bulges is actively forming
 stars with $M_{3.6}-M_{8.0}=0.8$~mag.\\
 {\bf NGC~1097} .SBS3../RSBbc(s) Pseudobulge. The bulge has a prominent
 nuclear ring. The nuclear ring is evident in the surface brightness
 profile from 5-15\" even in the near-IR. The decomposition yields
 $n_b=2.0$, and the bulges is actively forming stars with
 $M_{3.6}-M_{8.0}=1.3$~mag.\\
 {\bf NGC~1313} .SBS7../SBc(s) Pseudobulge. The bulge shows patchiness
 that is similar to what is seen in late-type disk galaxies. The
 decomposition yields $n_b=1.3$, and the bulges is inactive with
 $M_{3.6}-M_{8.0}=-0.7$~mag.\\
 {\bf NGC~1317} .SXR1../Sa Pseudobulge. The bulge has a prominent
 nuclear bar and spiral.  The decomposition yields $n_b=1.4$, and the
 bulges is actively forming stars with $M_{3.6}-M_{8.0}=0.4$~mag.\\
 {\bf NGC~1433} PSBR2../SBb(s) Pseudobulge. The bulge has a prominent
 nuclear spiral. The decomposition yields $n_b=0.8$, and the bulges is
 actively forming stars with $M_{3.6}-M_{8.0}=0.3$~mag.\\
 {\bf NGC~1512} .SBR1../SBb(rs) pec Pseudobulge. The bulge has a
 prominent nuclear ring and spiral. The decomposition yields $n_b=1.8$,
 and the bulges is actively forming stars with
 $M_{3.6}-M_{8.0}=0.2$~mag.\\
 {\bf NGC~1543} RLBS0../RSB0/a Pseudobulge. There is a prominent
 nuclear bar. The decomposition yields $n_b=1.5$, and the bulges is not
 actively forming stars with
 $M_{3.6}-M_{8.0}=-1.2$~mag.\\
 {\bf NGC~1559} .SBS6../SBc(s) Pseudobulge. A flocculent spiral extends
 all the way to the center of the galaxy. The decomposition yields
 $n_b=.7$, and the bulges is actively forming stars with
 $M_{3.6}-M_{8.0}=5.4$~mag.\\
 {\bf NGC~1566} .SXS4../Sbc(s) Pseudobulge. The bulge shows a bright
 two armed spiral at the center which becomes more diffuse around
 $r\sim 1-2$\" . The decomposition yields $n_b=1.6$, and the bulges is
 actively forming stars with $M_{3.6}-M_{8.0}=0.2$~mag. \\
 {\bf NGC~1617} .SBS1../Sa(s) Classical Bulge. There is a mild
 (non-spiral) dust lane slightly off center from the bulge center,
 otherwise the bulge has a smooth E-like morphology. The decomposition
 yields $n_b=2.1$, and the bulges is not forming stars with
 $M_{3.6}-M_{8.0}=-1.0$~mag.\\
 {\bf NGC~1637} .SXT5../SBc(s) Pseudobulge. A flocculent spiral extends
 all the way to the central arcsecond, where it becomes elongated and
 clumpy. The decomposition yields $n_b=1.6$, and the bulges is
 actively forming stars with $M_{3.6}-M_{8.0}=3.4$~mag. \\
 {\bf NGC~1672} .SBS3../Sb(rs) Pseudobulge. There is a bright spiral
 arm that almost completely wraps around the center of the bulge. Near
 $r\sim6$\" this spiral breaks into more spiral arms. The decomposition
 yields $n_b=2.1$, and the bulges is
 actively forming stars with $M_{3.6}-M_{8.0}=2.0$~mag. \\
 {\bf NGC~1744} .SBS7../SBcd(s) No bulge/Pseudobulge. The ``bulge'' is probably better described as a nuclear star cluster. Nonetheless, a decomposition yields $n_b=0.5$, and the bulges is actively forming stars with $M_{3.6}-M_{8.0}=5.2$~mag. \\
 {\bf NGC~1808} RSXS1../Sbc pec Pseudobulge.  The bulge has a very high
 surface brightness flocculent spiral. The decomposition yields
 $n_b=0.8$, and the bulges is actively forming stars with $M_{3.6}-M_{8.0}=3.3$~mag. \\
 {\bf NGC~2403} .SXS6../Sc(s) Pseudobulge. The bulge is comprised of
 several clumps of stars, reminiscent of a very late-type disk
 galaxy. The decomposition yields
 $n_b=0.7$, and the bulges is actively forming stars with $M_{3.6}-M_{8.0}=1.2$~mag. \\
 {\bf NGC~2500} .SBT7../Sc(s) Pseudobulge. The bulge is comprised of
 several clumps of stars, reminiscent of a very late-type disk
 galaxy. The decomposition yields
 $n_b=1.7$, and the bulges is actively forming stars with $M_{3.6}-M_{8.0}=0.4$~mag. \\
 {\bf NGC~2655} .SXS0../Sa pec Classical Bulge. This bulge has
 moderately prominent dust-lanes, however it is not clear if the dust
 is spiral. The decomposition yields
 $n_b=2.4$, and the bulges is not active with $M_{3.6}-M_{8.0}=-0.6$~mag. \\
 {\bf NGC~2685} RLB.+P./S0 pec Classical Bulge. There is a dust lane
 that is offset from the center of the bulge, it is possibly
 disk-contamination. Otherwise the bulge is smooth, and appears similar
 to an E-type. The decomposition yields
 $n_b=2.1$, and the bulges is not active with $M_{3.6}-M_{8.0}=-1.1$~mag. \\
 {\bf NGC~2775} .SAR2../Sa(r) Classical Bulge. There is a clear break
 in the disk near $r~22$\" , inside which the bulge is smooth and
 E-type. The decomposition yields $n_b=3.5$, and the bulges is not active with $M_{3.6}-M_{8.0}=-0.4$~mag. \\
 {\bf NGC~2841} .SAR3*./Sb Classical Bulge. At the very center of the bulge there is a small nuclear spiral, however the vast majority of bulge light is dominated by smooth isophotes, that appears similar to an E-type galaxy. The decomposition yields $n_b=3.6$, and the bulges is not active with $M_{3.6}-M_{8.0}=-0.8$~mag. \\
 {\bf NGC~2903} .SXT4../Sc(s) Pseudobulge. The bulge light is dominated
 by a flocculent spiral that breaks into clumps in the center.  The
 decomposition yields
 $n_b=0.5$, and the bulge is actively forming stars with $M_{3.6}-M_{8.0}=2.2$~mag. \\
 {\bf NGC~2950} RLBR0../RSB0 Pseudobulge. The bulge shows a nuclear
 bar, but otherwise shows little spiral structure.  The decomposition
 yields $n_b=1.3$, and the bulge is not active with $M_{3.6}-M_{8.0}=-1.3$~mag. \\
 {\bf NGC~2964} .SXR4*./Sc(s) The bulge has a strong nearly face-on
 nuclear spiral. The decomposition
 yields $n_b=1.0$, and the bulge is actively forming stars with $M_{3.6}-M_{8.0}=2.0$~mag. \\
 {\bf NGC~2976} .SA.5P./Sd No bulge/Pseudobulge. The center of the
 galaxy shows flocculent spiral structure. However, there is not a
 significant rise in surface brightness. Indeed, the decomposition
 shows $B/T\sim0.1$\%, thus it is probably best to say the galaxy has
 no bulge.  The decomposition
 yields $n_b=1.0$, and the bulge is actively forming stars with $M_{3.6}-M_{8.0}=1.4$~mag. \\
 {\bf NGC~2997} .SXT5../Sc(s) Pseudobulge. The bulge is comprised of a
 bright nuclear spiral that nearly forms a ring at $r\sim5$\" . The
 decomposition
 yields $n_b=1.0$, and the bulge is actively forming stars with $M_{3.6}-M_{8.0}=1.5$~mag. \\
 {\bf NGC~3031} .SAS2../Sb(r) Classical Bulge. There is a dust lane
 that extends to near the center of the galaxy, however most of the
 light is dominated by smooth isophotes that are resemble an E galaxy.
 The decomposition yields
 $n_b=3.9$, and the bulges is not active with $M_{3.6}-M_{8.0}=-0.5$~mag. \\
 {\bf NGC~3032} .LXR0../RSa pec Pseudobulge. The bulge has a prominent
 spiral that extends to the center of the galaxy. The decomposition
 yields $n_b=2.6$, and the bulge is actively forming stars with $M_{3.6}-M_{8.0}=1.4$~mag. \\
 {\bf NGC~3156} .L...*./ S0 Pseudobulge. There is a dust-lane that near
 the center becomes a nuclear spiral. The morphology is some what
 uncertain.  The decomposition
 yields $n_b=1.7$, and the bulge is not active with $M_{3.6}-M_{8.0}=-0.4$~mag. \\
 {\bf NGC~3184} .SXT6../Sc(r) Pseudobulge. The bulge has a prominent
 spiral that extends to the center.  The decomposition
 yields $n_b=1.7$, and the bulge is actively forming stars with $M_{3.6}-M_{8.0}=1.4$~mag. \\
 {\bf NGC~3185} RSBR1../SBa(s) Pseudobulge. The bulge has a nuclear
 ring that has radius of about $r\sim2$\" , with spiral arms extending
 off the ring.  The decomposition yields $n_b=1.0$, and the bulge is
 actively forming stars with
 $M_{3.6}-M_{8.0}=0.6$~mag. \\
 {\bf NGC~3190} .SAS1P/Sa Pseudobulge. The galaxy is somewhat inclined
 which makes morphology somewhat uncertain, however there appears to be
 spiral structure extending to the center of the galaxy.  The
 decomposition
 yields $n_b=2.0$, and the bulge is not active with $M_{3.6}-M_{8.0}=-0.4$~mag. \\
 {\bf NGC~3198} .SBT5../Sc(s) Pseudobulge. The bulge shows a prominent
 spiral. The decomposition yields $n_b=1.3$, and the bulge is actively
 forming stars with
 $M_{3.6}-M_{8.0}=1.2$~mag. \\
 {\bf NGC~3319} .SBT6./SBc(s) Pseudobulge. The bulge is very elongated,
 and broken into several star clusters. The decomposition yields
 $n_b=0.6$, and the bulge is actively forming stars with
 $M_{3.6}-M_{8.0}=1.2$~mag. \\
 {\bf NGC~3344} RSXR4./Sbc(rs) Classical Bulge. The bulge shows no sign
 of spiral structure, and is morphologically similar to an E
 galaxy. The decomposition
 yields $n_b=2.4$, and the bulge is not active with $M_{3.6}-M_{8.0}=-0.3$~mag. \\
 {\bf NGC~3351} .SBR3./SBb(r) Pseudobulge. The bulge has a nuclear
 spiral and a prominent nuclear ring. The decomposition yields
 $n_b=1.5$, and the bulge is actively forming stars with
 $M_{3.6}-M_{8.0}=0.9$~mag. \\
 {\bf NGC~3368} .SXT2./Sab(s) Pseudobulge. The bulge has a strong
 nuclear spiral.  The decomposition yields $n_b=1.6$, and the bulge is
 not active with
 $M_{3.6}-M_{8.0}=-0.2$~mag. \\
 {\bf NGC~3384} .LBS-*./SB0$_1$ Pseudobulge. The bulge has a nuclear
 bar. The decomposition yields $n_b=1.4$, and the bulge is not active
 with
 $M_{3.6}-M_{8.0}=-1.3$~mag. \\
 {\bf NGC~3412} .LBS0../SB0 Classical Bulge. The bulge shows no sign of
 spiral structure, and is morphologically similar to an E galaxy. The
 decomposition
 yields $n_b=2.6$, and the bulge is not active with $M_{3.6}-M_{8.0}=-1.3$~mag. \\
 {\bf NGC~3486} .SXR5./Sbc(r) Pseudobulge. The bulge shows spiral
 structure extending from the ring (at $r\sim 15$\") to the center of
 the galaxy.  The decomposition yields $n_b=1.6$, and the bulge is
 actively forming stars with
 $M_{3.6}-M_{8.0}=1.6$~mag. \\
 {\bf NGC~3489} .LXT+../S0-Sa Pseudobulge. The bulge has a nuclear
 spiral, which becomes a nuclear bar in the central few arcseconds. The
 decomposition yields $n_b=1.5$, and the bulge is not active
 $M_{3.6}-M_{8.0}=-0.7$~mag. \\
 {\bf NGC~3511} .SAS5./Sc(s) Pseudobulge. The decomposition yields
 $n_b=1.6$, and the bulge is actively forming stars
 $M_{3.6}-M_{8.0}=2.0$~mag. There is not optical HST image.\\
 {\bf NGC~3521} .SXT4./Sbc(s) Classical Bulge. There is a distinct
 change in morphology from the outer disk to the bulge, near $r\sim9$\"
 inside of which the bulge appears similar to an E-type galaxy.  The
 decomposition yields $n_b=2.6$, and the bulge is actively forming
 stars
 $M_{3.6}-M_{8.0}=0.4$~mag.\\
 {\bf NGC~3593} .SAS0*./Sa pec Pseudobulge. The bulge shows a prominent
 nuclear spiral. The decomposition yields $n_b=0.8$, and the bulge is
 actively forming stars
 $M_{3.6}-M_{8.0}=1.9$~mag.\\
 {\bf NGC~3621} .SAS7./Sc(s) Pseudobulge. The bulge light is dominated
 by a flocculent spiral.  The decomposition yields $n_b=2.8$, and the
 bulge is actively forming stars
 $M_{3.6}-M_{8.0}=0.8$~mag.\\
 {\bf NGC~3675} .SAS3./Sb(r) Pseudobulge. The bulge light is dominated
 by a flocculent spiral.  The decomposition yields $n_b=1.6$, and the
 bulge is actively forming stars
 $M_{3.6}-M_{8.0}=0.5$~mag.\\
 {\bf NGC~3726} .SXR5./Sbc(rs) Pseudobulge. The decomposition yields
 $n_b=0.8$, and the bulge is actively forming stars
 $M_{3.6}-M_{8.0}=1.3$~mag. There is no optical HST image. \\
 {\bf NGC~3906} .SBS7./... Pseudobulge. This bulge is broken into
 several nuclear star clusters, similar to a very late-type galaxy.
 The decomposition yields $n_b=1.0$, and the bulge is
 actively forming stars $M_{3.6}-M_{8.0}=1.0$~mag.\\
 {\bf NGC~3938} .SAS5./Sc(s) Pseudobulge. The bulge shows a prominent
 face-on flocculent spiral.  The decomposition yields $n_b=1.3$, and
 the bulge is
 actively forming stars $M_{3.6}-M_{8.0}=1.1$~mag.\\
 {\bf NGC~3941}.LBS0../ SB0/a. Pseudobulge. The decomposition yields
 $n_b=1.5$, and
 the bulge is not active $M_{3.6}-M_{8.0}=-1.2$~mag. There is no optical HST image.\\
 {\bf NGC~3945} RLBT+../RSB0 Pseudobulge. The bulge shows a strong
 nuclear bar.The decomposition yields $n_b=1.5$, and the bulge is not
 active $M_{3.6}-M_{8.0}=-1.3$~mag.\\
 {\bf NGC~3953} .SBR4./SBbc(r) Pseudobulge. The decomposition yields
 $n_b=1.5$, and the bulge is not
 active $M_{3.6}-M_{8.0}=-0.4$~mag. There is not optical HST image.\\
 {\bf NGC~3982} .SXR3*/... Pseudobulge. The bulge shows a prominent
 face-on nuclear spiral.  The decomposition yields
 $n_b=1.9$, and the bulge is actively forming stars $M_{3.6}-M_{8.0}=0.6$~mag. \\
 {\bf NGC~3990} .L..-* / S0-Sa Pseudobulge.  The decomposition yields
 $n_b=1.1$, and the bulge is actively forming stars $M_{3.6}-M_{8.0}=1.1$~mag. There is not optical HST image.\\
 {\bf NGC~4020} .SB.7? /... No Bulge/Pseudobulge. The decomposition
 yields
 $n_b=0.3$, and the bulge is actively forming stars $M_{3.6}-M_{8.0}=0.5$~mag. There is not optical HST image.\\
 {\bf NGC~4117} .L..0*./... Pseudobulge. The bulge light shows a spiral
 pattern.  The decomposition yields
 $n_b=1.4$, and the bulge is not active  $M_{3.6}-M_{8.0}=-0.22$~mag.\\
 {\bf NGC~4136} .SXR5./Sc(r) Pseudobulge. The decomposition yields
 $n_b=0.6$, and the bulge is actively forming stars  $M_{3.6}-M_{8.0}=0.94$~mag. There is not optical HST image.\\
 {\bf NGC~4138} .LAR+../Sa(r) Pseudobulge. The bulge morphology is
 somewhere between pseudo- and classical.  The decomposition yields
 $n_b=1.6$, and the bulge is not active  $M_{3.6}-M_{8.0}=-0.95$~mag.\\
 {\bf NGC~4150} .LAR0\$./S0-Sa Classical Bulge. There is a very weak
 spiral dust pattern, however the center of the spiral is not
 coincident with the center of the stellar light. Also, most of the
 bulge light appears to come from a classical like component in which
 the dust resides.  The decomposition yields
 $n_b=5.4$, and the bulge is slightly active $M_{3.6}-M_{8.0}=0.15$~mag.\\
 {\bf NGC~4203} .LX.-*./S0 Classical Bulge. Other than a thin-faint
 wisp of dust, the bulge light is smooth and looks very similar to an
 E-type galaxy.  The decomposition yields
 $n_b=2.8$, and the bulge is mildly active with $M_{3.6}-M_{8.0}=0.26$~mag.\\
 {\bf NGC~4237} .SXT4./Sc(r) Pseudobulge. The galaxy is a nearly
 face-on flocculent spiral that extends all the way into the resolution
 limit of HST.  The decomposition yields
 $n_b=1.3$, and the bulge is actively forming stars with $M_{3.6}-M_{8.0}=0.58$~mag.\\
 {\bf NGC~4254} .SAS5../Sc(s) Pseudobulge. The galaxy is a nearly
 face-on spiral that extends all the way into the resolution limit of
 HST. The decomposition yields
 $n_b=1.7$, and the bulge is actively forming stars with $M_{3.6}-M_{8.0}=1.51$~mag.\\
 {\bf NGC~4258} .SXS4./Sb(s) Pseudobulge (possibly Classical
 Bulge). The bulge has a nuclear spiral. Their is an opaque dust-lane
 that is off center that may affect morphology.  The decomposition
 yields
 $n_b=2.8$, and the bulge is mildly forming stars with $M_{3.6}-M_{8.0}=0.04$~mag.\\
 {\bf NGC~4267} .LBS-\$./... Classical Bulge. The bulge has smooth
 undisturbed isophotes, no sign of disk-like morphology. The
 decomposition yields
 $n_b=4.2$, and the bulge is not active with $M_{3.6}-M_{8.0}=-1.30$~mag.\\
 {\bf NGC~4274} RSBR2./Sa(sr) Pseudobulge. The bulge has a prominent
 nuclear spiral, and possibly a moderately inclined nuclear ring.  The
 decomposition yields
 $n_b=1.6$, and the bulge is mildly forming stars with $M_{3.6}-M_{8.0}=0.12$~mag.\\
 {\bf NGC~4293} RSBS0../Sa Pseudobulge.  The bulge is flattened and
 shows a patchy, moderately spiral pattern. The decomposition yields
 $n_b=1.9$, and the bulge is actively forming stars with
 $M_{3.6}-M_{8.0}=1.68$~mag.\\
 {\bf NGC~4294} .SBS6./SBc(s) Pseudobulge. The decomposition yields
 $n_b=1.7$, and the bulge is actively forming stars with
 $M_{3.6}-M_{8.0}=2.31$~mag. There is no optical HST image.\\
 {\bf NGC~4303} .SXT4../Sc(s) Pseudobulge. The bulge shows a tightly
 wound spiral, that extends down to a small nuclear bar. The
 decomposition yields $n_b=1.7$, and the bulge is actively forming
 stars with
 $M_{3.6}-M_{8.0}=1.00$~mag.\\
 {\bf NGC~4314} .SBT1./SBa(rs) pec Pseudobulge. The bulge has a very
 prominent nuclear ring, and nuclear spiral. The decomposition yields
 $n_b=2.9$, and the bulge is mildly forming stars with
 $M_{3.6}-M_{8.0}=0.35$~mag.\\
 {\bf NGC~4321} .SXS4../Sc(s) Pseudobulge. The bulge has a two very
 prominent spiral arms that almost form a ring.  The decomposition
 yields $n_b=1.2$, and the bulge is actively forming stars with
 $M_{3.6}-M_{8.0}=2.41$~mag.\\
 {\bf NGC~4371} .LBR+../SB0(r) Pseudobulge. This bulge shows
 little-to-no spiral structure, however there is a nuclear ring (near
 $r\sim5$\") and a nuclear bar. The decomposition yields $n_b=3.9$, and
 the bulge is not active with
 $M_{3.6}-M_{8.0}=-1.21$~mag.\\
 {\bf NGC~4380} .SAT3*/Sab(s) Pseudobulge. The galaxy has a flocculent
 spiral that extends into the central tens of parsecs. The
 decomposition yields $n_b=1.8$, and the bulge is actively forming
 stars with
 $M_{3.6}-M_{8.0}=0.46$~mag.\\
 {\bf NGC~4394} RSBR3./SBb(sr) Pseudobulge. There is a weak face-on
 spiral at that comprises the central $r\sim5$\" of the bulge. The
 decomposition yields $n_b=2.0$, and the bulge is not active with
 $M_{3.6}-M_{8.0}=-1.92$~mag.\\
 {\bf NGC~4413} PSBT2*/... Pseudobulge.  The decomposition yields
 $n_b=1.1$, and the bulge is actively forming stars with
 $M_{3.6}-M_{8.0}=1.83$~mag. There is no optical HST image.\\
 {\bf NGC~4414} .SAT5\$/Sc(sr) Pseudobulge. The bulge shows spiral
 structure, and there is little-to-no break from the spiral of the
 outer disk.  The decomposition yields $n_b=0.4$, and the bulge is
 actively forming stars with
 $M_{3.6}-M_{8.0}=1.11$~mag.\\
 {\bf NGC~4419} .SBS1./Sa Pseudobulge. The decomposition yields
 $n_b=1.9$, and the bulge is actively forming stars with
 $M_{3.6}-M_{8.0}=0.84$~mag. There is no optical HST image.\\
 {\bf NGC~4421} .SBS0../... Classical Bulge. The decomposition yields
 $n_b=2.7$, and the bulge is actively forming stars with
 $M_{3.6}-M_{8.0}=-1.18$~mag. There is no optical HST image.\\
 {\bf NGC~4424} .SBS1*/Sa pec Pseudobulge. The bulge is broken into
 several knots (or star clusters) threaded with a dust-lane.  The
 decomposition yields $n_b=0.7$, and the bulge is actively forming
 stars with
 $M_{3.6}-M_{8.0}=1.09$~mag. \\
 {\bf NGC~4442} .LBS0../SB0 Classical Bulge. The bulge isophotes are
 smooth and very much like an E-type galaxy. The decomposition yields
 $n_b=2.4$, and the bulge is not active with
 $M_{3.6}-M_{8.0}=-1.17$~mag. \\
 {\bf NGC~4448} .SBR2./Sa(late) Pseudobulge. The bulge shows a mild
 spiral pattern that continues to its center. The decomposition yields
 $n_b=1.2$, and the bulge is not active with
 $M_{3.6}-M_{8.0}=-0.29$~mag. \\
 {\bf NGC~4450} .SAS2../Sab pec Pseudobulge. The bulge shows a spiral
 pattern. The decomposition yields $n_b=2.1$, and the bulge is not
 active with
 $M_{3.6}-M_{8.0}=-0.66$~mag. \\
 {\bf NGC~4457} RSXS0../RSb(s) Pseudobulge.  The decomposition yields
 $n_b=1.4$, and the bulge is mildly active with
 $M_{3.6}-M_{8.0}=0.20$~mag.  There is no optical HST image.\\
 {\bf NGC~4491} .SBS1*/... Pseudobulge. The decomposition yields
 $n_b=0.9$, and the bulge is not active with
 $M_{3.6}-M_{8.0}=-2.8$~mag.  There is no optical HST image.\\
 {\bf NGC~4498} .SXS7../... Pseudobulge. The bulge is very thin,
 elongated, and broken into clumps of stars.  The decomposition yields
 $n_b=1.0$, and the bulge is actively forming stars with
 $M_{3.6}-M_{8.0}=1.48$~mag. \\
 {\bf NGC~4501} .SAT3../Sbc(s) Pseudobulge. The bulge is a nuclear
 spiral that becomes radial near $r\sim8$\" , and this may be
 indicating that the bulge also contains a nuclear bar.  The
 decomposition yields $n_b=0.9$, and the bulge is not active with
 $M_{3.6}-M_{8.0}=-0.37$~mag. \\
 {\bf NGC~4519} .SBT7./SBc(rs) Pseudobulge. The bulge is a spiral that
 breaks into several clumps near the center.  The decomposition yields
 $n_b=0.8$, and the bulge is actively forming stars with
 $M_{3.6}-M_{8.0}=1.76$~mag. \\
 {\bf NGC~4526} .LXS0*./S0 Classical Bulge. This is not a standard
 bulge. The bulge is very large and when viewed at high contrast looks
 very much like an E-type galaxy. However, in the central $\sim$10\"
 there is a spiral that is not continuing the spiral of the outer disk,
 but rather seems embedded within a smooth E-like bulge. We decide that
 most of the light is dominated by the component that appears
 classical.  The decomposition yields $n_b=3.5$, and the bulge is not
 very active with
 $M_{3.6}-M_{8.0}=0.03$~mag.\\
 {\bf NGC~4548} .SBT3./... Classical Bulge. The decomposition yields
 $n_b=2.9$, and the bulge is not active with
 $M_{3.6}-M_{8.0}=-0.66$~mag. There is no optical HST image.\\
 {\bf NGC~4559} .SXT6./Sc(s) Pseudobulge. The bulge is a nuclear spiral
 that breaks into a few star clusters in the center of the galaxy.  The
 decomposition yields $n_b=1.1$, and the bulge is actively forming
 stars with
 $M_{3.6}-M_{8.0}=1.07$~mag. \\
 {\bf NGC~4569} .SXT2./Sab(s) Pseudobulge. The bulge shows a nuclear
 spiral that becomes elongated near the center.  The decomposition
 yields $n_b=2.0$, and the bulge is actively forming stars with
 $M_{3.6}-M_{8.0}=1.38$~mag.\\
 {\bf NGC~4571} .SAR7./Sc(s) Pseudobulge. The bulge shows a prominent
 face-on flocculent spiral that extends from the outer disk all the way
 to the center of the galaxy. The
 decomposition yields $n_b=1.9$. There is no 8~$\mu$m data. \\
 {\bf NGC~4578} .LAR0*./S0 Classical Bulge. The bulge light is very
 smooth and featureless, much like an E-galaxy.  The decomposition
 yields $n_b=3.7$, and the bulge is not active with
 $M_{3.6}-M_{8.0}=-1.26$~mag.\\
 {\bf NGC~4580} .SXT1P/Sc(s)-Sa Pseudobulge.  The decomposition yields
 $n_b=0.6$, and the bulge is actively forming stars with
 $M_{3.6}-M_{8.0}=1.92$~mag. There is no available optical HST image.\\
 {\bf NGC~4605} .SBS5P/Sc(s) No bulge/Pseudobulge. The 'bulge' is
 really almost non-existent; the decomposition yields
 $B/T\sim0.1$\%. There morphology of the bulge light is broken into
 several star clusters, and there is no obvious center of the galaxy
 from looking at the image.  The decomposition yields $n_b=0.6$, and
 the bulge is actively forming stars with
 $M_{3.6}-M_{8.0}=1.47$~mag.\\
 {\bf NGC~4612} RLX.0../E5 Classical Bulge.  The decomposition yields
 $n_b=3.7$, and the bulge is not active with
 $M_{3.6}-M_{8.0}=-1.23$~mag. There is no optical HST image available.\\
 {\bf NGC~4639} .SXT4./SBb(r) Pseudobulge. The galaxy has a spiral that
 continues unbroken into the center of the bulge. The decomposition
 yields $n_b=1.3$, and the bulge is not active with
 $M_{3.6}-M_{8.0}=-0.27$~mag.\\
 {\bf NGC~4651} .SAT5./Sc(r) Pseudobulge. The bulge is a nuclear
 spiral.  The decomposition yields $n_b=1.6$, and the bulge is actively
 forming stars with
 $M_{3.6}-M_{8.0}=0.54$~mag.\\
 {\bf NGC~4654} .SXT6./SBc(rs) Pseudobulge. The bulge is a prominent
 spiral that breaks into clumps near the center. The decomposition
 yields $n_b=0.6$, and the bulge is actively forming stars with
 $M_{3.6}-M_{8.0}=1.72$~mag.\\
 {\bf NGC~4688} .SBS6./... Pseudobulge.  The decomposition yields
 $n_b=1.4$, and the bulge is not active with
 $M_{3.6}-M_{8.0}=-0.13$~mag. There is no optical HST image available.\\
 {\bf NGC~4689} .SAT4../Sa Pseudobulge.  The decomposition yields
 $n_b=1.2$, and the bulge is actively forming stars with
 $M_{3.6}-M_{8.0}=1.31$~mag.There is no optical HST image available.\\
 {\bf NGC~4698} .SAS2./Sa Classical Bulge. The bulge show a very faint
 spiral structure, however the bulge light is dominated by smooth
 E-like isophotes. The decomposition yields $n_b=3.2$, and the bulge is
 not active with
 $M_{3.6}-M_{8.0}=-1.05$~mag.\\
 {\bf NGC~4701} .SAS1./... Pseudobulge. The bulge light is a strong
 multiple arm spiral that becomes elongated near the center, possibly
 indicating a nuclear bar. The decomposition yields $n_b=1.8$, and the
 bulge is actively forming stars with
 $M_{3.6}-M_{8.0}=1.55$~mag.\\
 {\bf NGC~4713} .SXT7./... No Bulge/Pseudobulge. The 'bulge' is almost
 non-existent ($B/T\sim0.6$\%). The spiral pattern of the disk extends
 into the central arcsecond of the galaxy. The decomposition yields
 $n_b=1.6$, and the bulge is actively forming stars with
 $M_{3.6}-M_{8.0}=0.73$~mag.\\
 {\bf NGC~4725} .SXR2P/SBb(r) Classical Bulge. The bulge has some
 patchy non-spiral dust, but is mostly dominated by a smooth component
 that is similar to E-type galaxies. The decomposition yields
 $n_b=3.6$, and the bulge is not active with
 $M_{3.6}-M_{8.0}=-0.66$~mag.\\
 {\bf NGC~4736} RSAR2./RSab(s) Pseudobulge. The bulge has a strong
 nuclear spiral and a nuclear bar.The decomposition yields $n_b=1.3$,
 and the bulge is not active with $M_{3.6}-M_{8.0}=-0.08$~mag.\\
 {\bf NGC~4808} .SAS6*/Sc(s) Pseudobulge.  The decomposition yields
 $n_b=1.0$, and the bulge is actively forming stars with
 $M_{3.6}-M_{8.0}=1.22$~mag. There is no optical HST image available.\\
 {\bf NGC~4941} RSXR2*/Sab(s) Pseudobulge. There is a nuclear spiral
 that extends to the central few arcseconds. Inside of that there
 appears to be a nuclear bar. The decomposition yields $n_b=1.93$,
 and the bulge is mildly active with $M_{3.6}-M_{8.0}=0.49$~mag.\\
 {\bf NGC~4984} RSXR2*/Sab(s) Pseudobulge. The decomposition yields
 $n_b=1.76$,
 and the bulge is actively forming stars with $M_{3.6}-M_{8.0}=1.18$~mag. There is no optical HST image available. \\
 {\bf NGC~5005} .SXT4./Sb(s) Pseudobulge. This bulge has a prominent
 two armed spiral in the central $r\sim5$\" . The becomes radial most
 likely indicating a nuclear bar, and finally there is a small nuclear
 cluster that is not spherical in the optical images. The decomposition
 yields $n_b=1.2$, and the bulge is mildly active with
 $M_{3.6}-M_{8.0}=0.36$~mag.\\
 {\bf NGC~5033} .SAS5./Sb(s) Pseudobulge. The bulge light is dominated
 by a flocculent spiral. There is no break in the spiral pattern of the
 outer disk to that of the bulge. The decomposition yields $n_b=1.7$,
 and the bulge is actively forming stars with $M_{3.6}-M_{8.0}=1.10$~mag. \\
 {\bf NGC~5055} .SAT4./Sbc(s) Pseudobulge. The bulge light is dominated
 by a flocculent spiral. The spiral of the outer disk extends into the
 central few arcseconds. The light in the central arcsecond is
 dominated by a nuclear star cluster.  The decomposition yields
 $n_b=1.3$,
 and the bulge is actively forming stars with $M_{3.6}-M_{8.0}=1.20$~mag. \\
 {\bf NGC~5068} .SXT6./SBc(s) No bulge/ Pseudobulge. The bulge is
 almost non-existent with $B/T\sim0.1$\%. The bulge is very elongated
 and comprised of several star clusters. The decomposition yields
 $n_b=0.3$,
 and the bulge is actively forming stars with $M_{3.6}-M_{8.0}=2.08$~mag. \\
 {\bf NGC~5128} .L...P./S0 + S pec Classical Bulge. The bulge light is
 dominated by a smooth component that is very much like an E-type
 galaxy, however there is a strong dust lane going across the center of
 the galaxy.  The decomposition yields $n_b=2.6$, and the bulge is not
 active with
 $M_{3.6}-M_{8.0}=-0.15$~mag.\\
 {\bf NGC~5194} .SAS4P/Sbc(s) Pseudobulge. Though this galaxy is
 clearly interacting with NGC~5194B we include it because it is well
 studied, and may provide useful in understanding how interactions in
 the outer disk relate to pseudobulge growth. The bulge light is
 dominated by a strong nuclear spiral. The decomposition yields
 $n_b=0.5$, and the bulge is not active with
 $M_{3.6}-M_{8.0}=-0.13$~mag.\\
 {\bf NGC~5236} .SXS5./SBc(s) Pseudobulge. The bulge light is broken
 into several large clumps. There is a roughly spiral dust pattern in
 the central 2-4\" .The decomposition yields $n_b=0.4$, and the bulge
 is actively forming stars with
 $M_{3.6}-M_{8.0}=1.93$~mag.\\
 {\bf NGC~5248} .SXT4./Sbc(s) Pseudobulge. The bulge light is comprised
 of a nuclear spiral that also forms a ring near $r\sim4$\" . The
 decomposition yields $n_b=0.7$, and the bulge is actively forming
 stars with $M_{3.6}-M_{8.0}=1.62$~mag.\\
 {\bf NGC~5273} .LAS0../ S0/a Classical Bulge. The very center of the
 bulge has a small dust pattern that is almost spiral, however the vast
 majority of th bulge light is smooth and appears similar to an E-type
 galaxy. The decomposition yields $n_b=4.1$, and the bulge is not
 active
 with $M_{3.6}-M_{8.0}=-0.09$~mag.\\
 {\bf NGC~5338} .LB..*./... Bulge Type Uncertain. The decomposition
 yields $n_b=3.3$, and the bulge is active
 with $M_{3.6}-M_{8.0}=-0.09$~mag. There is no optical HST image available.\\
 {\bf NGC~5457} .SXT6./Sc(s) Pseudobulge. The bulge is spiral structure
 of the disk extends to the center of the galaxy. A bright nuclear star
 cluster dominates the central arcsecond. The decomposition yields
 $n_b=1.51$, and the bulge is mildly active
 with $M_{3.6}-M_{8.0}=0.27$~mag.\\
 {\bf NGC~5474} .SAS6P/Scd Pseudobulge. The bulge is a set of star
 clusters roughly arranged in a spiral pattern. The decomposition
 yields $n_b=0.7$, and the bulge is not active
 with $M_{3.6}-M_{8.0}=-0.093$~mag.\\
 {\bf NGC~5585} .SXS7./Sd(s) Pseudobulge. The bulge is a set of star
 clusters roughly arranged in a spiral pattern. The decomposition
 yields $n_b=0.9$, and the bulge is mildly active
 with $M_{3.6}-M_{8.0}=-0.02$~mag.\\
 {\bf NGC~5643} .SXT5./SBc(s) Pseudobulge. The bulge is a nuclear
 spiral that extends to the center of HST image. The decomposition
 yields $n_b=3.0$, and the bulge is actively forming stars
 with $M_{3.6}-M_{8.0}=1.10$~mag.\\
 {\bf NGC~5832} .SBT3\$/... Pseudobulge. The decomposition yields
 $n_b=1.6$, and the bulge is not active
 with $M_{3.6}-M_{8.0}=-0.14$~mag. There is no optical HST image available.\\
 {\bf NGC~5879} .SAT4*/Sb(s) Pseudobulge. The flocculent spiral of the
 outer disk shows now break and continues to the central sub-arcsecond
 of the galaxy.  The decomposition yields $n_b=1.3$, and the bulge is
 actively forming stars
 with $M_{3.6}-M_{8.0}=1.13$~mag.\\
 {\bf NGC~5949} .SAR4\$/Sc Pseudobulge. The decomposition yields
 $n_b=1.7$, and the bulge is actively forming stars
 with $M_{3.6}-M_{8.0}=0.66$~mag. There is no optical HST image available. \\
 {\bf NGC~6207} .SAS5./Sc(s) Pseudobulge.  The flocculent spiral of the
 outer disk shows now break and continues to the central sub-arcsecond
 of the galaxy.  The decomposition yields $n_b=1.4$, and the bulge is
 actively forming stars
 with $M_{3.6}-M_{8.0}=0.66$~mag.\\
 {\bf NGC~6300} .SBT3./SBb(s) pec Pseudobulge. The bulge light contains
 a nuclear spiral that extends to the bulge center. The decomposition
 yields $n_b=0.5$, and the bulge is actively forming stars
 with $M_{3.6}-M_{8.0}=1.08$~mag.\\
 {\bf NGC~6503} .SAS6./Sc(s) Pseudobulge. The flocculent spiral of the
 outer disk shows now break and continues to the central sub-arcsecond
 of the galaxy.  The decomposition yields $n_b=1.0$, and the bulge is
 slightly active with $M_{3.6}-M_{8.0}=0.26$~mag.\\
 {\bf NGC~6684} PLBS0../SBa(s) Classical Bulge. The bulge light is for
 the most part smooth and similar to an E-type galaxy. There is
 possibly a nuclear bar near $r\sim3$\" . The decomposition yields
 $n_b=3.5$, and the bulge is
 not active with $M_{3.6}-M_{8.0}=-1.21$~mag.\\
 {\bf NGC~6744} .SXR4./Sbc(r) Classical Bulge. There is a mild dust
 lane that passes near the center, otherwise the bulge light is smooth
 and very similar to an E-type galaxy. The decomposition yields
 $n_b=3.1$, and the bulge is
 not active with $M_{3.6}-M_{8.0}=-0.57$~mag.\\
 {\bf NGC~7177} .SXR3./Sab(r) Pseudobulge. The bulge light is comprised
 of a bar that passes through the center and ends with a ring near
 $r\sim10$\" . The decomposition yields $n_b=1.8$, and the bulge is
 not active with $M_{3.6}-M_{8.0}=-0.11$~mag.\\
 {\bf NGC~7217} RSAR2./Sb(r) Classical Bulge. There is an abrupt break
 in the spiral pattern of the outer disk near $r\sim8$\" , inside of
 this the bulge is mostly smooth and similar to an E-type galaxy.  The
 decomposition yields $n_b=2.2$, and the bulge is
 not active with $M_{3.6}-M_{8.0}=-0.10$~mag.\\
 {\bf NGC~7331} .SAS3./Sb(rs) Classical Bulge. The disk is inclined
 making morphology difficult, and not trustworthy. There is possibly a
 nuclear spiral in the bulge.  The decomposition yields $n_b=5.7$, and
 the bulge is
 not active with $M_{3.6}-M_{8.0}=-1.13$~mag.\\
 {\bf NGC~7457} .LAT-\$./S01 Classical Bulge. The bulge isophotes are
 smooth and look very much like an E-type galaxy. The decomposition
 yields $n_b=2.7$, and the bulge is
 not active with $M_{3.6}-M_{8.0}=-1.22$~mag.\\
 {\bf NGC~7713} .SBR7*/Sc(s) No bulge/Pseudobulge. The bulge is very
 small with $B/T\sim0.5$\%. The flocculent spiral of the disk extends
 all the way into the center of the galaxy until a nuclear star cluster
 dominates the light in the sub-arcsecond region.  The decomposition
 yields $n_b=1.1$, and the bulge is
 actively forming stars with $M_{3.6}-M_{8.0}=0.87$~mag.\\
 {\bf NGC~7741} .SBS6./SBc(s) Pseudobulge. There is a weak nuclear
 spiral in the center of the bulge.  The decomposition yields
 $n_b=0.5$, and the bulge is
 actively forming stars with $M_{3.6}-M_{8.0}=1.42$~mag.\\
 {\bf NGC~7793} .SAS7./Sd(s) Pseudobulge. The flocculent spiral of the
 outer disk shows now break and continues to the central sub-arcsecond
 of the galaxy.  The decomposition yields $n_b=1.1$, and the bulge is
 actively forming stars with $M_{3.6}-M_{8.0}=1.09$~mag.\\
 {\bf UGC~10445} .S..6?/... Pseudobulge. The bulge light is comprised
 of diffuse low surface brightness light and a set of bright star
 clusters near the center. The decomposition yields $n_b=1.3$, and the
 bulge is
 not active with $M_{3.6}-M_{8.0}=-0.36$~mag.\\

 \section{Decomposition Results}

 \begin{deluxetable}{lccccccccc}
   \tablewidth{0pt} \tablecaption{Derived Parameters}
   \tablehead{\colhead{Galaxy} & \colhead{Bulge\tablenotemark{a}} & \colhead{$n_b$}  & \colhead{M$_{3.6}^{Sersic}$} & \colhead{log($r_e$)} & \colhead{$\langle \mu_{e,3.6} \rangle$} & 
 \colhead{$\mu_{0,3.6}^{disk}$} & \colhead{ log(h)} & \colhead{M$_{8,dust}$} &
  \colhead{M$_{3.6}$-M$_{8.0}$} \\
     \colhead{ } & \colhead {Morphology} & \colhead{ } & \colhead{mag} & \colhead{log(pc)}  & \colhead{mag arcsec$^{-2}$} & \colhead{mag arcsec$^{-2}$} & \colhead{log(pc)} & \colhead{mag} & \colhead{mag} } 
 \startdata
 IC0342 & P & 1.78 $\pm$ 0.37 & -17.17 $\pm$ 0.70 & 1.86 $\pm$ 0.38 & 14.92 $\pm$ 0.70 & 19.88 $\pm$ 0.08 & 3.46 $\pm$ 0.02 & -19.89 & 1.99 $\pm$ 0.16 \\
 IC0749 & ... & 1.44 $\pm$ 1.06 & -13.40 $\pm$ 1.27 & 2.31 $\pm$ 0.86 & 20.94 $\pm$ 1.16 & 20.21 $\pm$ 0.15 & 3.14 $\pm$ 0.03 & -15.05 & 0.98 $\pm$ 0.54 \\
 NGC0300 & P & 1.64 $\pm$ 0.49 & -10.91 $\pm$ 1.01 & 1.32 $\pm$ 0.54 & 18.49 $\pm$ 1.01 & 20.74 $\pm$ 0.04 & 3.29 $\pm$ 0.01 & -9.63 & -1.46 $\pm$ 1.40 \\
 NGC0404 & C & 3.74 $\pm$ 0.33 & -17.66 $\pm$ 0.44 & 2.25 $\pm$ 0.32 & 16.43 $\pm$ 0.44 & 17.31 $\pm$ 0.13 & 2.51 $\pm$ 0.02 & -16.09 & -0.46 $\pm$ 0.05 \\
 NGC0628 & P & 1.55 $\pm$ 0.17 & -18.53 $\pm$ 0.17 & 2.92 $\pm$ 0.10 & 18.88 $\pm$ 0.16 & 20.13 $\pm$ 0.06 & 3.58 $\pm$ 0.01 & -18.80 & 0.18 $\pm$ 0.01 \\
 NGC0672 & P & 1.10 $\pm$ 1.14 & -10.86 $\pm$ 1.22 & 1.23 $\pm$ 0.35 & 18.11 $\pm$ 1.20 & 20.04 $\pm$ 0.14 & 3.26 $\pm$ 0.02 & -14.84 & 2.20 $\pm$ 1.17 \\
 NGC0925 & P & 0.74 $\pm$ 0.32 & -17.68 $\pm$ 0.38 & 3.05 $\pm$ 0.14 & 20.37 $\pm$ 0.33 & 20.46 $\pm$ 0.10 & 3.53 $\pm$ 0.01 & -20.41 & 2.86 $\pm$ 0.06 \\
 NGC1023 & C & 2.47 $\pm$ 0.34 & -20.90 $\pm$ 0.53 & 2.89 $\pm$ 0.21 & 16.35 $\pm$ 0.53 & 19.02 $\pm$ 0.12 & 3.55 $\pm$ 0.02 & -18.11 & -1.17 $\pm$ 0.08 \\
 NGC1058 & P & 1.09 $\pm$ 0.46 & -14.74 $\pm$ 1.11 & 1.89 $\pm$ 0.30 & 17.53 $\pm$ 1.10 & 19.73 $\pm$ 0.05 & 3.04 $\pm$ 0.01 & -16.20 & 0.82 $\pm$ 0.08 \\
 NGC1097 & P & 1.96 $\pm$ 0.16 & -21.47 $\pm$ 0.17 & 3.10 $\pm$ 0.12 & 16.83 $\pm$ 0.17 & 19.17 $\pm$ 0.11 & 3.68 $\pm$ 0.02 & -22.59 & 1.31 $\pm$ 0.01 \\
 NGC1313 & P & 1.32 $\pm$ 1.22 & -13.67 $\pm$ 1.00 & 1.97 $\pm$ 0.28 & 18.99 $\pm$ 0.97 & 18.42 $\pm$ 0.07 & 3.03 $\pm$ 0.01 & -13.71 & -0.72 $\pm$ 0.11 \\
 NGC1317 & P & 1.44 $\pm$ 0.12 & -21.75 $\pm$ 0.18 & 2.70 $\pm$ 0.09 & 14.55 $\pm$ 0.18 & 18.06 $\pm$ 0.10 & 3.38 $\pm$ 0.02 & -21.95 & 0.40 $\pm$ 0.01 \\
 NGC1433 & P & 0.80 $\pm$ 0.10 & -18.91 $\pm$ 0.22 & 2.48 $\pm$ 0.07 & 16.30 $\pm$ 0.22 & 19.64 $\pm$ 0.05 & 3.51 $\pm$ 0.01 & -19.06 & 0.26 $\pm$ 0.01 \\
 NGC1512 & P & 1.79 $\pm$ 0.62 & -19.11 $\pm$ 0.61 & 2.76 $\pm$ 0.22 & 17.51 $\pm$ 0.61 & 20.06 $\pm$ 0.76 & 3.45 $\pm$ 0.12 & -18.96 & 0.21 $\pm$ 0.00 \\
 NGC1543 & P & 1.51 $\pm$ 0.16 & -19.73 $\pm$ 0.26 & 2.64 $\pm$ 0.14 & 16.28 $\pm$ 0.26 & 19.40 $\pm$ 0.09 & 3.37 $\pm$ 0.01 & -17.52 & -1.27 $\pm$ 0.01 \\
 NGC1559 & P & 0.68 $\pm$ 0.29 & -20.38 $\pm$ 0.28 & 2.91 $\pm$ 0.09 & 16.96 $\pm$ 0.27 & 17.59 $\pm$ 1.69 & 3.28 $\pm$ 0.11 & -25.22 & 5.50 $\pm$ 0.03 \\
 NGC1566 & P & 1.60 $\pm$ 0.57 & -19.21 $\pm$ 0.96 & 2.61 $\pm$ 0.27 & 16.66 $\pm$ 0.96 & 18.81 $\pm$ 0.11 & 3.39 $\pm$ 0.01 & -19.56 & 0.21 $\pm$ 0.03 \\
 NGC1617 & C & 2.05 $\pm$ 0.22 & -19.41 $\pm$ 0.27 & 2.58 $\pm$ 0.16 & 16.29 $\pm$ 0.27 & 18.82 $\pm$ 0.11 & 3.36 $\pm$ 0.01 & -17.60 & -1.04 $\pm$ 0.05 \\
 NGC1637 & P & 1.62 $\pm$ 0.53 & -16.77 $\pm$ 0.86 & 2.08 $\pm$ 0.22 & 16.47 $\pm$ 0.86 & 19.50 $\pm$ 0.11 & 3.05 $\pm$ 0.01 & -20.76 & 3.42 $\pm$ 0.06 \\
 NGC1672 & P & 2.05 $\pm$ 0.17 & -20.51 $\pm$ 0.24 & 2.69 $\pm$ 0.14 & 15.73 $\pm$ 0.23 & 19.44 $\pm$ 0.08 & 3.62 $\pm$ 0.01 & -22.61 & 1.99 $\pm$ 0.00 \\
 NGC1744 & P & 0.54 $\pm$ 1.23 & -15.29 $\pm$ 1.00 & 2.33 $\pm$ 1.03 & 19.19 $\pm$ 0.95 & 19.54 $\pm$ 0.65 & 3.55 $\pm$ 0.23 & -17.00 & 5.16 $\pm$ 2.36 \\
 NGC1808 & P & 0.77 $\pm$ 0.05 & -22.31 $\pm$ 0.12 & 2.49 $\pm$ 0.04 & 12.93 $\pm$ 0.12 & 17.05 $\pm$ 0.04 & 3.33 $\pm$ 0.01 & -25.17 & 3.27 $\pm$ 0.01 \\
 NGC2403 & P & 0.66 $\pm$ 0.33 & -16.82 $\pm$ 0.33 & 2.75 $\pm$ 0.09 & 19.74 $\pm$ 0.32 & 20.04 $\pm$ 0.81 & 3.28 $\pm$ 0.15 & -17.92 & 1.20 $\pm$ 0.02 \\
 NGC2500 & P & 1.73 $\pm$ 0.76 & -15.27 $\pm$ 0.55 & 2.52 $\pm$ 0.16 & 20.14 $\pm$ 0.54 & 19.15 $\pm$ 0.08 & 3.01 $\pm$ 0.02 & -15.92 & 0.38 $\pm$ 0.13 \\
 NGC2655 & C & 2.36 $\pm$ 0.35 & -20.65 $\pm$ 0.55 & 2.76 $\pm$ 0.37 & 15.98 $\pm$ 0.55 & 17.88 $\pm$ 0.13 & 3.19 $\pm$ 0.01 & -19.82 & -0.55 $\pm$ 0.10 \\
 NGC2685 & C & 2.12 $\pm$ 0.30 & -18.96 $\pm$ 0.35 & 2.51 $\pm$ 0.23 & 16.39 $\pm$ 0.35 & 18.17 $\pm$ 0.16 & 2.96 $\pm$ 0.02 & -16.79 & -1.00 $\pm$ 0.00 \\
 NGC2775 & C & 3.51 $\pm$ 0.35 & -20.61 $\pm$ 0.45 & 3.16 $\pm$ 0.26 & 17.99 $\pm$ 0.44 & 19.55 $\pm$ 0.21 & 3.46 $\pm$ 0.03 & -20.18 & -0.38 $\pm$ 0.08 \\
 NGC2841 & C & 3.60 $\pm$ 0.27 & -20.33 $\pm$ 0.29 & 2.95 $\pm$ 0.36 & 17.24 $\pm$ 0.29 & 18.95 $\pm$ 0.10 & 3.48 $\pm$ 0.02 & -18.30 & -0.79 $\pm$ 0.01 \\
 NGC2903 & P & 0.46 $\pm$ 0.04 & -18.95 $\pm$ 0.09 & 2.33 $\pm$ 0.02 & 15.49 $\pm$ 0.09 & 17.99 $\pm$ 0.04 & 3.25 $\pm$ 0.01 & -20.92 & 2.17 $\pm$ 0.01 \\
 NGC2950 & P & 1.32 $\pm$ 0.60 & -22.11 $\pm$ 0.88 & 2.62 $\pm$ 0.17 & 13.82 $\pm$ 0.88 & 19.38 $\pm$ 0.25 & 3.61 $\pm$ 0.03 & -19.45 & -1.25 $\pm$ 0.01 \\
 NGC2964 & P & 1.05 $\pm$ 0.22 & -18.42 $\pm$ 0.32 & 2.25 $\pm$ 0.12 & 15.62 $\pm$ 0.31 & 18.44 $\pm$ 0.11 & 3.19 $\pm$ 0.01 & -20.45 & 1.96 $\pm$ 0.06 \\
 NGC2976 & P & 1.06 $\pm$ 0.14 & -10.53 $\pm$ 0.19 & 0.65 $\pm$ 0.08 & 15.53 $\pm$ 0.18 & 19.68 $\pm$ 0.03 & 2.91 $\pm$ 0.01 & -14.11 & 1.44 $\pm$ 0.10 \\
 NGC2997 & P & 1.02 $\pm$ 0.12 & -18.87 $\pm$ 0.18 & 2.58 $\pm$ 0.07 & 16.85 $\pm$ 0.18 & 20.13 $\pm$ 0.07 & 3.83 $\pm$ 0.04 & -20.57 & 1.49 $\pm$ 0.01 \\
 NGC3031 & C & 3.88 $\pm$ 0.23 & -21.21 $\pm$ 0.36 & 3.10 $\pm$ 0.28 & 17.11 $\pm$ 0.36 & 19.07 $\pm$ 0.11 & 3.52 $\pm$ 0.01 & -20.25 & -0.51 $\pm$ 0.01 \\
 NGC3032 & P & 2.60 $\pm$ 0.29 & -18.94 $\pm$ 0.66 & 2.67 $\pm$ 0.53 & 17.21 $\pm$ 0.66 & 20.35 $\pm$ 0.34 & 3.16 $\pm$ 0.04 & -20.21 & 1.40 $\pm$ 0.02 \\
 NGC3156 & P & 1.65 $\pm$ 0.54 & -19.51 $\pm$ 0.68 & 2.65 $\pm$ 0.13 & 16.57 $\pm$ 0.68 & 18.26 $\pm$ 0.41 & 3.07 $\pm$ 0.05 & -18.97 & -0.38 $\pm$ 0.00 \\
 NGC3184 & P & 1.65 $\pm$ 0.42 & -16.20 $\pm$ 0.61 & 2.28 $\pm$ 0.29 & 18.03 $\pm$ 0.60 & 20.12 $\pm$ 0.09 & 3.37 $\pm$ 0.01 & -17.70 & 1.38 $\pm$ 1.80 \\
 NGC3185 & P & 1.02 $\pm$ 0.76 & -18.57 $\pm$ 2.02 & 2.69 $\pm$ 0.45 & 17.71 $\pm$ 2.00 & 18.85 $\pm$ 0.79 & 3.19 $\pm$ 0.11 & -19.37 & 0.65 $\pm$ 0.00 \\
 NGC3190 & P & 1.99 $\pm$ 0.65 & -20.47 $\pm$ 0.95 & 2.70 $\pm$ 0.27 & 15.84 $\pm$ 0.95 & 18.61 $\pm$ 0.23 & 3.38 $\pm$ 0.02 & -19.17 & -0.42 $\pm$ 1.97 \\
 NGC3198 & P & 1.30 $\pm$ 0.56 & -16.60 $\pm$ 1.00 & 2.41 $\pm$ 0.27 & 18.26 $\pm$ 1.00 & 19.91 $\pm$ 0.09 & 3.35 $\pm$ 0.01 & -18.23 & 1.18 $\pm$ 1.10 \\
 NGC3319 & P & 0.61 $\pm$ 0.13 & -17.39 $\pm$ 0.12 & 2.98 $\pm$ 0.05 & 20.33 $\pm$ 0.11 & 22.25 $\pm$ 0.09 & 3.61 $\pm$ 0.02 & -17.36 & 1.20 $\pm$ 0.18 \\
 NGC3344 & C & 2.39 $\pm$ 0.33 & -17.20 $\pm$ 0.50 & 2.34 $\pm$ 0.18 & 17.32 $\pm$ 0.50 & 19.70 $\pm$ 0.07 & 3.20 $\pm$ 0.01 & -16.69 & -0.31 $\pm$ 6.02 \\
 NGC3351 & P & 1.54 $\pm$ 0.22 & -18.87 $\pm$ 0.24 & 2.47 $\pm$ 0.16 & 16.32 $\pm$ 0.24 & 19.28 $\pm$ 0.09 & 3.31 $\pm$ 0.01 & -19.93 & 0.91 $\pm$ 0.03 \\
 NGC3368 & P & 1.63 $\pm$ 0.18 & -19.59 $\pm$ 0.22 & 2.50 $\pm$ 0.12 & 15.72 $\pm$ 0.22 & 18.30 $\pm$ 0.12 & 3.15 $\pm$ 0.02 & -19.16 & -0.23 $\pm$ 0.43 \\
 NGC3384 & P & 1.42 $\pm$ 0.20 & -19.52 $\pm$ 0.34 & 2.38 $\pm$ 0.15 & 15.21 $\pm$ 0.34 & 19.22 $\pm$ 0.13 & 3.29 $\pm$ 0.03 & -16.79 & -1.28 $\pm$ 0.01 \\
 NGC3412 & C & 2.65 $\pm$ 0.31 & -18.60 $\pm$ 0.62 & 2.35 $\pm$ 0.54 & 15.97 $\pm$ 0.62 & 19.22 $\pm$ 0.15 & 3.02 $\pm$ 0.01 & -15.73 & -1.35 $\pm$ 0.02 \\
 NGC3486 & P & 1.62 $\pm$ 0.56 & -16.79 $\pm$ 0.77 & 2.22 $\pm$ 0.25 & 17.14 $\pm$ 0.77 & 19.06 $\pm$ 0.31 & 2.93 $\pm$ 0.05 & -19.67 & 1.57 $\pm$ 0.19 \\
 NGC3489 & P & 1.47 $\pm$ 0.28 & -18.90 $\pm$ 0.54 & 2.29 $\pm$ 0.22 & 15.36 $\pm$ 0.54 & 18.35 $\pm$ 0.11 & 2.93 $\pm$ 0.01 & -17.53 & -0.69 $\pm$ 0.08 \\
 NGC3511 & ... & 1.55 $\pm$ 0.58 & -17.90 $\pm$ 0.60 & 2.77 $\pm$ 0.25 & 18.75 $\pm$ 0.59 & 19.31 $\pm$ 0.12 & 3.49 $\pm$ 0.02 & -20.20 & 1.97 $\pm$ 0.01 \\
 NGC3521 & C & 2.60 $\pm$ 0.82 & -19.61 $\pm$ 1.17 & 2.54 $\pm$ 0.49 & 15.88 $\pm$ 1.16 & 17.93 $\pm$ 0.21 & 3.27 $\pm$ 0.02 & -20.12 & 0.43 $\pm$ 0.10 \\
 NGC3593 & P & 0.81 $\pm$ 0.11 & -20.15 $\pm$ 0.17 & 2.71 $\pm$ 0.06 & 16.22 $\pm$ 0.17 & 19.58 $\pm$ 0.13 & 3.25 $\pm$ 0.01 & -20.87 & 1.95 $\pm$ 0.00 \\
 NGC3621 & P & 2.79 $\pm$ 0.48 & -17.65 $\pm$ 0.62 & 2.58 $\pm$ 0.49 & 18.06 $\pm$ 0.61 & 17.26 $\pm$ 0.05 & 3.31 $\pm$ 0.01 & -18.84 & 0.81 $\pm$ 0.17 \\
 NGC3675 & P & 1.62 $\pm$ 0.66 & -19.27 $\pm$ 1.02 & 2.67 $\pm$ 0.43 & 16.88 $\pm$ 1.01 & 18.28 $\pm$ 0.16 & 3.26 $\pm$ 0.01 & -19.92 & 0.55 $\pm$ 0.32 \\
 NGC3726 & ... & 0.83 $\pm$ 0.16 & -17.03 $\pm$ 0.22 & 2.51 $\pm$ 0.08 & 18.34 $\pm$ 0.22 & 19.97 $\pm$ 0.07 & 3.47 $\pm$ 0.01 & -18.26 & 1.25 $\pm$ 0.03 \\
 NGC3906 & P & 0.97 $\pm$ 0.46 & -18.20 $\pm$ 0.41 & 2.90 $\pm$ 0.20 & 19.12 $\pm$ 0.29 & 20.03 $\pm$ 0.60 & 3.44 $\pm$ 0.17 & -15.62 & 0.96 $\pm$ 0.90 \\
 NGC3938 & P & 1.35 $\pm$ 0.12 & -17.48 $\pm$ 0.13 & 2.53 $\pm$ 0.07 & 17.99 $\pm$ 0.13 & 19.60 $\pm$ 0.04 & 3.33 $\pm$ 0.01 & -18.94 & 1.11 $\pm$ 1.00 \\
 NGC3941 & ... & 1.53 $\pm$ 0.47 & -19.09 $\pm$ 0.86 & 2.32 $\pm$ 0.30 & 15.32 $\pm$ 0.86 & 17.49 $\pm$ 0.14 & 2.92 $\pm$ 0.01 & -17.43 & -1.18 $\pm$ 0.01 \\
 NGC3945 & P & 1.54 $\pm$ 0.14 & -20.88 $\pm$ 0.19 & 2.90 $\pm$ 0.11 & 16.44 $\pm$ 0.19 & 20.71 $\pm$ 0.37 & 3.64 $\pm$ 0.10 & -18.13 & -1.26 $\pm$ 0.01 \\
 NGC3953 & ... & 1.54 $\pm$ 0.61 & -18.93 $\pm$ 1.04 & 2.59 $\pm$ 0.33 & 16.84 $\pm$ 1.04 & 18.91 $\pm$ 0.11 & 3.41 $\pm$ 0.01 & -18.89 & -0.42 $\pm$ 0.07 \\
 NGC3982 & P & 1.95 $\pm$ 0.19 & -16.82 $\pm$ 0.20 & 2.21 $\pm$ 0.11 & 17.03 $\pm$ 0.20 & 18.60 $\pm$ 0.11 & 2.91 $\pm$ 0.01 & -17.02 & 0.58 $\pm$ 5.24 \\
 NGC3990 & ... & 1.08 $\pm$ 0.57 & -17.81 $\pm$ 1.16 & 2.32 $\pm$ 0.22 & 16.59 $\pm$ 1.16 & 18.64 $\pm$ 0.20 & 2.75 $\pm$ 0.01 & -18.70 & 1.03 $\pm$ 0.01 \\
 NGC4020 & ... & 0.30 $\pm$ 0.93 & -13.92 $\pm$ 1.86 & 2.03 $\pm$ 0.17 & 19.05 $\pm$ 1.81 & 18.46 $\pm$ 0.06 & 3.16 $\pm$ 0.02 & -15.07 & 0.50 $\pm$ 0.77 \\
 NGC4117 & P & 1.36 $\pm$ 0.47 & -17.34 $\pm$ 0.62 & 2.52 $\pm$ 0.24 & 18.08 $\pm$ 0.62 & 19.74 $\pm$ 2.23 & 2.84 $\pm$ 0.11 & -16.67 & -0.22 $\pm$ 0.02 \\
 NGC4136 & ... & 0.64 $\pm$ 0.56 & -14.21 $\pm$ 0.75 & 2.19 $\pm$ 0.15 & 19.53 $\pm$ 0.74 & 20.34 $\pm$ 0.06 & 3.16 $\pm$ 0.01 & -16.03 & 0.94 $\pm$ 0.19 \\
 NGC4138 & P & 1.65 $\pm$ 0.47 & -19.94 $\pm$ 0.76 & 2.26 $\pm$ 0.21 & 14.18 $\pm$ 0.76 & 17.39 $\pm$ 0.19 & 3.07 $\pm$ 0.02 & -18.98 & -0.95 $\pm$ 0.95 \\
 NGC4150 & C & 5.37 $\pm$ 0.29 & -19.36 $\pm$ 0.56 & 2.81 $\pm$ 0.49 & 17.49 $\pm$ 0.56 & 19.67 $\pm$ 0.23 & 2.98 $\pm$ 0.03 & -17.98 & 0.16 $\pm$ 0.11 \\
 NGC4203 & C & 2.76 $\pm$ 0.29 & -20.45 $\pm$ 0.57 & 2.84 $\pm$ 0.40 & 16.55 $\pm$ 0.57 & 20.46 $\pm$ 0.19 & 3.60 $\pm$ 0.04 & -20.32 & 0.27 $\pm$ 0.05 \\
 NGC4237 & P & 1.35 $\pm$ 0.41 & -19.85 $\pm$ 0.40 & 2.80 $\pm$ 0.18 & 16.97 $\pm$ 0.39 & 18.26 $\pm$ 0.33 & 3.12 $\pm$ 0.03 & -18.83 & 0.58 $\pm$ 0.10 \\
 NGC4254 & P & 1.72 $\pm$ 0.56 & -20.85 $\pm$ 0.50 & 2.87 $\pm$ 0.21 & 16.31 $\pm$ 0.49 & 17.39 $\pm$ 0.17 & 3.43 $\pm$ 0.02 & -22.99 & 1.51 $\pm$ 0.03 \\
 NGC4258 & P & 2.80 $\pm$ 0.28 & -19.72 $\pm$ 0.42 & 2.77 $\pm$ 0.05 & 16.96 $\pm$ 0.42 & 18.82 $\pm$ 0.06 & 3.50 $\pm$ 0.01 & -19.42 & 0.04 $\pm$ 0.30 \\
 NGC4267 & C & 4.23 $\pm$ 0.27 & -21.80 $\pm$ 0.36 & 3.01 $\pm$ 0.33 & 16.04 $\pm$ 0.35 & 20.10 $\pm$ 0.51 & 3.50 $\pm$ 0.07 & -18.81 & -1.30 $\pm$ 0.01 \\
 NGC4274 & P & 1.60 $\pm$ 0.35 & -19.65 $\pm$ 0.56 & 2.65 $\pm$ 0.27 & 16.44 $\pm$ 0.56 & 19.56 $\pm$ 0.14 & 3.52 $\pm$ 0.02 & -19.80 & 0.12 $\pm$ 0.01 \\
 NGC4293 & P & 1.93 $\pm$ 0.36 & -18.07 $\pm$ 0.48 & 2.17 $\pm$ 0.28 & 15.59 $\pm$ 0.48 & 19.00 $\pm$ 0.10 & 3.51 $\pm$ 0.02 & -19.18 & 1.68 $\pm$ 0.64 \\
 NGC4294 & ... & 1.31 $\pm$ 0.89 & -13.50 $\pm$ 0.72 & 2.32 $\pm$ 0.25 & 20.92 $\pm$ 0.69 & 19.44 $\pm$ 0.04 & 3.09 $\pm$ 0.01 & -15.96 & 2.31 $\pm$ 0.12 \\
 \enddata

  \tablenotetext{a}{P- Pseudobulge C- Classical Bulge}

  \end{deluxetable}

 \begin{deluxetable}{lccccccccc}
   \tablewidth{0pt} \tablecaption{Derived Parameters}
   \tablehead{\colhead{Galaxy} & \colhead{Bulge\tablenotemark{a}} & \colhead{$n_b$}  & \colhead{M$_{3.6}^{Sersic}$} & \colhead{log($r_e$)} & \colhead{$\langle \mu_{e,3.6} \rangle$} & 
 \colhead{$\mu_{0,3.6}^{disk}$} & \colhead{ log(h)} & \colhead{M$_{8,dust}$} &
  \colhead{M$_{3.6}$-M$_{8.0}$} \\
     \colhead{ } & \colhead {Morphology} & \colhead{ } & \colhead{mag} & \colhead{log(pc)}  & \colhead{mag arcsec$^{-2}$} & \colhead{mag arcsec$^{-2}$} & \colhead{log(pc)} & \colhead{mag} & \colhead{mag} } 
 \startdata
 NGC4303 & P & 1.75 $\pm$ 0.21 & -21.26 $\pm$ 0.13 & 2.43 $\pm$ 0.05 & 13.69 $\pm$ 0.13 & 17.23 $\pm$ 0.04 & 3.50 $\pm$ 0.01 & -22.60 & 1.00 $\pm$ 0.01 \\
 NGC4314 & P & 3.00 $\pm$ 0.30 & -19.13 $\pm$ 0.26 & 2.78 $\pm$ 0.19 & 17.57 $\pm$ 0.26 & 18.94 $\pm$ 0.15 & 3.30 $\pm$ 0.03 & -19.13 & 0.35 $\pm$ 0.09 \\
 NGC4321 & P & 1.20 $\pm$ 0.19 & -19.93 $\pm$ 0.24 & 2.79 $\pm$ 0.13 & 16.85 $\pm$ 0.23 & 20.33 $\pm$ 0.24 & 3.76 $\pm$ 0.04 & -21.87 & 2.41 $\pm$ 0.07 \\
 NGC4371 & P & 3.92 $\pm$ 0.31 & -22.11 $\pm$ 0.32 & 3.21 $\pm$ 0.59 & 16.73 $\pm$ 0.32 & 19.33 $\pm$ 0.42 & 3.50 $\pm$ 0.04 & -19.32 & -1.21 $\pm$ 0.10 \\
 NGC4380 & P & 1.85 $\pm$ 0.28 & -17.56 $\pm$ 0.39 & 2.65 $\pm$ 0.28 & 18.49 $\pm$ 0.38 & 19.97 $\pm$ 0.07 & 3.33 $\pm$ 0.01 & -17.77 & 0.46 $\pm$ 0.18 \\
 NGC4394 & P & 1.99 $\pm$ 0.12 & -20.70 $\pm$ 0.19 & 2.85 $\pm$ 0.10 & 16.37 $\pm$ 0.18 & 18.96 $\pm$ 0.04 & 3.55 $\pm$ 0.01 & -17.08 & -1.91 $\pm$ 0.05 \\
 NGC4413 & ... & 1.09 $\pm$ 0.49 & -17.46 $\pm$ 1.01 & 2.45 $\pm$ 0.26 & 17.62 $\pm$ 1.01 & 20.03 $\pm$ 0.15 & 3.20 $\pm$ 0.04 & -18.65 & 1.83 $\pm$ 0.17 \\
 NGC4414 & P & 0.45 $\pm$ 0.49 & -17.78 $\pm$ 1.57 & 2.17 $\pm$ 0.14 & 15.91 $\pm$ 1.57 & 17.18 $\pm$ 0.04 & 3.07 $\pm$ 0.00 & -19.48 & 1.11 $\pm$ 0.93 \\
 NGC4419 & ... & 1.88 $\pm$ 0.60 & -19.07 $\pm$ 0.99 & 2.40 $\pm$ 0.55 & 15.76 $\pm$ 0.99 & 17.37 $\pm$ 0.10 & 3.06 $\pm$ 0.01 & -20.42 & 0.84 $\pm$ 0.05 \\
 NGC4421 & ... & 2.70 $\pm$ 0.36 & -20.56 $\pm$ 0.37 & 2.94 $\pm$ 0.20 & 16.96 $\pm$ 0.37 & 19.03 $\pm$ 0.31 & 3.33 $\pm$ 0.03 & -17.59 & -1.17 $\pm$ 0.03 \\
 NGC4424 & P & 0.65 $\pm$ 0.21 & -18.82 $\pm$ 0.23 & 2.75 $\pm$ 0.09 & 17.75 $\pm$ 0.22 & 19.01 $\pm$ 0.09 & 3.24 $\pm$ 0.01 & -19.64 & 1.09 $\pm$ 0.01 \\
 NGC4442 & C & 2.36 $\pm$ 0.14 & -20.63 $\pm$ 0.17 & 2.84 $\pm$ 0.10 & 16.38 $\pm$ 0.17 & 19.27 $\pm$ 0.12 & 3.38 $\pm$ 0.01 & -17.82 & -1.17 $\pm$ 0.03 \\
 NGC4448 & P & 1.19 $\pm$ 0.45 & -18.40 $\pm$ 0.81 & 2.55 $\pm$ 0.22 & 17.16 $\pm$ 0.81 & 18.95 $\pm$ 0.12 & 3.22 $\pm$ 0.01 & -18.17 & -0.30 $\pm$ 0.61 \\
 NGC4450 & P & 2.06 $\pm$ 0.42 & -19.12 $\pm$ 0.58 & 2.56 $\pm$ 0.29 & 16.48 $\pm$ 0.58 & 18.74 $\pm$ 0.19 & 3.35 $\pm$ 0.02 & -19.33 & -0.66 $\pm$ 0.15 \\
 NGC4457 & ... & 1.43 $\pm$ 0.40 & -19.02 $\pm$ 0.82 & 2.35 $\pm$ 0.15 & 15.52 $\pm$ 0.82 & 19.82 $\pm$ 0.24 & 3.18 $\pm$ 0.04 & -19.05 & 0.20 $\pm$ 0.06 \\
 NGC4491 & ... & 0.91 $\pm$ 0.44 & -13.47 $\pm$ 0.79 & 1.82 $\pm$ 0.18 & 18.46 $\pm$ 0.79 & 19.31 $\pm$ 0.03 & 2.60 $\pm$ 0.01 & 999.00 & -2.86 $\pm$ 0.07 \\
 NGC4498 & P & 1.00 $\pm$ 0.63 & -16.19 $\pm$ 0.89 & 2.70 $\pm$ 0.23 & 20.10 $\pm$ 0.88 & 19.74 $\pm$ 0.13 & 3.15 $\pm$ 0.01 & -17.58 & 1.48 $\pm$ 1.07 \\
 NGC4501 & P & 0.89 $\pm$ 0.65 & -19.16 $\pm$ 1.32 & 2.56 $\pm$ 0.24 & 16.43 $\pm$ 1.10 & 18.09 $\pm$ 0.09 & 3.46 $\pm$ 0.02 & -20.46 & -0.37 $\pm$ 0.03 \\
 NGC4519 & P & 0.84 $\pm$ 0.15 & -17.53 $\pm$ 0.19 & 2.47 $\pm$ 0.07 & 17.64 $\pm$ 0.19 & 18.49 $\pm$ 0.04 & 3.10 $\pm$ 0.00 & -19.60 & 1.76 $\pm$ 1.04 \\
 NGC4526 & C & 3.53 $\pm$ 0.14 & -21.78 $\pm$ 0.12 & 3.21 $\pm$ 0.10 & 17.09 $\pm$ 0.12 & 19.49 $\pm$ 0.19 & 3.62 $\pm$ 0.03 & -20.30 & 0.03 $\pm$ 0.07 \\
 NGC4548 & ... & 2.89 $\pm$ 0.32 & -19.94 $\pm$ 0.51 & 2.86 $\pm$ 0.36 & 17.16 $\pm$ 0.51 & 19.58 $\pm$ 0.12 & 3.63 $\pm$ 0.02 & -18.81 & -0.66 $\pm$ 0.02 \\
 NGC4559 & P & 1.07 $\pm$ 0.53 & -17.62 $\pm$ 0.55 & 2.89 $\pm$ 0.20 & 19.65 $\pm$ 0.52 & 19.80 $\pm$ 0.09 & 3.55 $\pm$ 0.01 & -18.42 & 1.07 $\pm$ 0.12 \\
 NGC4569 & P & 1.97 $\pm$ 0.38 & -19.76 $\pm$ 0.60 & 2.43 $\pm$ 0.28 & 15.22 $\pm$ 0.60 & 19.49 $\pm$ 0.12 & 3.66 $\pm$ 0.01 & -21.09 & 1.38 $\pm$ 1.22 \\
 NGC4571 & P & 1.88 $\pm$ 0.46 & -17.85 $\pm$ 0.33 & 3.10 $\pm$ 0.17 & 20.45 $\pm$ 0.32 & 20.69 $\pm$ 0.23 & 3.49 $\pm$ 0.03 & ... & ... $\pm$ ... \\
 NGC4578 & C & 3.66 $\pm$ 0.25 & -21.69 $\pm$ 0.33 & 3.08 $\pm$ 0.43 & 16.51 $\pm$ 0.33 & 19.17 $\pm$ 0.22 & 3.42 $\pm$ 0.02 & -19.08 & -1.26 $\pm$ 0.01 \\
 NGC4580 & ... & 0.56 $\pm$ 0.13 & -14.80 $\pm$ 0.17 & 2.20 $\pm$ 0.05 & 19.00 $\pm$ 0.15 & 19.16 $\pm$ 0.03 & 3.10 $\pm$ 0.00 & -17.96 & 1.92 $\pm$ 0.12 \\
 NGC4605 & P & 0.61 $\pm$ 0.13 & -12.13 $\pm$ 0.18 & 0.89 $\pm$ 0.06 & 15.15 $\pm$ 0.17 & 18.62 $\pm$ 0.09 & 2.99 $\pm$ 0.01 & -16.04 & 1.47 $\pm$ 0.86 \\
 NGC4612 & ... & 3.71 $\pm$ 0.40 & -21.28 $\pm$ 0.71 & 3.04 $\pm$ 0.30 & 16.75 $\pm$ 0.70 & 19.96 $\pm$ 1.31 & 3.43 $\pm$ 0.21 & -18.72 & -1.23 $\pm$ 0.03 \\
 NGC4639 & P & 1.32 $\pm$ 0.76 & -18.38 $\pm$ 1.17 & 2.60 $\pm$ 0.43 & 17.44 $\pm$ 1.17 & 20.03 $\pm$ 0.30 & 3.26 $\pm$ 0.03 & -17.61 & -0.27 $\pm$ 0.21 \\
 NGC4651 & P & 1.58 $\pm$ 0.74 & -18.49 $\pm$ 0.99 & 2.67 $\pm$ 0.53 & 17.68 $\pm$ 0.99 & 19.07 $\pm$ 0.55 & 3.25 $\pm$ 0.07 & -18.73 & 0.54 $\pm$ 0.53 \\
 NGC4654 & P & 0.57 $\pm$ 0.09 & -18.28 $\pm$ 0.14 & 2.67 $\pm$ 0.04 & 17.89 $\pm$ 0.14 & 18.90 $\pm$ 0.03 & 3.38 $\pm$ 0.01 & -19.89 & 1.72 $\pm$ 0.13 \\
 NGC4688 & ... & 1.43 $\pm$ 0.32 & -16.16 $\pm$ 0.32 & 2.83 $\pm$ 0.15 & 20.78 $\pm$ 0.32 & 19.89 $\pm$ 0.09 & 3.05 $\pm$ 0.01 & -14.74 & -0.13 $\pm$ 0.32 \\
 NGC4689 & ... & 1.24 $\pm$ 0.21 & -18.81 $\pm$ 0.21 & 3.02 $\pm$ 0.10 & 19.12 $\pm$ 0.21 & 20.10 $\pm$ 0.05 & 3.72 $\pm$ 0.01 & -21.13 & 1.30 $\pm$ 0.03 \\
 NGC4698 & C & 3.19 $\pm$ 0.29 & -20.08 $\pm$ 0.44 & 3.04 $\pm$ 0.24 & 17.92 $\pm$ 0.44 & 20.50 $\pm$ 0.15 & 3.45 $\pm$ 0.02 & -17.63 & -1.05 $\pm$ 0.38 \\
 NGC4701 & P & 1.79 $\pm$ 0.31 & -18.95 $\pm$ 0.44 & 2.70 $\pm$ 0.20 & 17.37 $\pm$ 0.44 & 19.75 $\pm$ 0.86 & 3.01 $\pm$ 0.07 & -19.68 & 1.54 $\pm$ 0.18 \\
 NGC4713 & P & 1.60 $\pm$ 0.80 & -15.75 $\pm$ 2.26 & 2.48 $\pm$ 0.22 & 19.46 $\pm$ 0.55 & 17.98 $\pm$ 0.10 & 3.15 $\pm$ 0.01 & -16.57 & 0.72 $\pm$ 0.18 \\
 NGC4725 & C & 3.61 $\pm$ 0.22 & -20.51 $\pm$ 0.28 & 3.00 $\pm$ 0.22 & 17.31 $\pm$ 0.28 & 19.65 $\pm$ 0.13 & 3.66 $\pm$ 0.01 & -19.22 & -0.67 $\pm$ 0.02 \\
 NGC4736 & P & 1.30 $\pm$ 0.20 & -19.40 $\pm$ 0.28 & 2.20 $\pm$ 0.14 & 14.40 $\pm$ 0.28 & 17.19 $\pm$ 0.20 & 2.76 $\pm$ 0.04 & -19.02 & -0.08 $\pm$ 0.03 \\
 NGC4808 & ... & 1.01 $\pm$ 0.46 & -15.33 $\pm$ 0.69 & 2.27 $\pm$ 0.29 & 18.82 $\pm$ 0.67 & 18.50 $\pm$ 0.04 & 2.96 $\pm$ 0.01 & -17.65 & 1.22 $\pm$ 0.37 \\
 NGC4941 & P & 1.93 $\pm$ 0.34 & -16.46 $\pm$ 0.50 & 1.97 $\pm$ 0.23 & 16.20 $\pm$ 0.50 & 19.80 $\pm$ 0.10 & 2.96 $\pm$ 0.01 & -16.35 & 0.49 $\pm$ 0.21 \\
 NGC4984 & ... & 1.76 $\pm$ 0.23 & -20.49 $\pm$ 0.41 & 2.68 $\pm$ 0.14 & 15.72 $\pm$ 0.41 & 21.87 $\pm$ 0.17 & 3.80 $\pm$ 0.04 & -21.12 & 1.18 $\pm$ 0.02 \\
 NGC5005 & P & 1.20 $\pm$ 0.27 & -20.21 $\pm$ 0.31 & 2.25 $\pm$ 0.14 & 13.85 $\pm$ 0.31 & 18.70 $\pm$ 0.17 & 3.50 $\pm$ 0.03 & -20.38 & 0.36 $\pm$ 0.96 \\
 NGC5033 & P & 1.70 $\pm$ 0.55 & -19.02 $\pm$ 0.51 & 2.24 $\pm$ 0.18 & 14.98 $\pm$ 0.51 & 20.46 $\pm$ 0.10 & 3.65 $\pm$ 0.02 & -20.49 & 1.10 $\pm$ 0.24 \\
 NGC5055 & P & 1.33 $\pm$ 0.68 & -20.14 $\pm$ 0.96 & 2.98 $\pm$ 0.33 & 17.59 $\pm$ 0.96 & 18.91 $\pm$ 1.68 & 3.48 $\pm$ 0.14 & -21.49 & 1.20 $\pm$ 0.07 \\
 NGC5068 & P & 0.30 $\pm$ 0.11 & -11.78 $\pm$ 0.35 & 1.67 $\pm$ 0.06 & 19.39 $\pm$ 0.33 & 19.80 $\pm$ 0.05 & 3.07 $\pm$ 0.02 & -14.35 & 2.08 $\pm$ 0.38 \\
 NGC5128 & C & 2.63 $\pm$ 0.46 & -20.13 $\pm$ 0.41 & 2.66 $\pm$ 0.48 & 15.97 $\pm$ 0.41 & 17.25 $\pm$ 0.24 & 3.16 $\pm$ 0.04 & -18.93 & -0.15 $\pm$ 0.03 \\
 NGC5194 & P & 0.50 $\pm$ 0.13 & -19.58 $\pm$ 0.22 & 2.78 $\pm$ 0.06 & 17.14 $\pm$ 0.21 & 18.98 $\pm$ 0.10 & 3.47 $\pm$ 0.02 & -19.58 & 0.05 $\pm$ 0.02 \\
 NGC5236 & P & 0.44 $\pm$ 0.06 & -19.95 $\pm$ 0.11 & 2.42 $\pm$ 0.03 & 14.97 $\pm$ 0.10 & 18.16 $\pm$ 0.07 & 3.49 $\pm$ 0.01 & -22.17 & 1.93 $\pm$ 0.02 \\
 NGC5248 & P & 0.69 $\pm$ 0.27 & -19.83 $\pm$ 0.50 & 2.74 $\pm$ 0.10 & 16.70 $\pm$ 0.50 & 19.24 $\pm$ 0.18 & 3.42 $\pm$ 0.02 & -21.41 & 1.62 $\pm$ 0.00 \\
 NGC5273 & C & 4.07 $\pm$ 0.37 & -18.64 $\pm$ 0.51 & 2.63 $\pm$ 0.63 & 17.31 $\pm$ 0.51 & 19.87 $\pm$ 0.15 & 3.24 $\pm$ 0.02 & -18.23 & -0.08 $\pm$ 0.03 \\
 NGC5338 & ... & 3.26 $\pm$ 0.39 & -17.22 $\pm$ 0.53 & 2.88 $\pm$ 0.89 & 20.01 $\pm$ 0.53 & 21.67 $\pm$ 0.28 & 3.17 $\pm$ 0.03 & -17.34 & 0.95 $\pm$ 0.02 \\
 NGC5457 & P & 1.52 $\pm$ 0.92 & -16.77 $\pm$ 1.34 & 2.57 $\pm$ 0.51 & 18.90 $\pm$ 1.34 & 20.05 $\pm$ 0.09 & 3.47 $\pm$ 0.01 & -18.14 & 0.27 $\pm$ 1.25 \\
 NGC5474 & P & 0.74 $\pm$ 0.20 & -16.75 $\pm$ 0.22 & 2.50 $\pm$ 0.08 & 18.54 $\pm$ 0.22 & 20.04 $\pm$ 0.24 & 3.09 $\pm$ 0.06 & -16.67 & -0.09 $\pm$ 0.03 \\
 NGC5585 & P & 0.89 $\pm$ 0.16 & -16.00 $\pm$ 0.19 & 2.39 $\pm$ 0.07 & 18.77 $\pm$ 0.18 & 19.81 $\pm$ 0.06 & 3.09 $\pm$ 0.01 & -16.51 & 0.03 $\pm$ 0.03 \\
 NGC5643 & P & 2.96 $\pm$ 0.18 & -18.89 $\pm$ 0.30 & 2.30 $\pm$ 0.13 & 15.44 $\pm$ 0.30 & 18.97 $\pm$ 0.07 & 3.40 $\pm$ 0.01 & -20.06 & 1.10 $\pm$ 0.16 \\
 NGC5832 & ... & 1.63 $\pm$ 1.28 & -15.24 $\pm$ 1.23 & 2.31 $\pm$ 0.53 & 19.11 $\pm$ 1.22 & 19.32 $\pm$ 0.17 & 3.17 $\pm$ 0.03 & -14.48 & -0.14 $\pm$ 0.03 \\
 NGC5879 & P & 1.34 $\pm$ 0.88 & -18.28 $\pm$ 0.85 & 2.53 $\pm$ 0.23 & 17.16 $\pm$ 0.85 & 20.04 $\pm$ 0.37 & 3.27 $\pm$ 0.07 & -19.78 & 1.13 $\pm$ 0.43 \\
 NGC5949 & ... & 1.67 $\pm$ 0.93 & -14.58 $\pm$ 1.53 & 2.29 $\pm$ 1.05 & 19.69 $\pm$ 1.51 & 19.54 $\pm$ 0.18 & 3.18 $\pm$ 0.02 & -16.58 & 0.66 $\pm$ 0.14 \\
 NGC6207 & P & 1.40 $\pm$ 0.80 & -16.61 $\pm$ 0.60 & 2.71 $\pm$ 0.56 & 19.75 $\pm$ 0.57 & 19.04 $\pm$ 0.95 & 3.08 $\pm$ 0.11 & -16.77 & 0.66 $\pm$ 0.62 \\
 NGC6300 & P & 0.52 $\pm$ 0.19 & -18.53 $\pm$ 0.34 & 2.70 $\pm$ 0.08 & 17.80 $\pm$ 0.34 & 18.74 $\pm$ 0.05 & 3.39 $\pm$ 0.01 & -20.40 & 1.08 $\pm$ 0.04 \\
 NGC6503 & P & 0.95 $\pm$ 0.77 & -13.88 $\pm$ 1.33 & 1.72 $\pm$ 0.22 & 17.54 $\pm$ 1.33 & 18.36 $\pm$ 0.06 & 2.95 $\pm$ 0.01 & -14.72 & 0.26 $\pm$ 0.20 \\
 NGC6684 & C & 3.45 $\pm$ 0.41 & -19.39 $\pm$ 0.52 & 2.72 $\pm$ 0.80 & 17.03 $\pm$ 0.51 & 19.24 $\pm$ 0.19 & 3.14 $\pm$ 0.02 & -17.55 & -1.21 $\pm$ 0.02 \\
 NGC6744 & C & 3.06 $\pm$ 0.53 & -20.06 $\pm$ 0.50 & 3.18 $\pm$ 0.39 & 18.68 $\pm$ 0.49 & 20.03 $\pm$ 0.45 & 3.79 $\pm$ 0.14 & -18.21 & -0.57 $\pm$ 0.12 \\
 NGC7177 & P & 1.79 $\pm$ 0.22 & -19.90 $\pm$ 0.23 & 2.73 $\pm$ 0.10 & 16.58 $\pm$ 0.23 & 18.66 $\pm$ 0.14 & 3.17 $\pm$ 0.01 & -19.67 & -0.10 $\pm$ 0.06 \\
 NGC7217 & C & 2.16 $\pm$ 0.30 & -21.35 $\pm$ 0.37 & 3.27 $\pm$ 0.13 & 17.82 $\pm$ 0.36 & 19.89 $\pm$ 4.89 & 3.55 $\pm$ 0.50 & -20.04 & -0.10 $\pm$ 0.03 \\
 NGC7331 & C & 5.73 $\pm$ 0.34 & -22.75 $\pm$ 0.48 & 3.73 $\pm$ 0.30 & 18.69 $\pm$ 0.48 & 18.06 $\pm$ 0.23 & 3.48 $\pm$ 0.03 & -19.23 & -1.13 $\pm$ 0.65 \\
 NGC7457 & C & 2.72 $\pm$ 0.40 & -19.59 $\pm$ 0.48 & 2.85 $\pm$ 0.23 & 17.46 $\pm$ 0.48 & 19.02 $\pm$ 0.17 & 3.49 $\pm$ 0.04 & -17.32 & -1.23 $\pm$ 0.02 \\
 NGC7713 & P & 1.13 $\pm$ 0.95 & -15.53 $\pm$ 0.72 & 2.32 $\pm$ 0.21 & 18.90 $\pm$ 0.70 & 19.01 $\pm$ 0.40 & 3.33 $\pm$ 0.06 & -14.34 & 0.87 $\pm$ 0.80 \\
 NGC7741 & P & 0.52 $\pm$ 0.09 & -17.44 $\pm$ 0.12 & 2.83 $\pm$ 0.03 & 19.50 $\pm$ 0.11 & 20.57 $\pm$ 0.07 & 3.37 $\pm$ 0.01 & -18.23 & 1.42 $\pm$ 0.19 \\
 NGC7793 & P & 1.08 $\pm$ 0.41 & -15.99 $\pm$ 0.50 & 2.26 $\pm$ 0.17 & 18.15 $\pm$ 0.49 & 18.55 $\pm$ 0.06 & 2.99 $\pm$ 0.01 & -17.38 & 1.09 $\pm$ 0.06 \\
 UGC10445 & P & 1.30 $\pm$ 0.62 & -14.95 $\pm$ 0.56 & 2.77 $\pm$ 0.26 & 21.71 $\pm$ 0.55 & 21.79 $\pm$ 0.25 & 3.26 $\pm$ 0.03 & -14.49 & -0.36 $\pm$ 0.23 \\
 \enddata

  \tablenotetext{a}{P- Pseudobulge C- Classical Bulge}
  \end{deluxetable}

  Here we show figures of all bulge-disk decompositions used in this
  paper. For each galaxy the points of the surface brightness profile
  that are included in the fitting are represented by filled circles,
  those points not included in the fit are represented by crosses. The
  fits are represented by three solid red lines, one for the S\'ersic
  function, another for the exponential outer disk, and the third is
  the sum of the bulge and disk light. We plot all profiles against
  $r^{1/4}$ because to facilitate identification of structural features
  such as bars and rings.

 \begin{figure*}
 \includegraphics[width=0.99\textwidth]{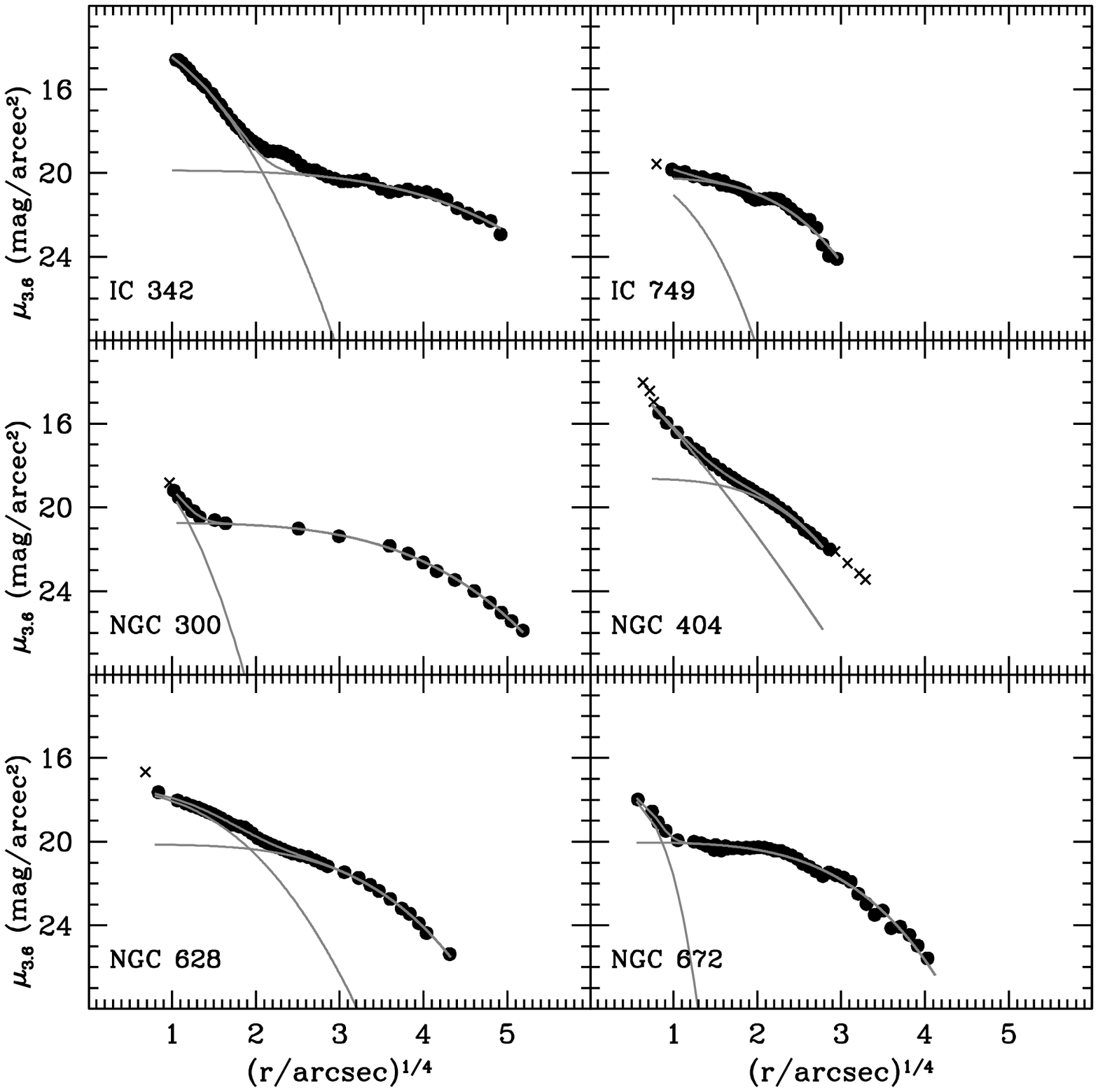}
 \caption{Galaxy surface brightness profiles and decompositions from
   Table 2. Solid circles mark data included in the fit, crosses
   are points not included in the fit. The three red lines represent
   the S\'ersic function, exponential disk, and sum of the two which
   results from bulge-disk decomposition. \label{fig:profiles}}
 \end{figure*}

 \setcounter{figure}{0}
 \begin{figure*}
 \includegraphics[width=0.99\textwidth]{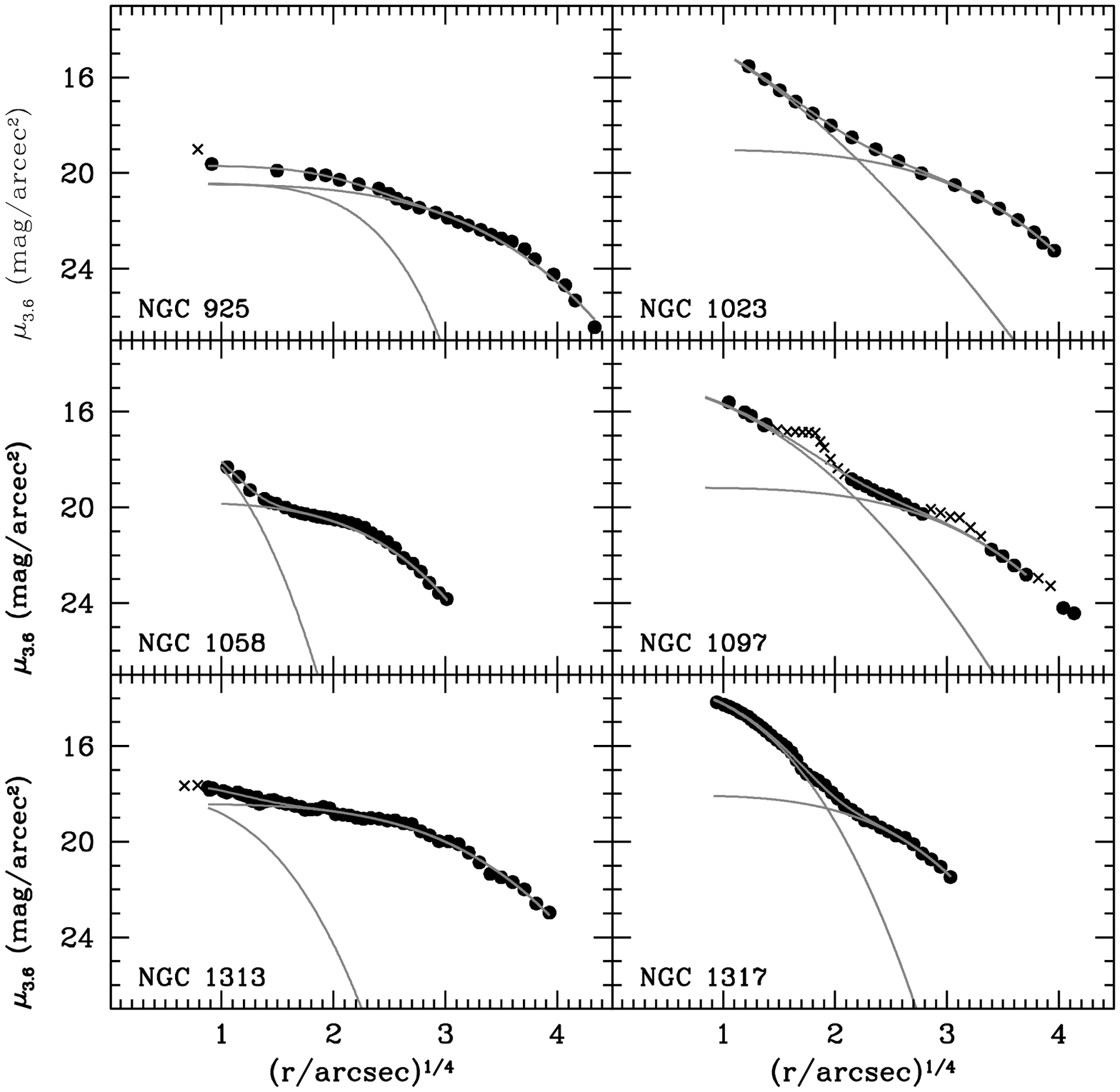}
 \caption{Galaxy surface brightness profiles and decompositions from Table 2. Solid circles are those included in the fitting, crosses are points not included in the fit. The three red lines represent the S\'ersic function, exponential disk, and sum of the two which results from bulge-disk decomposition. \label{fig:profiles}}
 \end{figure*}

 \setcounter{figure}{0}
 \begin{figure*}
 \includegraphics[width=0.99\textwidth]{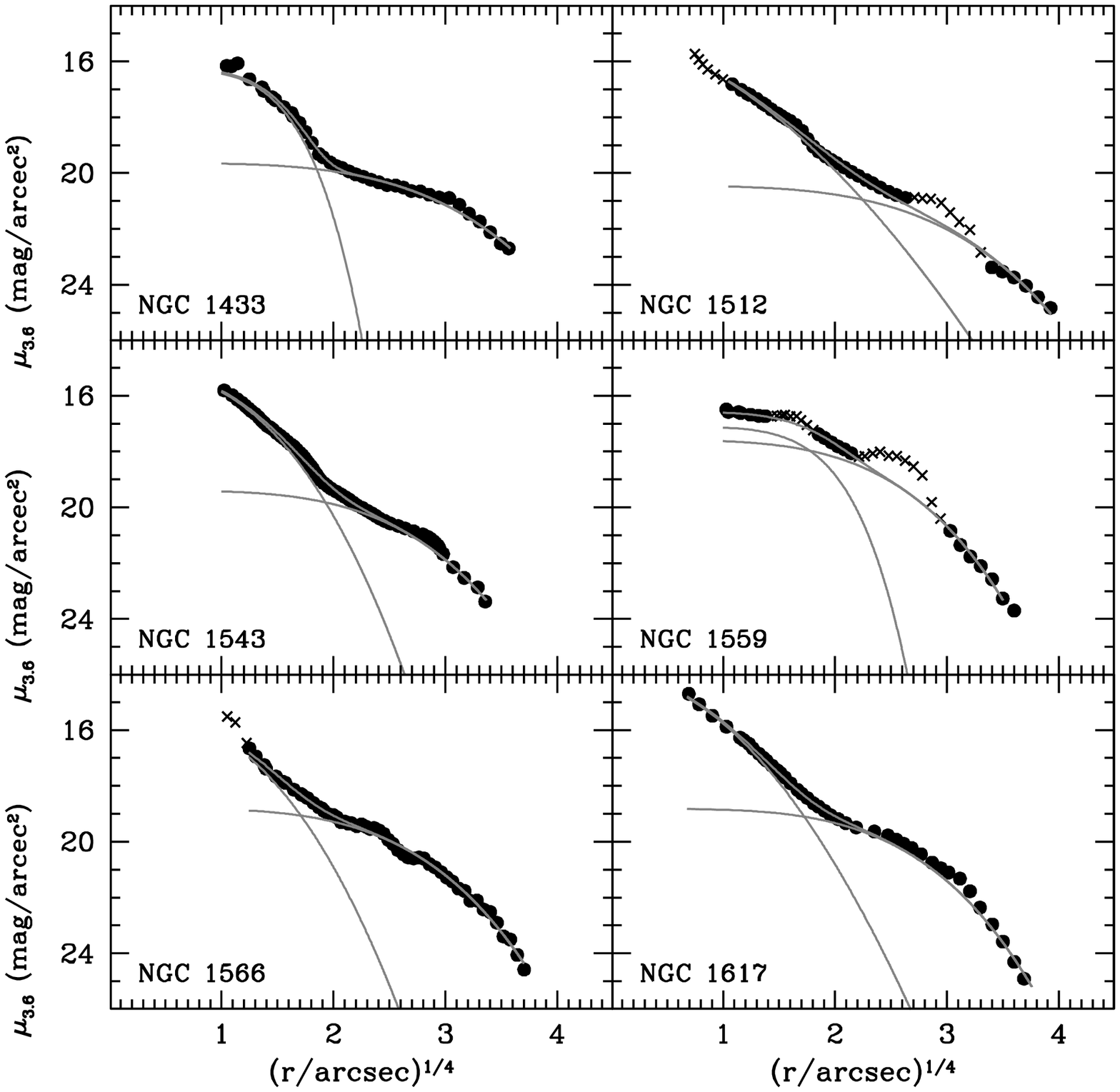}
 \caption{Galaxy surface brightness profiles and decompositions from Table 2. Solid circles are those included in the fitting, crosses are points not included in the fit. The three red lines represent the S\'ersic function, exponential disk, and sum of the two which results from bulge-disk decomposition. \label{fig:profiles}}
 \end{figure*}

 \setcounter{figure}{0}
 \begin{figure*}
 \includegraphics[width=0.99\textwidth]{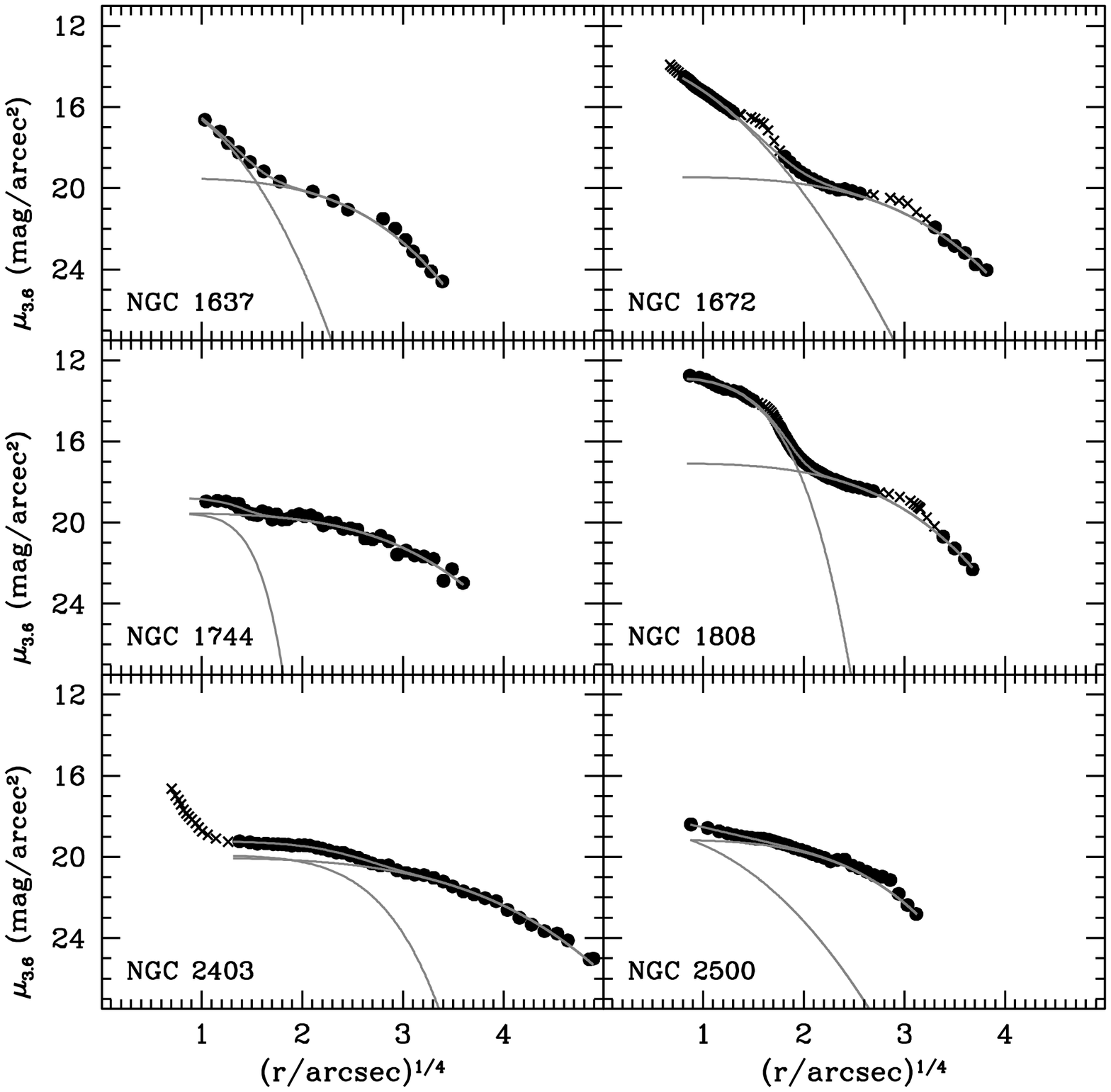}
 \caption{Galaxy surface brightness profiles and decompositions from Table 2. Solid circles are those included in the fitting, crosses are points not included in the fit. The three red lines represent the S\'ersic function, exponential disk, and sum of the two which results from bulge-disk decomposition. \label{fig:profiles}}
 \end{figure*}

 \setcounter{figure}{0}
 \begin{figure*}
 \includegraphics[width=0.99\textwidth]{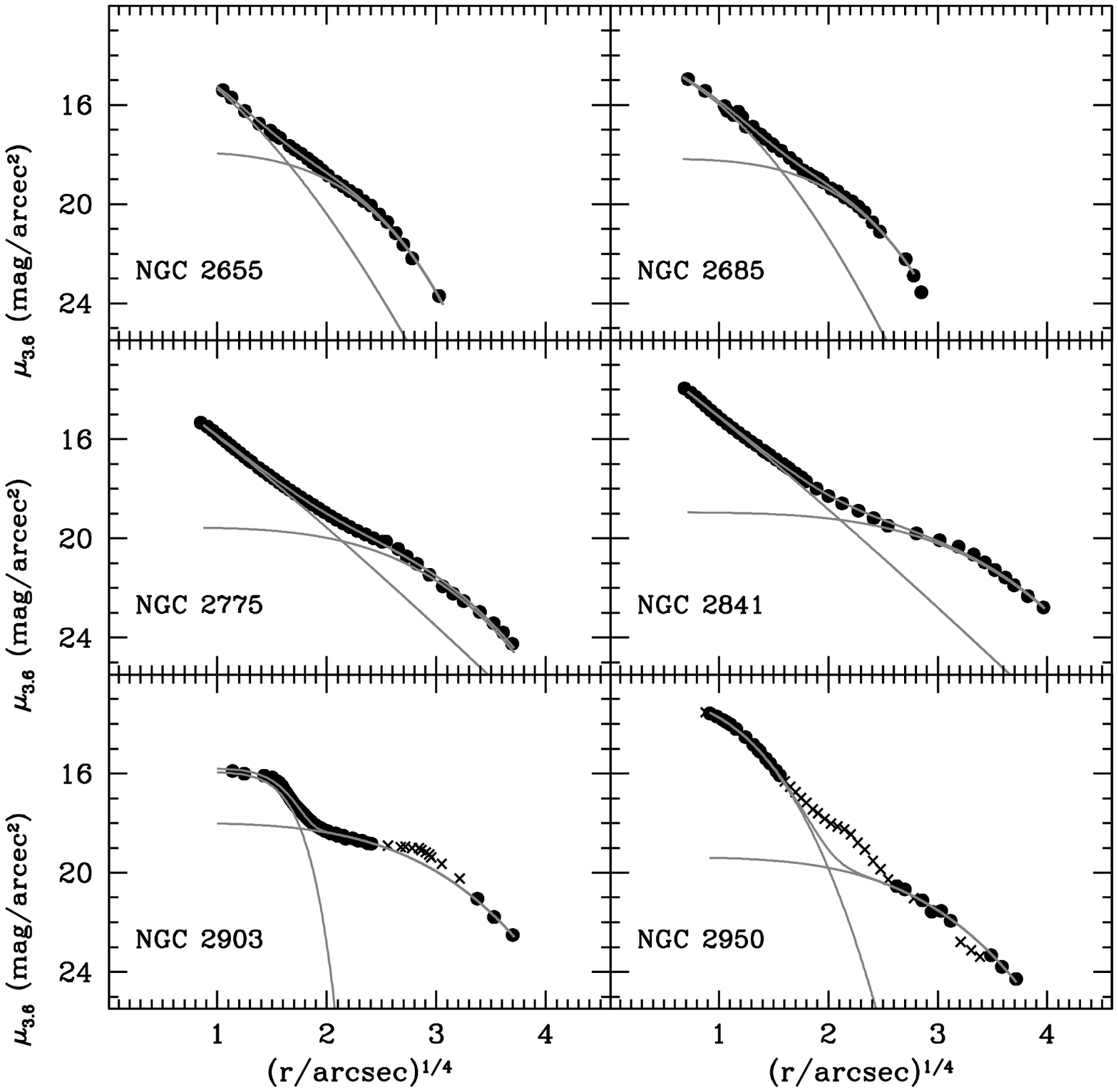}
 \caption{Galaxy surface brightness profiles and decompositions from Table 2. Solid circles are those included in the fitting, crosses are points not included in the fit. The three red lines represent the S\'ersic function, exponential disk, and sum of the two which results from bulge-disk decomposition. \label{fig:profiles}}
 \end{figure*}

 \setcounter{figure}{0}
 \begin{figure*}
 \includegraphics[width=0.99\textwidth]{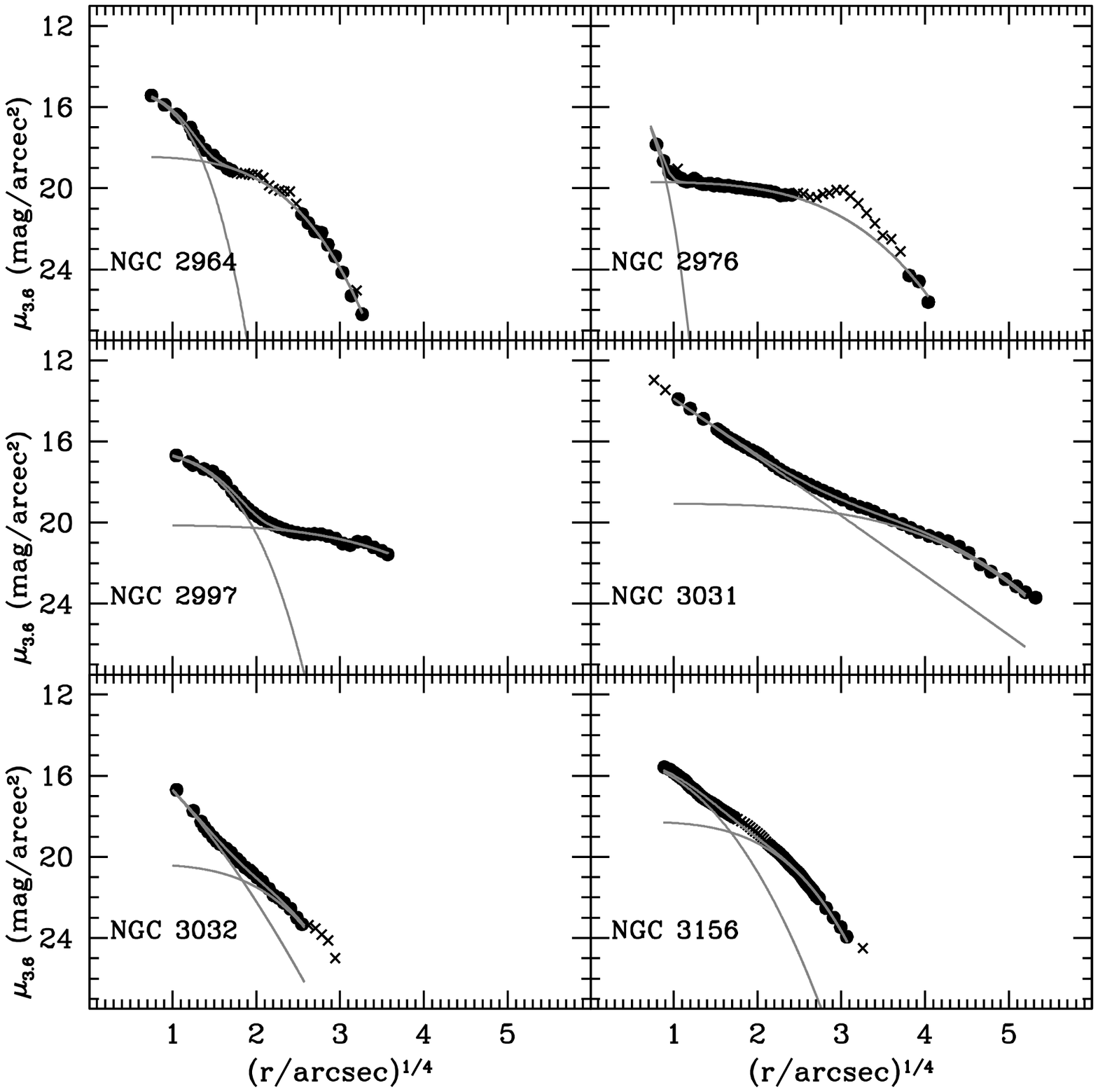}
 \caption{Galaxy surface brightness profiles and decompositions from Table 2. Solid circles are those included in the fitting, crosses are points not included in the fit. The three red lines represent the S\'ersic function, exponential disk, and sum of the two which results from bulge-disk decomposition. \label{fig:profiles}}
 \end{figure*}

 \setcounter{figure}{0}
 \begin{figure*}
 \includegraphics[width=0.99\textwidth]{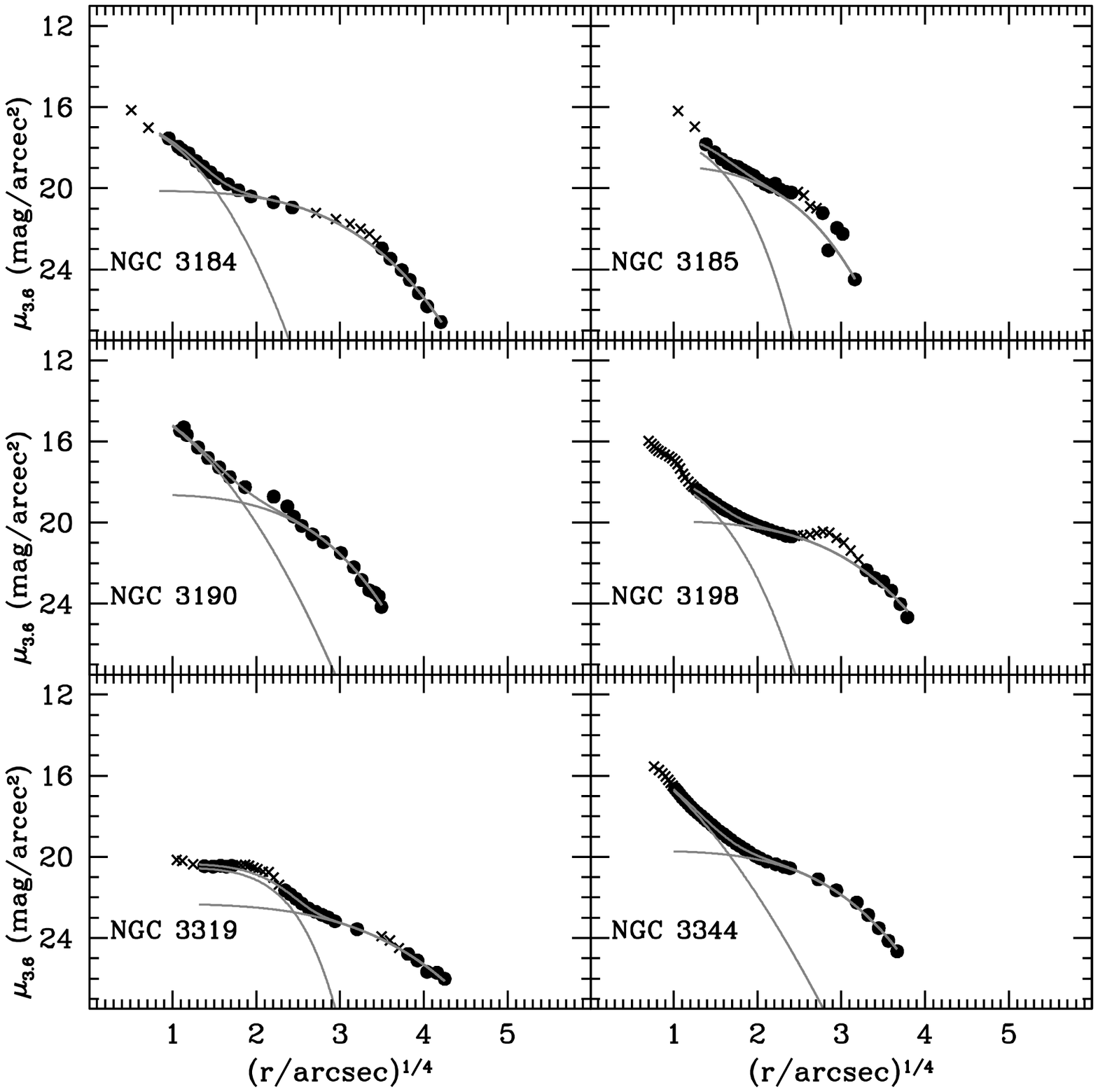}
 \caption{Galaxy surface brightness profiles and decompositions from Table 2. Solid circles are those included in the fitting, crosses are points not included in the fit. The three red lines represent the S\'ersic function, exponential disk, and sum of the two which results from bulge-disk decomposition. \label{fig:profiles}}
 \end{figure*}

 \setcounter{figure}{0}
 \begin{figure*}
 \includegraphics[width=0.99\textwidth]{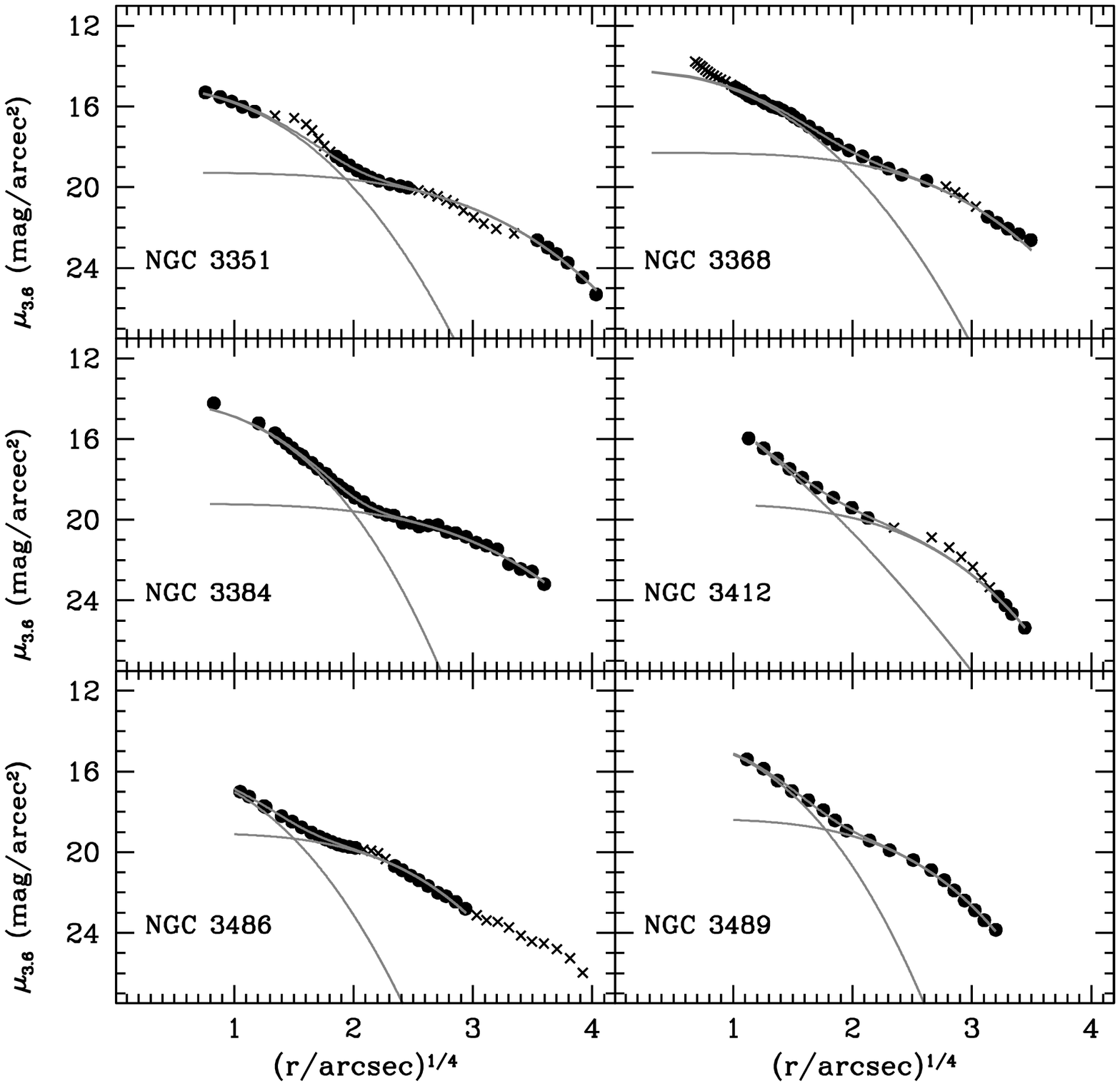}
 \caption{Galaxy surface brightness profiles and decompositions from Table 2. Solid circles are those included in the fitting, crosses are points not included in the fit. The three red lines represent the S\'ersic function, exponential disk, and sum of the two which results from bulge-disk decomposition. \label{fig:profiles}}
 \end{figure*}

 \setcounter{figure}{0}
 \begin{figure*}
 \includegraphics[width=0.99\textwidth]{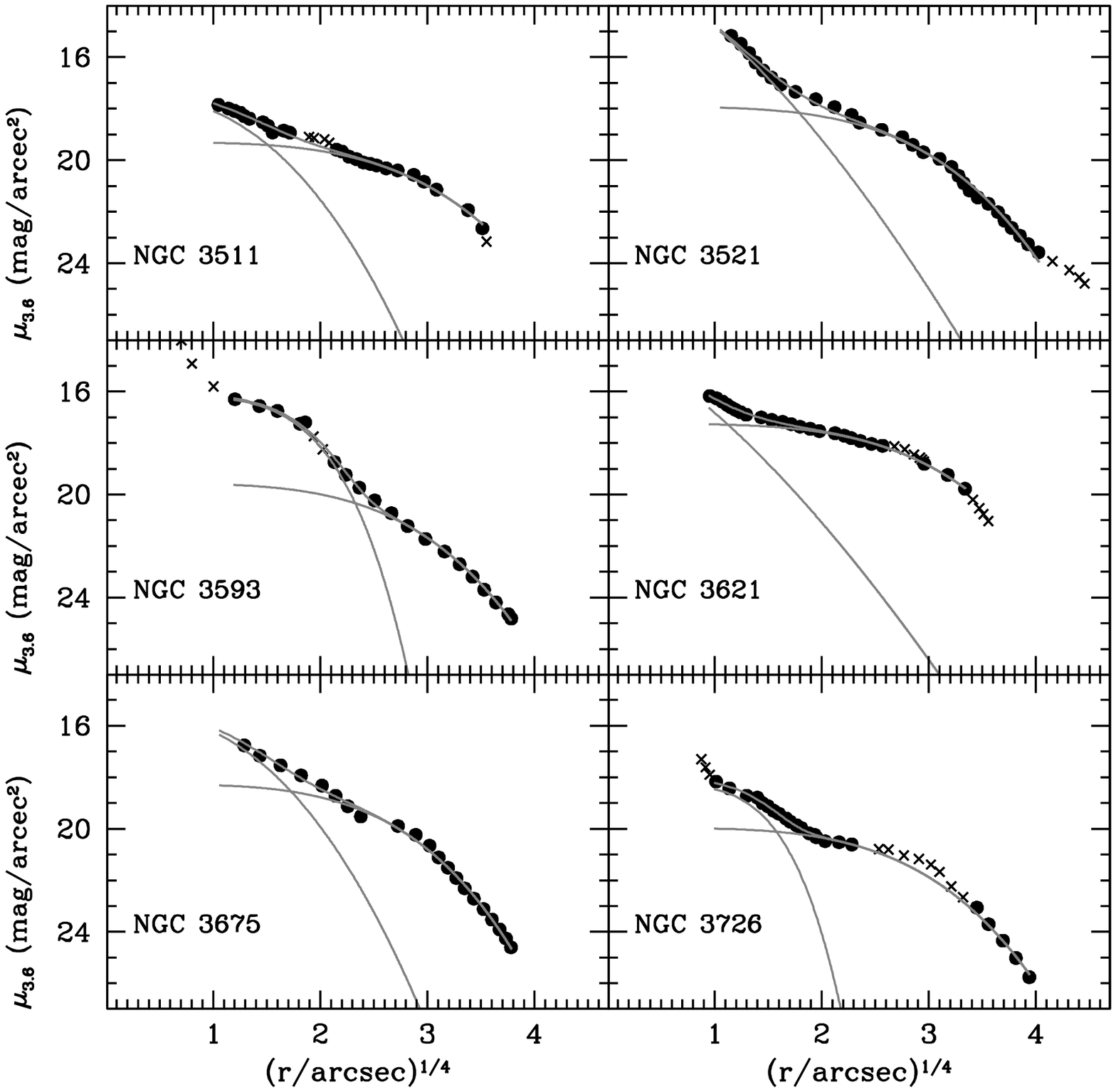}
 \caption{Galaxy surface brightness profiles and decompositions from Table 2. Solid circles are those included in the fitting, crosses are points not included in the fit. The three red lines represent the S\'ersic function, exponential disk, and sum of the two which results from bulge-disk decomposition. \label{fig:profiles}}
 \end{figure*}

 \setcounter{figure}{0}
 \begin{figure*}
 \includegraphics[width=0.99\textwidth]{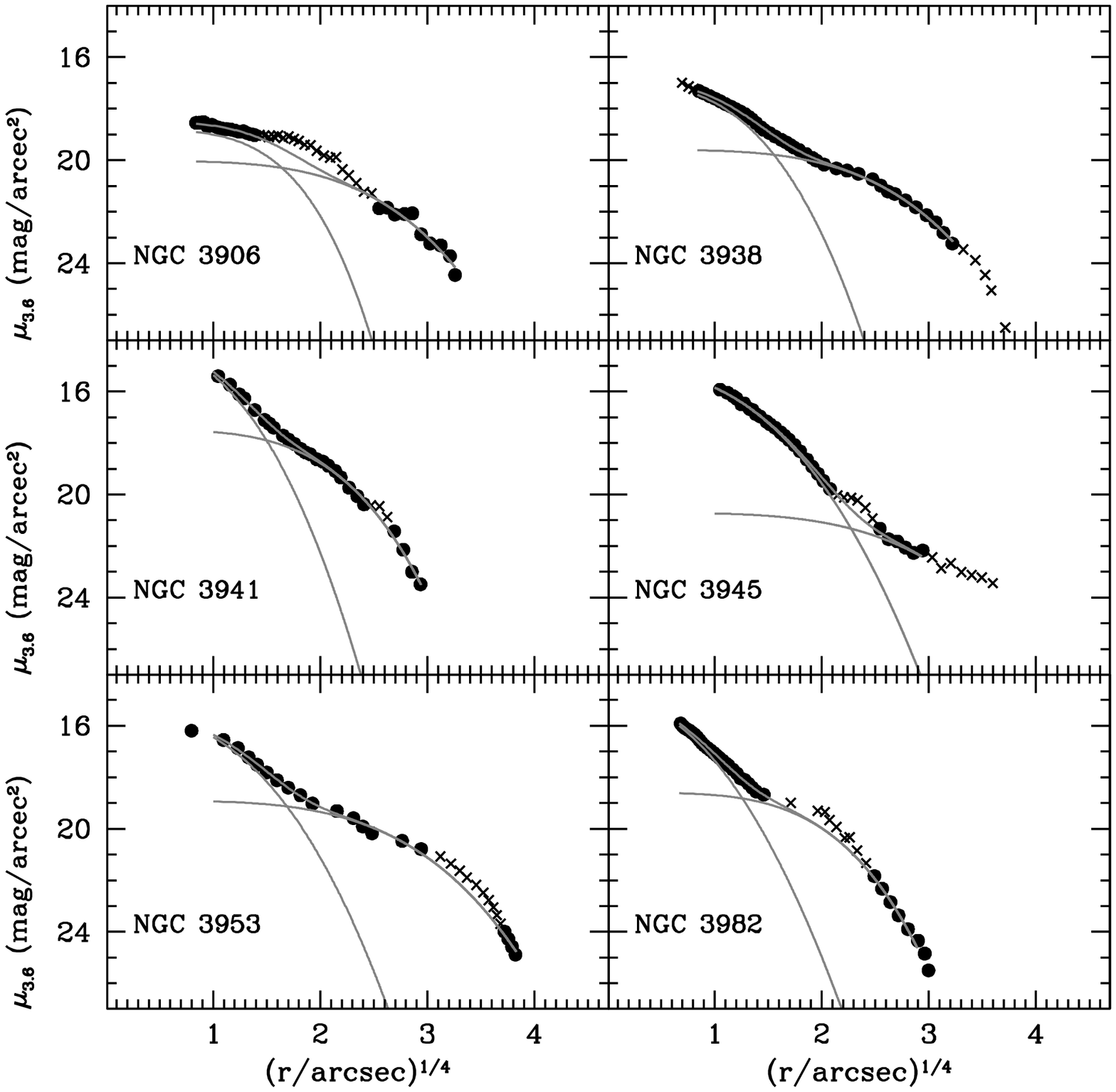}
 \caption{Galaxy surface brightness profiles and decompositions from Table 2. Solid circles are those included in the fitting, crosses are points not included in the fit. The three red lines represent the S\'ersic function, exponential disk, and sum of the two which results from bulge-disk decomposition. \label{fig:profiles}}
 \end{figure*}

 \setcounter{figure}{0}
 \begin{figure*}
 \includegraphics[width=0.99\textwidth]{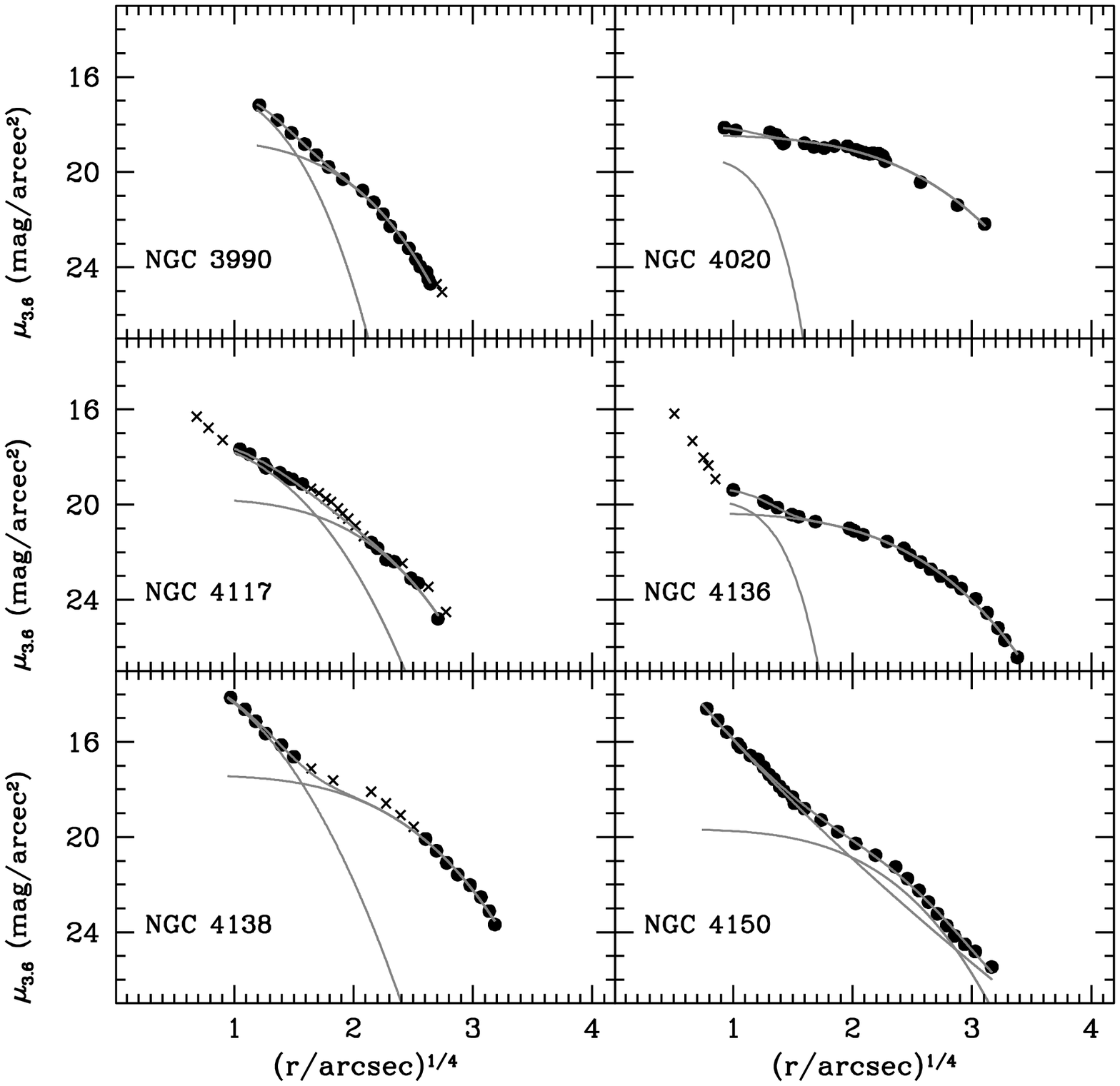}
 \caption{Galaxy surface brightness profiles and decompositions from Table 2. Solid circles are those included in the fitting, crosses are points not included in the fit. The three red lines represent the S\'ersic function, exponential disk, and sum of the two which results from bulge-disk decomposition. \label{fig:profiles}}
 \end{figure*}

 \setcounter{figure}{0}
 \begin{figure*}
 \includegraphics[width=0.99\textwidth]{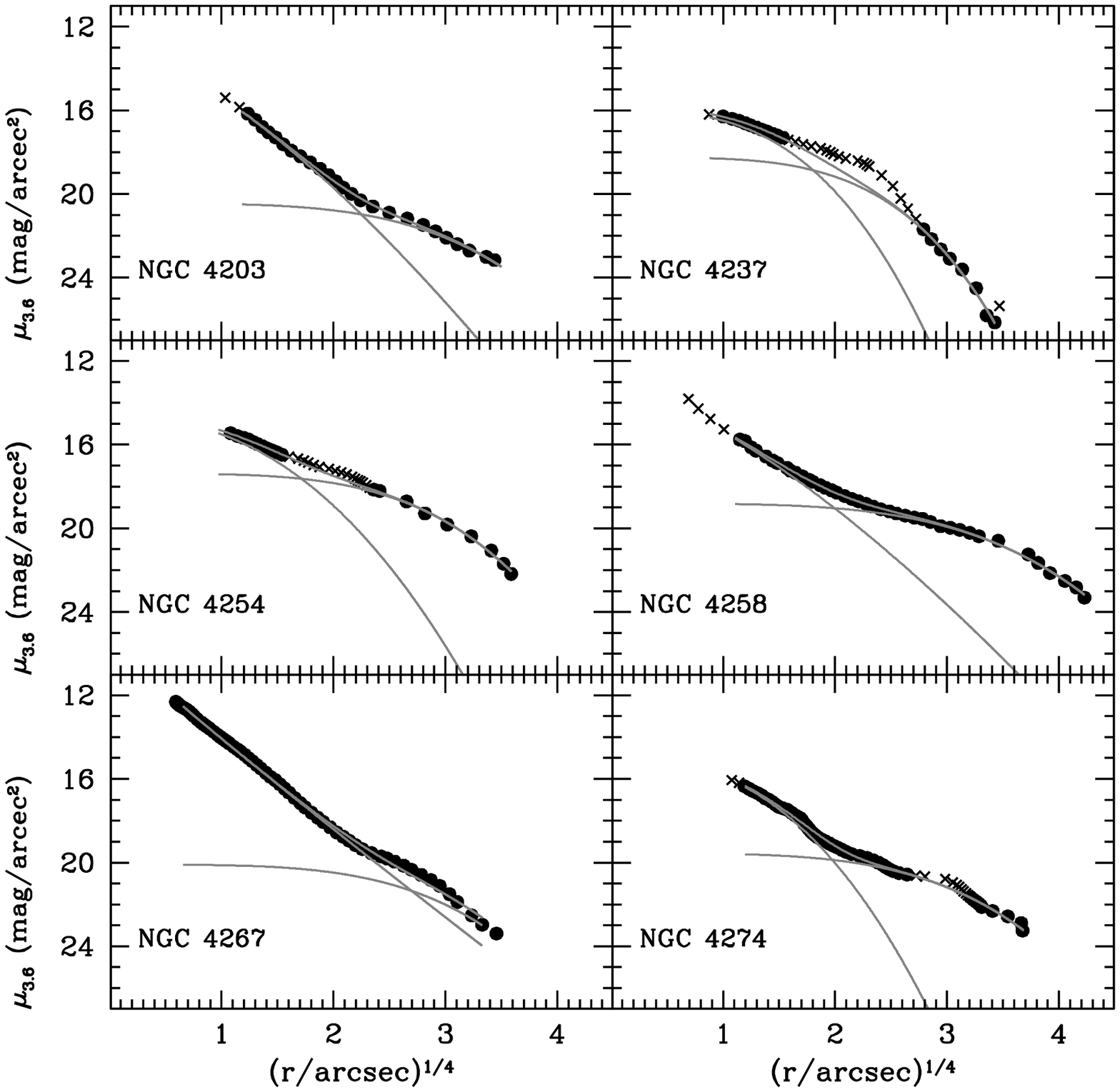}
 \caption{Galaxy surface brightness profiles and decompositions from Table 2. Solid circles are those included in the fitting, crosses are points not included in the fit. The three red lines represent the S\'ersic function, exponential disk, and sum of the two which results from bulge-disk decomposition. \label{fig:profiles}}
 \end{figure*}
 \clearpage

 \setcounter{figure}{0}
 \begin{figure*}
 \includegraphics[width=0.99\textwidth]{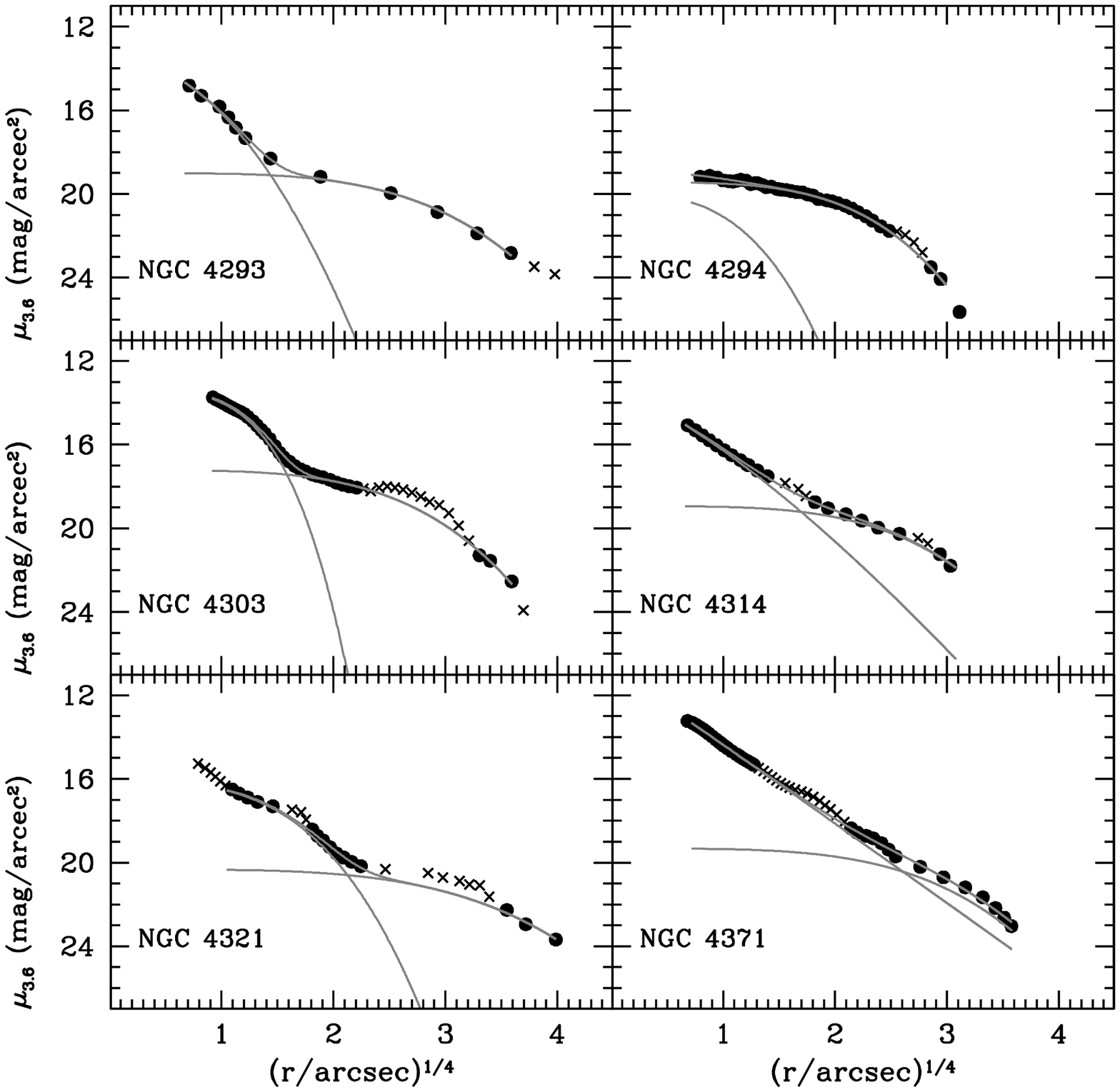}
 \caption{Galaxy surface brightness profiles and decompositions from Table 2. Solid circles are those included in the fitting, crosses are points not included in the fit. The three red lines represent the S\'ersic function, exponential disk, and sum of the two which results from bulge-disk decomposition. \label{fig:profiles}}
 \end{figure*}

 \setcounter{figure}{0}
 \begin{figure*}
 \includegraphics[width=0.99\textwidth]{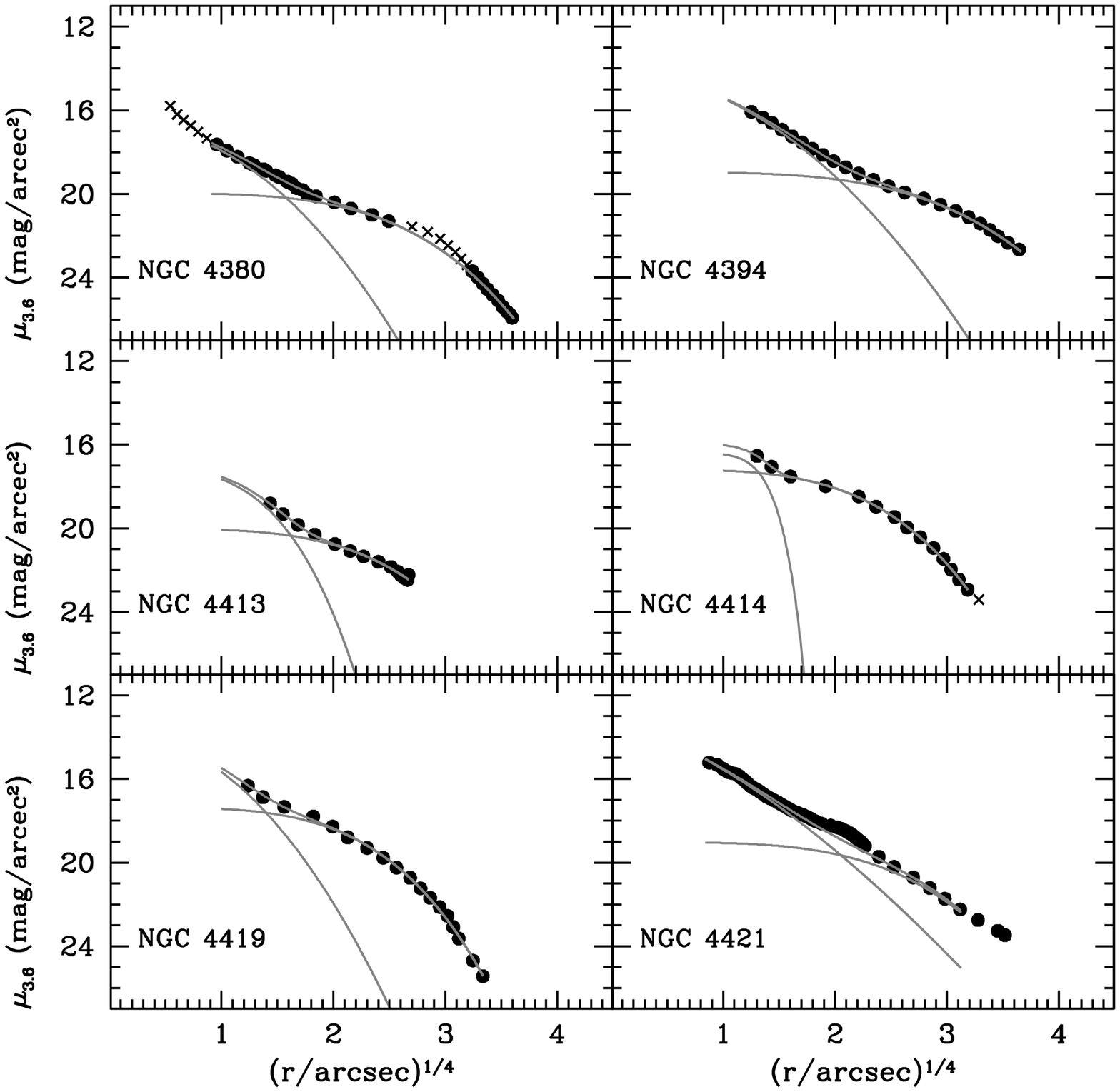}
 \caption{Galaxy surface brightness profiles and decompositions from Table 2. Solid circles are those included in the fitting, crosses are points not included in the fit. The three red lines represent the S\'ersic function, exponential disk, and sum of the two which results from bulge-disk decomposition. \label{fig:profiles}}
 \end{figure*}
 \clearpage

 \setcounter{figure}{0}
 \begin{figure*}
 \includegraphics[width=0.99\textwidth]{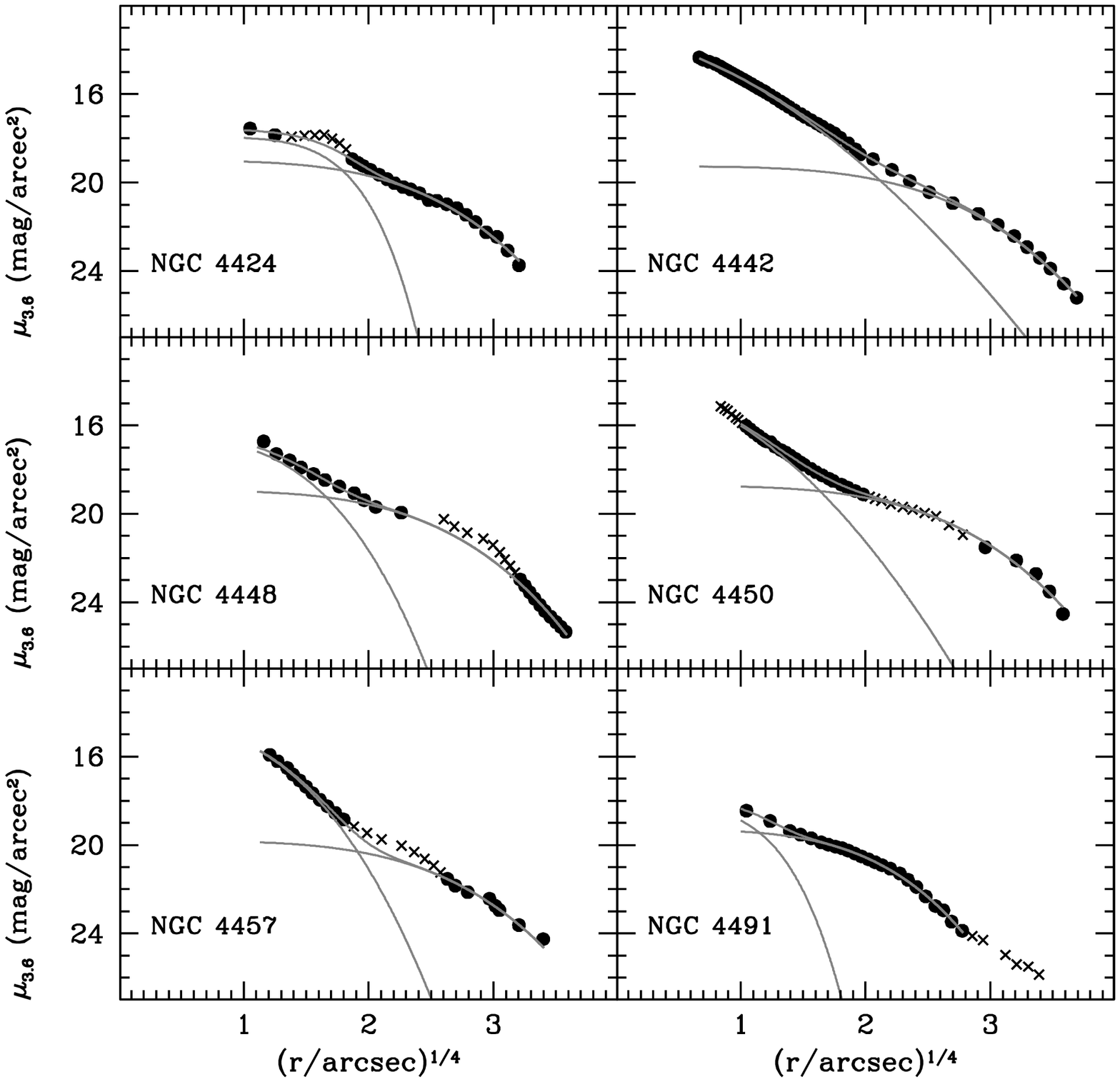}
 \caption{Galaxy surface brightness profiles and decompositions from Table 2. Solid circles are those included in the fitting, crosses are points not included in the fit. The three red lines represent the S\'ersic function, exponential disk, and sum of the two which results from bulge-disk decomposition. \label{fig:profiles}}
 \end{figure*}

 \setcounter{figure}{0}
 \begin{figure*}
 \includegraphics[width=0.99\textwidth]{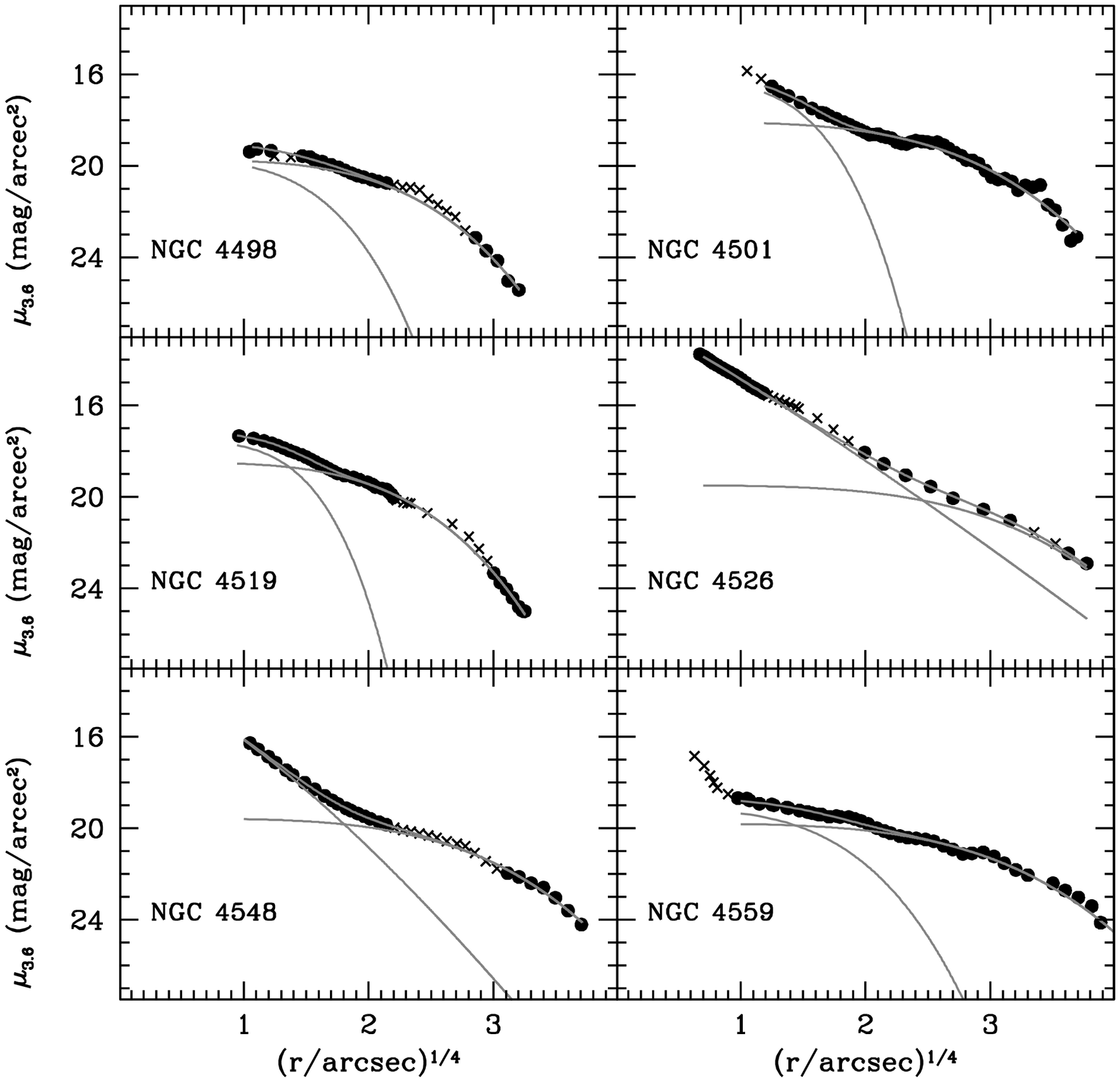}
 \caption{Galaxy surface brightness profiles and decompositions from Table 2. Solid circles are those included in the fitting, crosses are points not included in the fit. The three red lines represent the S\'ersic function, exponential disk, and sum of the two which results from bulge-disk decomposition. \label{fig:profiles}}
 \end{figure*}

 \setcounter{figure}{0}
 \begin{figure*}
 \includegraphics[width=0.99\textwidth]{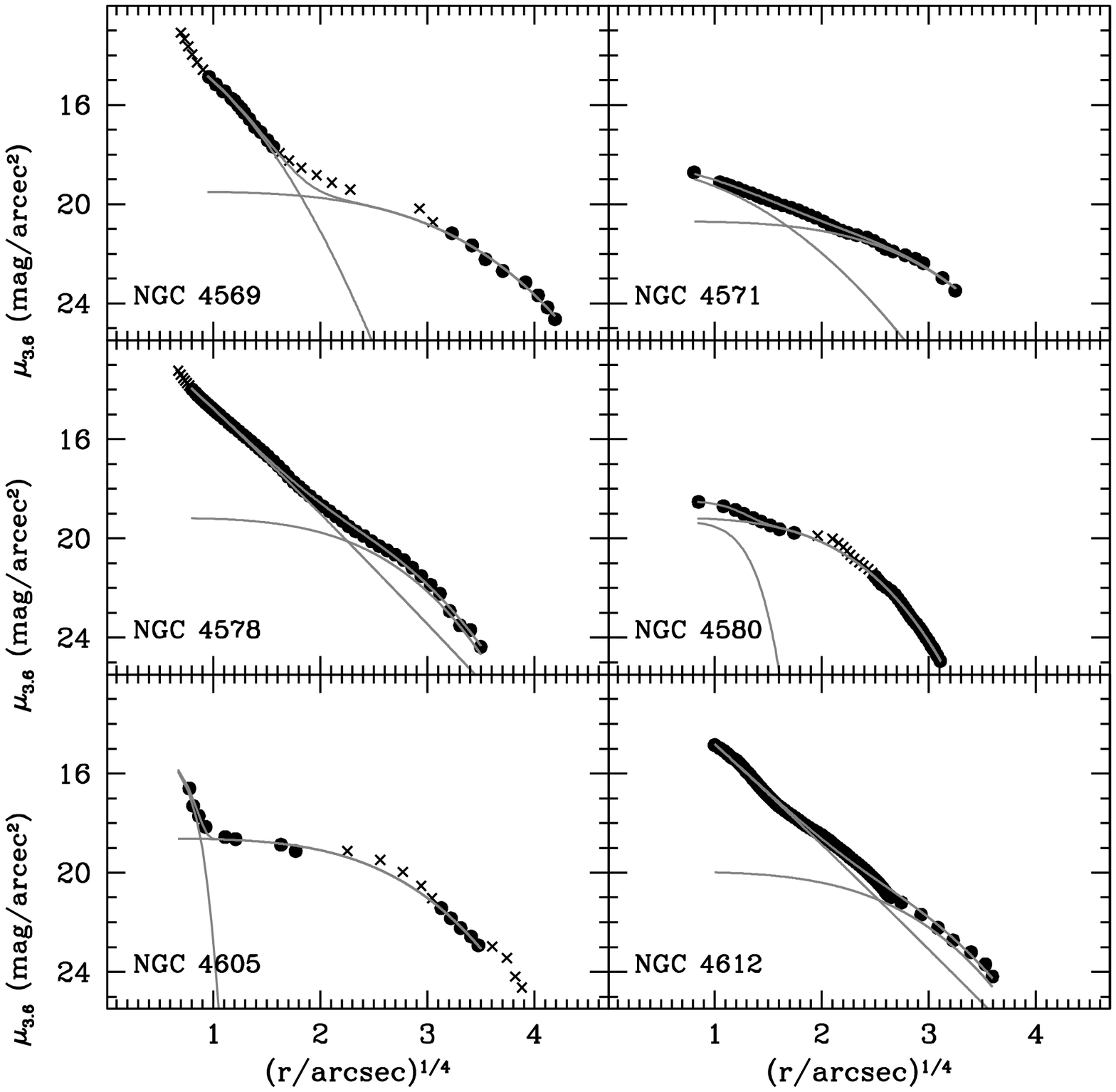}
 \caption{Galaxy surface brightness profiles and decompositions from Table 2. Solid circles are those included in the fitting, crosses are points not included in the fit. The three red lines represent the S\'ersic function, exponential disk, and sum of the two which results from bulge-disk decomposition. \label{fig:profiles}}
 \end{figure*}

 \setcounter{figure}{0}
 \begin{figure*}
 \includegraphics[width=0.99\textwidth]{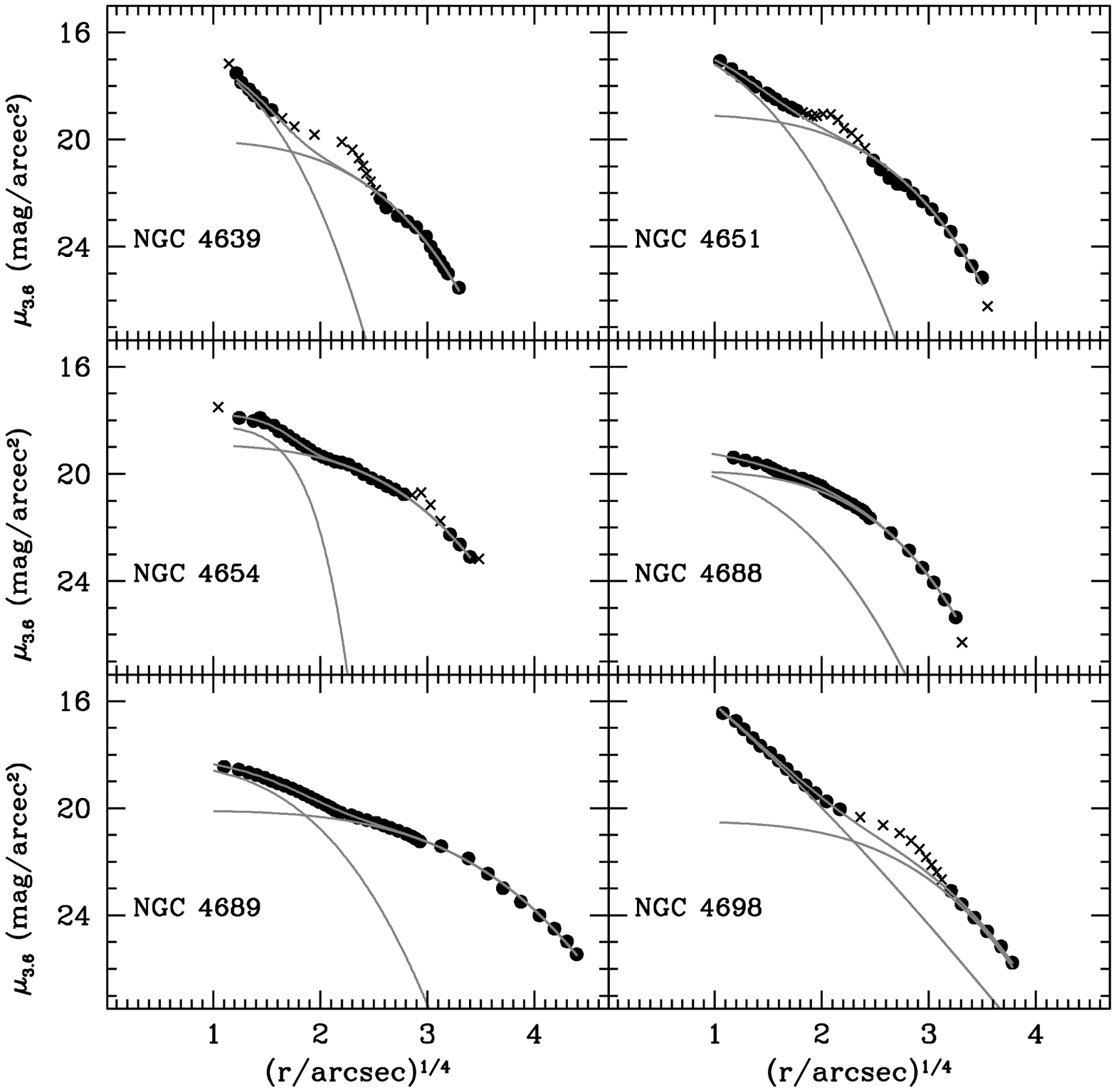}
 \caption{Galaxy surface brightness profiles and decompositions from Table 2. Solid circles are those included in the fitting, crosses are points not included in the fit. The three red lines represent the S\'ersic function, exponential disk, and sum of the two which results from bulge-disk decomposition. \label{fig:profiles}}
 \end{figure*}
 \clearpage

 \setcounter{figure}{0}
 \begin{figure*}
 \includegraphics[width=0.99\textwidth]{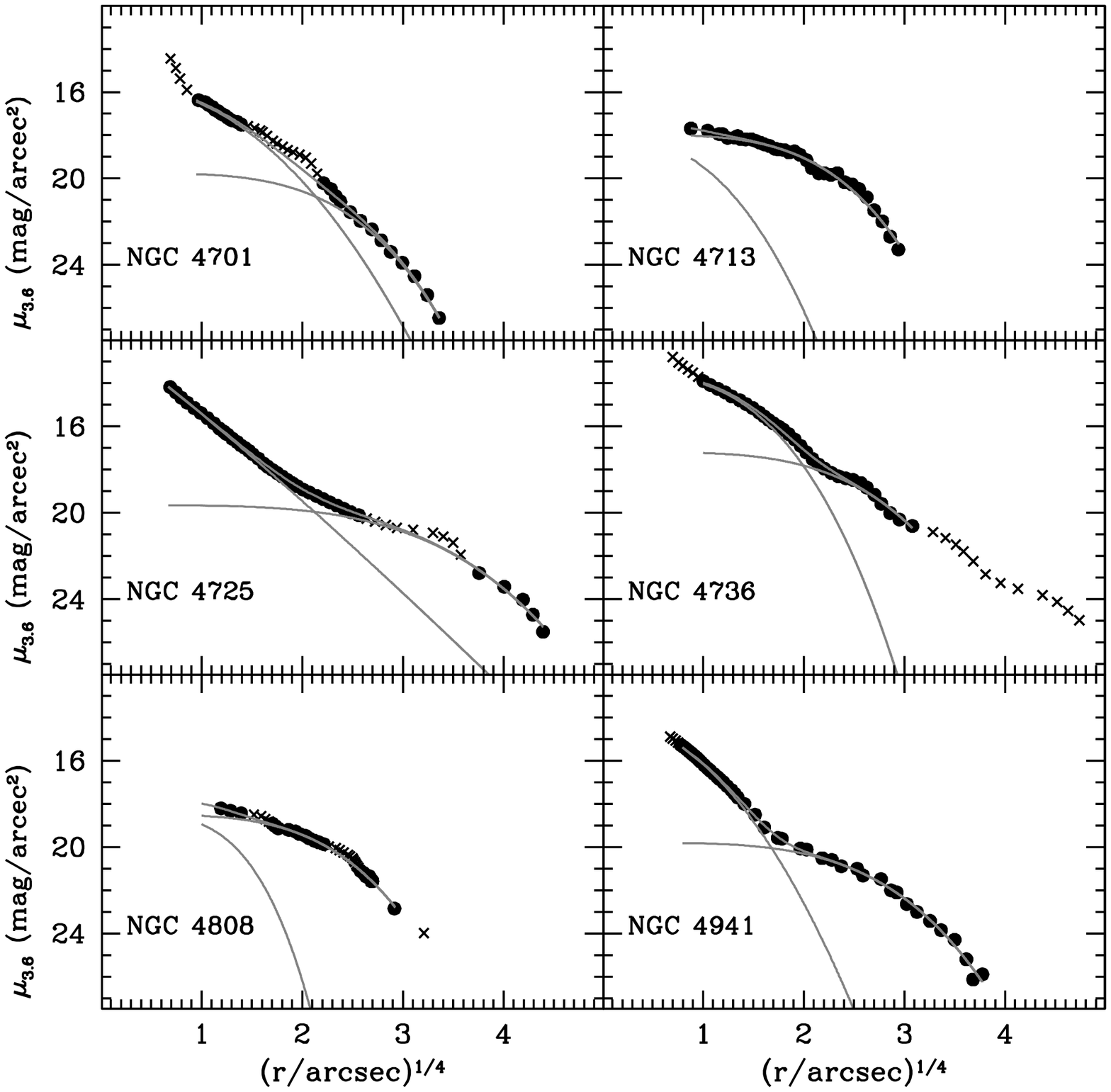}
 \caption{Galaxy surface brightness profiles and decompositions from Table 2. Solid circles are those included in the fitting, crosses are points not included in the fit. The three red lines represent the S\'ersic function, exponential disk, and sum of the two which results from bulge-disk decomposition. \label{fig:profiles}}
 \end{figure*}

 \setcounter{figure}{0}
 \begin{figure*}
 \includegraphics[width=0.99\textwidth]{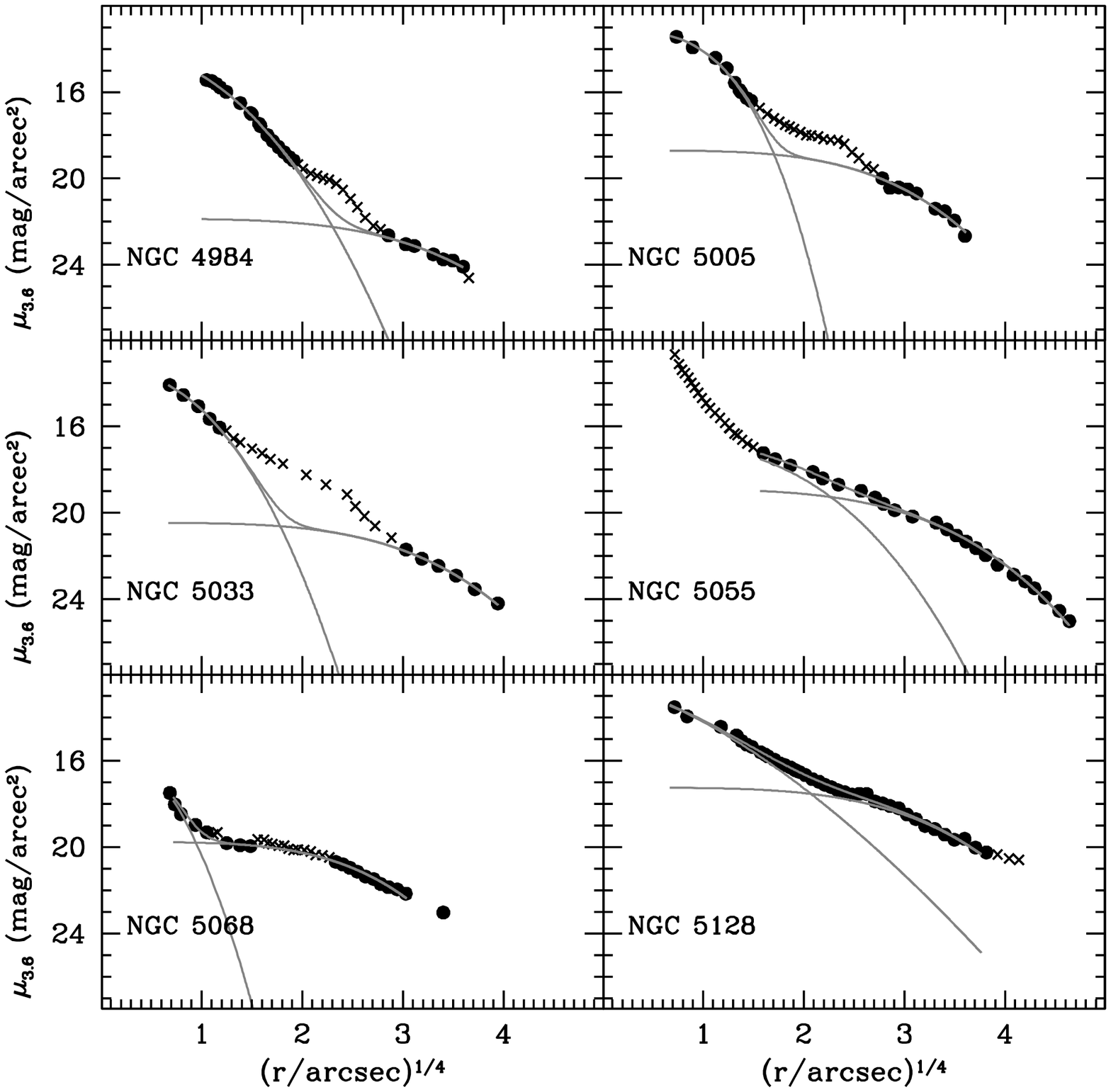}
 \caption{Galaxy surface brightness profiles and decompositions from Table 2. Solid circles are those included in the fitting, crosses are points not included in the fit. The three red lines represent the S\'ersic function, exponential disk, and sum of the two which results from bulge-disk decomposition. \label{fig:profiles}}
 \end{figure*}

 \setcounter{figure}{0}
 \begin{figure*}
 \includegraphics[width=0.99\textwidth]{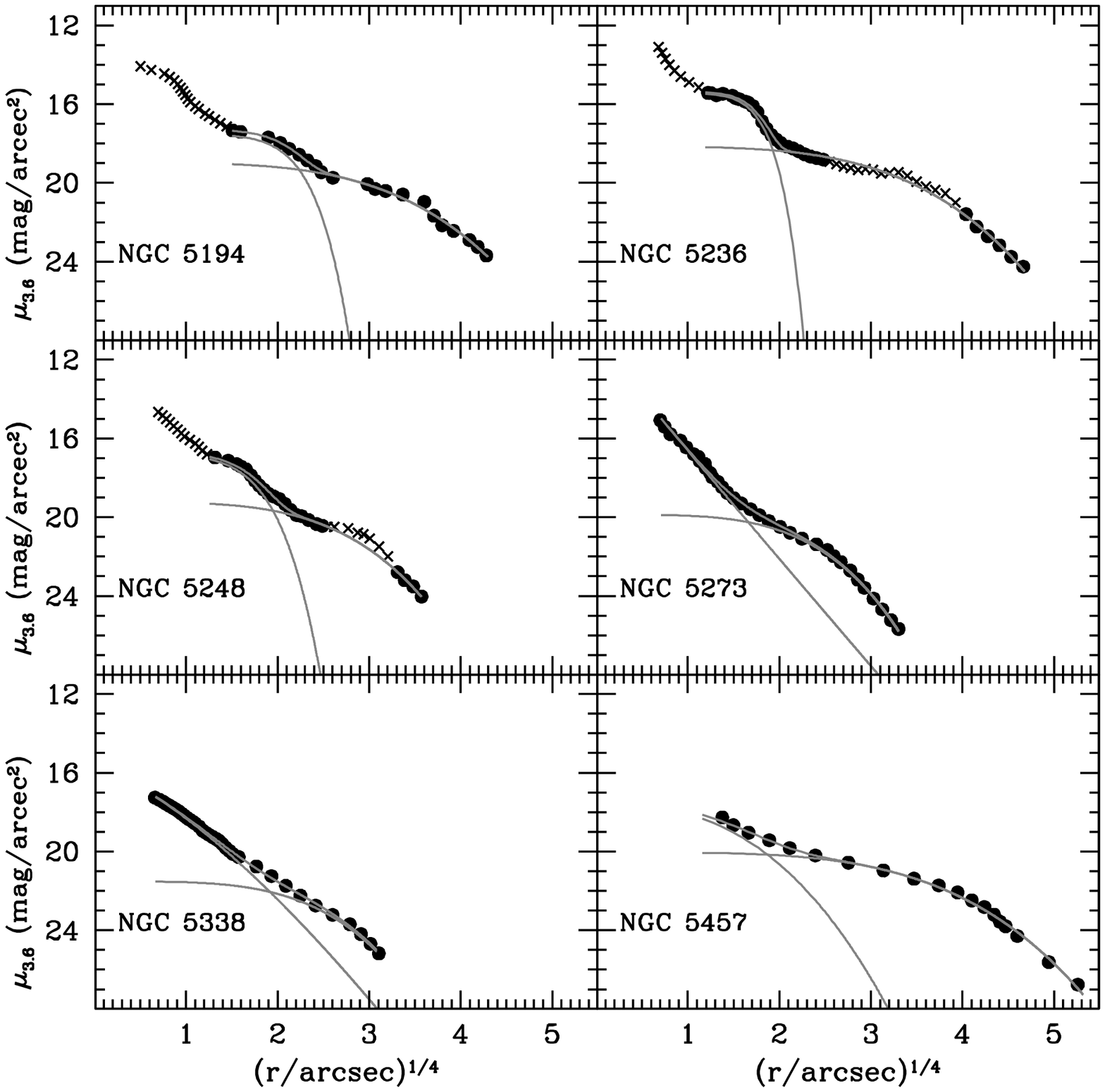}
 \caption{Galaxy surface brightness profiles and decompositions from Table 2. Solid circles are those included in the fitting, crosses are points not included in the fit. The three red lines represent the S\'ersic function, exponential disk, and sum of the two which results from bulge-disk decomposition. \label{fig:profiles}}
 \end{figure*}

 \setcounter{figure}{0}
 \begin{figure*}
 \includegraphics[width=0.99\textwidth]{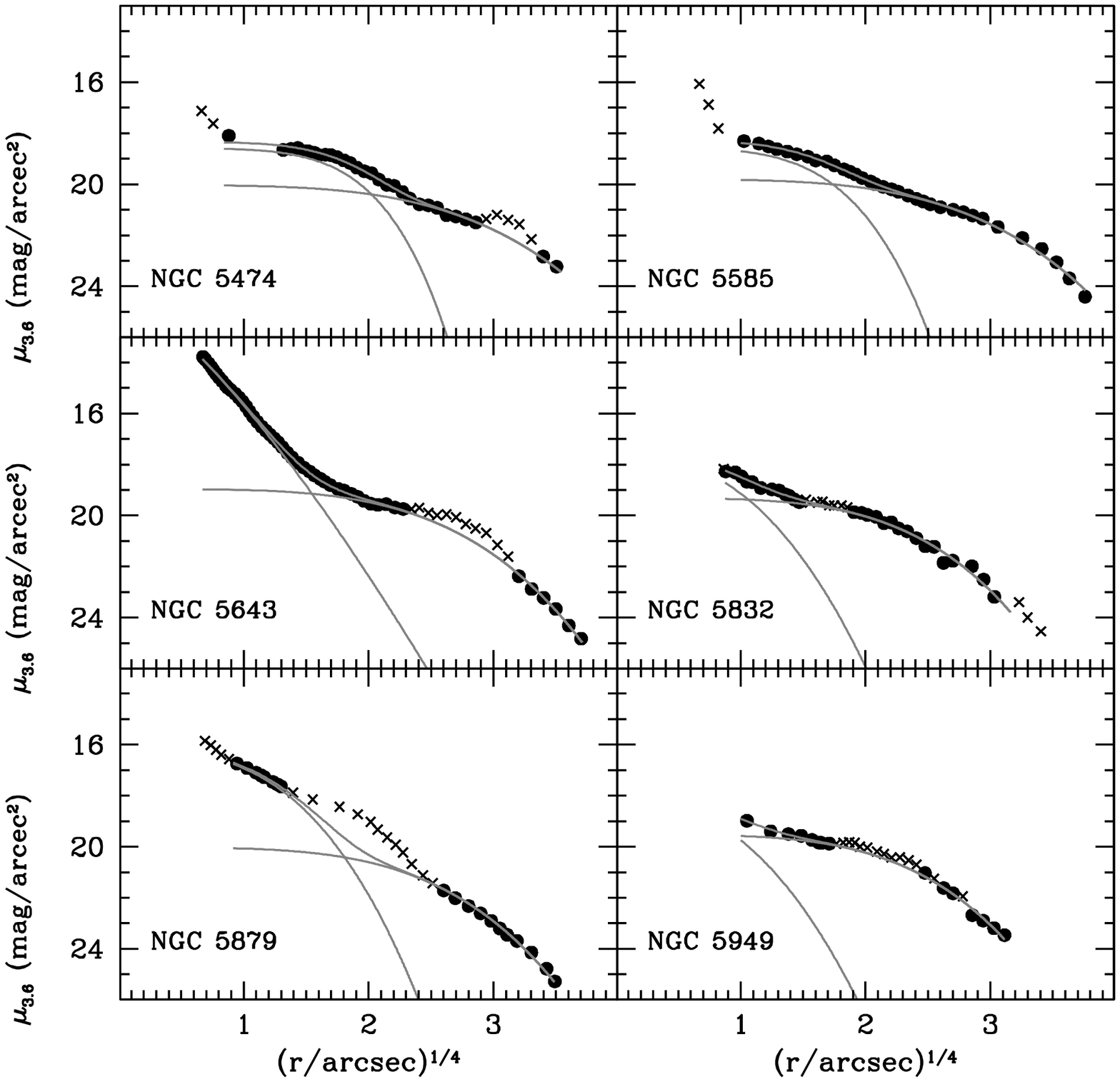}
 \caption{Galaxy surface brightness profiles and decompositions from Table 2. Solid circles are those included in the fitting, crosses are points not included in the fit. The three red lines represent the S\'ersic function, exponential disk, and sum of the two which results from bulge-disk decomposition. \label{fig:profiles}}
 \end{figure*}

 \setcounter{figure}{0}
 \begin{figure*}
 \includegraphics[width=0.99\textwidth]{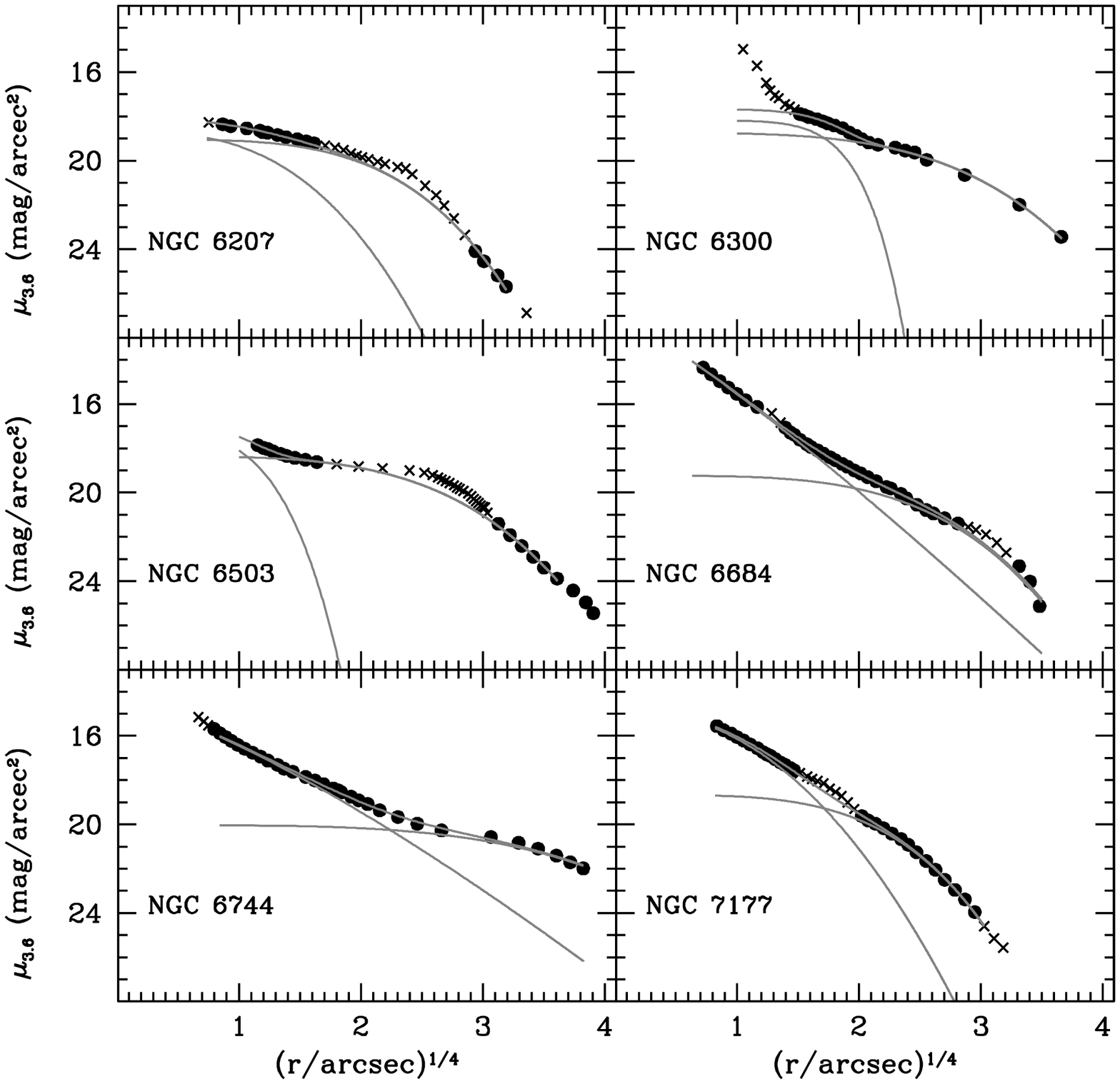}
 \caption{Galaxy surface brightness profiles and decompositions from Table 2. Solid circles are those included in the fitting, crosses are points not included in the fit. The three red lines represent the S\'ersic function, exponential disk, and sum of the two which results from bulge-disk decomposition. \label{fig:profiles}}
 \end{figure*}

 \setcounter{figure}{0}
 \begin{figure*}
 \includegraphics[width=0.99\textwidth]{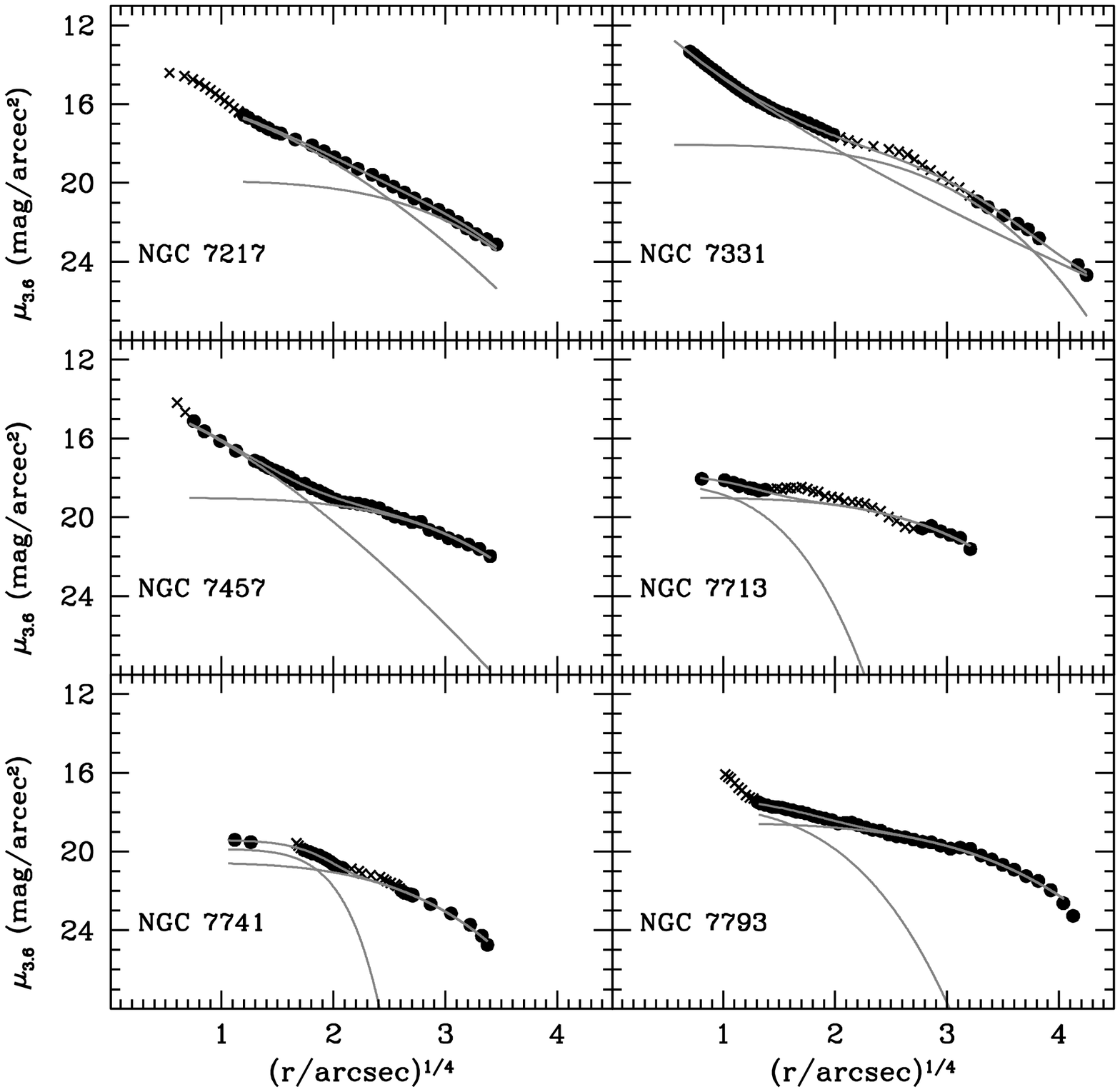}
 \caption{Galaxy surface brightness profiles and decompositions from Table 2. Solid circles are those included in the fitting, crosses are points not included in the fit. The three red lines represent the S\'ersic function, exponential disk, and sum of the two which results from bulge-disk decomposition. \label{fig:profiles}}
 \end{figure*}

\setcounter{figure}{0}
\begin{figure*}
\includegraphics[width=0.99\textwidth]{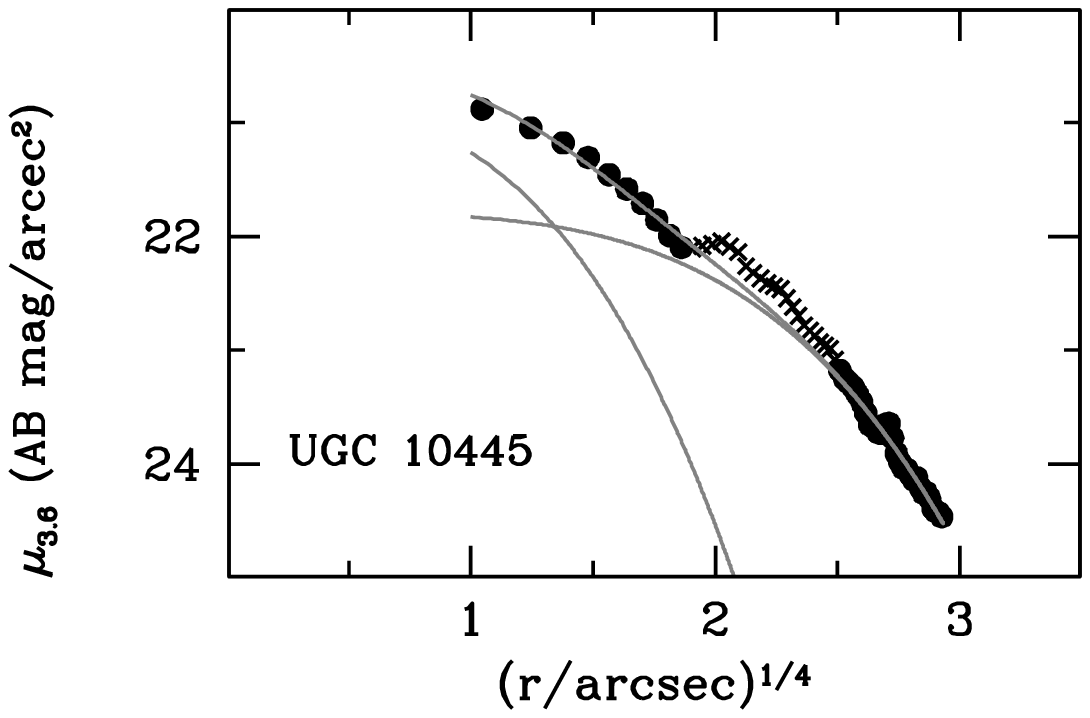}
\caption{Galaxy surface brightness profiles and decompositions from Table 2. Solid circles are those included in the fitting, crosses are points not included in the fit. The three red lines represent the S\'ersic function, exponential disk, and sum of the two which results from bulge-disk decomposition. \label{fig:profiles}}
\end{figure*}

\end{document}